\documentclass[aps,prx,reprint,superscriptaddress,nofootinbib]{revtex4-2}

\usepackage{graphicx}
\usepackage{xcolor}
\usepackage{amsmath}
\usepackage{subfigure}
 \usepackage{comment} 

\begin{document}


\title{Strong ensemble nonequivalence in systems with local constraints}


\author{Qi Zhang}
\email[]{zhang@lorentz.leidenuniv.nl}
\affiliation{Lorentz Institute for Theoretical Physics, Leiden University, Niels Bohrweg 2, 2333 CA Leiden (The Netherlands)}
\author{Diego Garlaschelli}
\email[]{garlaschelli@lorentz.leidenuniv.nl}
\affiliation{Lorentz Institute for Theoretical Physics, Leiden University, Niels Bohrweg 2, 2333 CA Leiden (The Netherlands)}
\affiliation{IMT School for Advanced Studies, Piazza San Francesco 19, 55100 Lucca (Italy)}

\date{\today}

\begin{abstract}
The asymptotic equivalence of canonical and microcanonical ensembles is a central concept in statistical physics, with important consequences for both theoretical research and practical applications. However, this property breaks down under certain circumstances. The most studied violation of ensemble equivalence requires phase transitions, in which case it has a `restricted' (i.e. confined to a certain region in parameter space) but `strong' (i.e. characterized by a difference between the entropies of the two ensembles that is of the same order as the entropies themselves) form. However, recent research on networks has shown that the presence of an extensive number of local constraints can lead to ensemble nonequivalence even in the absence of phase transitions. This occurs in a `weak' (i.e. leading to a subleading entropy difference) but remarkably `unrestricted' (i.e. valid in the entire parameter space) form.
Here we look for more general manifestations of ensemble nonequivalence in arbitrary ensembles of matrices with given margins.
These models have widespread applications in the study of spatially heterogeneous and/or temporally nonstationary systems, with consequences for the analysis of multivariate financial and neural time-series, multi-platform social activity, gene expression profiles and other Big Data.
We confirm that ensemble nonequivalence appears in `unrestricted' form throughout the entire parameter space due to the extensivity of local constraints. Surprisingly, at the same time it can also exhibit the `strong' form.
This novel, simultaneously `strong and unrestricted' form of nonequivalence is very robust and imposes a principled choice of the ensemble. We calculate the proper mathematical quantities to be used in real-world applications.
\end{abstract}


\maketitle


\section{Introduction}
In statistical physics, systems with different constraints can be described by different ensembles. For example, systems with fixed energy can be described by the \emph{microcanonical ensemble}, where all microscopic configurations have precisely the same value of the energy and are equiprobable, thereby modelling large isolated systems. In this case, the energy is treated as a `hard' constraint enforced separately on each configuration.
By contrast, systems with fixed temperature (which is the `dual' thermodynamic quantity conjugated with the energy) can be described by the \emph{canonical ensemble}, where individual microscopic configurations can have different values of the energy and are assigned different probabilities, but in such a way that the average value of the energy coincides with the one defining the corresponding microcanonical ensemble~\cite{boltzmann2012lectures}.
This ensemble represents systems that can exchange energy with their environment, and the energy is in fact treated as a `soft' constraint which is enforced only as an ensemble average.

When the size of the system is finite, the two ensembles are necessarily different. However, in the simplest and most traditional situation, the microcanonical description as a function of the energy becomes equivalent with the canonical description as a function of the temperature in the \emph{thermodynamic limit} (i.e., when the number of particles in the system tends to infinity).
This phenomenon is called \emph{ensemble equivalence} (EE) and is a basic concept in statistical mechanics as already established by Gibbs~\cite{gibbs1902elementary}.
The property of EE justifies the replacement of the (typically unfeasible) asymptotic calculations in the microcanonical ensemble with the corresponding (much easier) calculations in the canonical ensemble, i.e. to choose the ensemble based on mathematical convenience.

However, over the past decades, the breakdown of EE has been observed in various physical systems, including models of gravitation, fluid turbulence, quantum phase separation, and networks~\cite{ellis2002nonequivalent,kastner2010nonequivalence,squartini2015breaking}. When the system is under \emph{ensemble nonequivalence} (EN), the microcanonical description can no longer be replaced by the canonical description in the thermodynamic limit. In this situation, many assumptions and calculations that are based on EE in statistical mechanics do not hold anymore.
Thus checking for the breaking of EE is important for both practical applications and theoretical research.
Quantitatively, EN can be defined as a nonvanishing relative entropy density between the microcanonical and canonical probability distributions of microscopic configurations~\cite{touchette2015equivalence,squartini2015breaking}.
This is equivalent to a nonvanishing difference between the canonical and microcanonical entropy densities~\cite{squartini2017reconnecting}. Technically, this is the so-called \emph{measure-level} EN, which (under mild assumptions) has been shown to coincide with other definitions as well~\cite{touchette2015equivalence}.
Importantly, the traditional criterion for EE based on the vanishing of the relative canonical fluctuations of the contraints has been recently found to break down when the latter are local in nature and extensive in number~\cite{zhang2020ensemble}.

Indeed, for the most studied systems in statistical physics, the number of constraints defining the ensembles of interest is finite. Traditional physical examples are \emph{global} constraints such as the total energy and the total number of particles.
Non-physical examples of systems under global constraints have also been considered, e.g. networks with given total numbers of edges and triangles~\cite{hollander2018breaking}.
In order to observe the breakdown of EE in these systems, one typically needs to introduce long-range interactions implying the non-additivity of the energy and possibly associated with the onset of phase transitions~\cite{campa2009statistical} (in the example of networks, the underlying mechanism is a sort of `frustration' in the simultaneous realizability of the desired numbers of edges and triangles~\cite{hollander2018breaking}).
In this form of EN, the relative entropy between canonical and microcanonical ensembles is of the same order as the canonical entropy itself~\cite{touchette2015equivalence,hollander2018breaking}.
This is what we will refer to as a `strong' form of EN. At the same time, this form of EN is also `restricted', because it is confined to a selected (e.g. critical) region of the space of parameters.
Outside this region, EE is restored.

Recently, a new manifestation of EN has been observed in a different class of network ensembles, where a constraint is enforced on the \emph{degree} (number of links) and/or the \emph{strength} (total weight of all incident links) of each node~\cite{squartini2015breaking,garlaschelli2016ensemble,squartini2015unbiased,zhang2020ensemble}: unlike systems with global constraints, in these models the number of constraints grows linearly in the number of nodes.
This crucial difference implies that, at variance with the more `traditional' situation described above, the onset of EN in this class of models is completely unrelated to phase transitions and is instead the result of the presence of an \emph{extensive} number of \emph{local} constraints~\cite{squartini2017reconnecting,garlaschelli2018covariance,roccaverde2019breaking}.
This situation is by far less studied, because systems with local constraints are not the traditional focus of statistical physics and have attracted attention only recently as models of complex systems with built-in spatial heterogenetity~\cite{cimini2019statistical} and/or temporal non-stationarity~\cite{almog2014binary}.
In this different form of EN, the relative entropy between microcanonical and canonical ensembles is, at least for the cases studied so far, of lower order (i.e. subleading) with respect to the canonical entropy.
For this reason, we may refer to this situation as a `weak' (i.e. weaker than the one found in the presence of phase transitions) form of EN.
However, this form of EN is `unrestricted', precisely because it is not confined to specific values of the control parameters and holds in the entire parameter space.
Rather than a property of a phase (or a phase boundary), in this case EN appears to be an intrinsic property of the system itself.
In these models, no parameter value can restore EE.

The above results indicate that, so far, EN has manifested itself either in a `strong but restricted' form (under a finite number of global constraints, but in presence of phase transitions), or in a `unrestricted but weak' form (under an extensive number of local constraints, but without phase transitions).
Clearly, a number of questions remain open. How general is the manifestation of EN under local constraints, both in terms of the underlying mechanism and in terms of the strength of the resulting effect?
Besides networks, can the breaking of EE be observed in additional systems characterized by local constraints? If so, do these systems necessarily exhibit only the weak form of EN, or can the strong form appear as well?
Finally, is there a (possibly modified) way to exploit the canonical ensemble in order to bypass the challenge of unfeasible microcanonical calculations \emph{even when EE breaks down}, i.e. even when the two ensembles can no longer be treated interchangeably according to mathematical convenience?

In this paper, we will address these problems by exploring the effects of the presence of an extensive number of local constraints on more general ensembles than the ones that have been considered so far to model random networks with given node degrees~\cite{squartini2015breaking,garlaschelli2016ensemble,garlaschelli2018covariance,roccaverde2019is}. 
In particular, we consider the general setting where each of the $n$ units of the system has a number $m$ of `state variables' (or `degrees of freedom'), and where constraints are defined as sums over these state variables.
Surprisingly, besides confirming the onset of an unrestricted form of EN in the thermodynamic limit where $n$ diverges, we also find its simultaneous manifestation in strong form.
This happens when each element of the system retains only a finite number $m$ of degrees of freedom in the thermodynamic limit. For brevity, we will denote this situation as the `strong and unrestricted' form of EN.
To the best of our knowledge, this finding provides the first evidence that EN, \emph{even in its strong form}, does not need phase transitions and can appear in the entire parameter space as an intrinsic property of the system, if the latter is subject to an extensive number of local constraints.
This simultaneously `strong and unrestricted' form of EN is the most robust among the ones studied so far. Spatial heterogeneity and temporal non-stationarity are simple candidate mechanisms that can lead to this phenomenon.

To emphasize the general and important consequences of this form of EN for a diverse range of practical applications, we consider generic ensembles of random matrices with fixed margins. These ensembles, which include matrices with $0/1$ (or equivalently $\pm1$) and non-negative integer entries subject to global or local constraints, arise for instance in studies of multi-cell gene expression profiles~\cite{semrau2017dynamics}, multiplex (online) social activity~\cite{socialmultiplex}, multi-channel communication systems~\cite{cover2012elements}, complex networks~\cite{squartini2017maximum}, and multivariate time series in finance~\cite{almog2014binary}, neuroscience~\cite{almog2019uncovering} or other disciplines.
Our results imply that, in many practical situations, the assumption of EE is incorrect and leads to mathematically wrong conclusions.
For the benefit of the aforementioned applications, we compensate for the `disconnection' between the two ensembles by calculating explicitly the correct canonical and microcanonical quantities of interest via a generalized relationship that is either analytically computable or asymptotically determined by the covariance matrix of the constraints in the canonical ensemble.
These calculations represent a practical tool for properly dealing with the consequences of EN in all real-world situations.

\section{General formalism}

\subsection{Matrix ensembles}
A discrete $n\times m$ matrix ensemble is a  set  $\mathcal{G}$ of available configurations for an $n\times m$ integer-valued matrix $\mathbf{G}$, endowed with a suitably chosen probability distribution $P(\mathbf{G})$ over such configurations, such that $\sum_{\mathbf{G}\in\mathcal{G}}P(\mathbf{G})=1$.
An entry of the matrix is denoted by $g_{ij}$ (with $1\leq i\leq n$, $1\leq j\leq m$).
We distinguish two main cases, the \emph{binary} case where $g_{ij}$ takes one of the two values $\{0,1\}$ and the \emph{weighted} case where $g_{ij}$ takes a value in the set $\{0,1,2,\dots\}$ of non-negative integer values.
The number $n$ of rows in each matrix represents the number of elements (i.e. the \emph{size}) of the system being modelled.
The number $m$ of columns represents instead the number of state variables, or degrees of freedom, for each element.

In general, each matrix $\mathbf{G}$ can represent one of the possible states of a (large) real-world system.
For instance, $\mathbf{G}$ may represent the realization of a multivariate time series, where $n$ is the number of units (e.g. brain regions, financial stocks, etc.) emitting signals, and $m$ is the number of time steps during which the signals are recorded.
$\mathbf{G}$ may also represent a multi-cell array of gene expression profiles, where $n$ is the number of cells and $m$ the number of genes for which expression levels are being measured.
Similarly, $\mathbf{G}$ may represent the state of a multi-channel communication systems, where $n$ is the length of the sequences being transmitted from sender to receiver (in information theory, such length defines the `size' of the communication process) and $m$ is the number of channels.
Finally, $\mathbf{G}$ may represent the adjacency matrix of a \emph{bipartite graph}, where $n$ is the number of nodes in the layer of interest (e.g. people in a co-affiliation network), while $m$ is the number of possible dimensions where nodes can co-occur (e.g. work, family, sport, friendship, etc.). In the special case $m=n$, the network can also be interpreted as a (binary or weighted) \emph{directed unipartite graph}, i.e. one where there is a single set of $n$ nodes  that can be linked to each other via directed edges (note that, by contrast, undirected unipartite graphs are associated with a symmetric adjacency matrix, a property that we do not enforce in this paper; the nonequivalence of ensembles of binary or weighted undirected graphs with given constraints has been studied previously in~\cite{squartini2015breaking,garlaschelli2018covariance,roccaverde2019breaking,zhang2020ensemble}).

\begin{figure*}
	\centering
	\includegraphics[width=14cm]{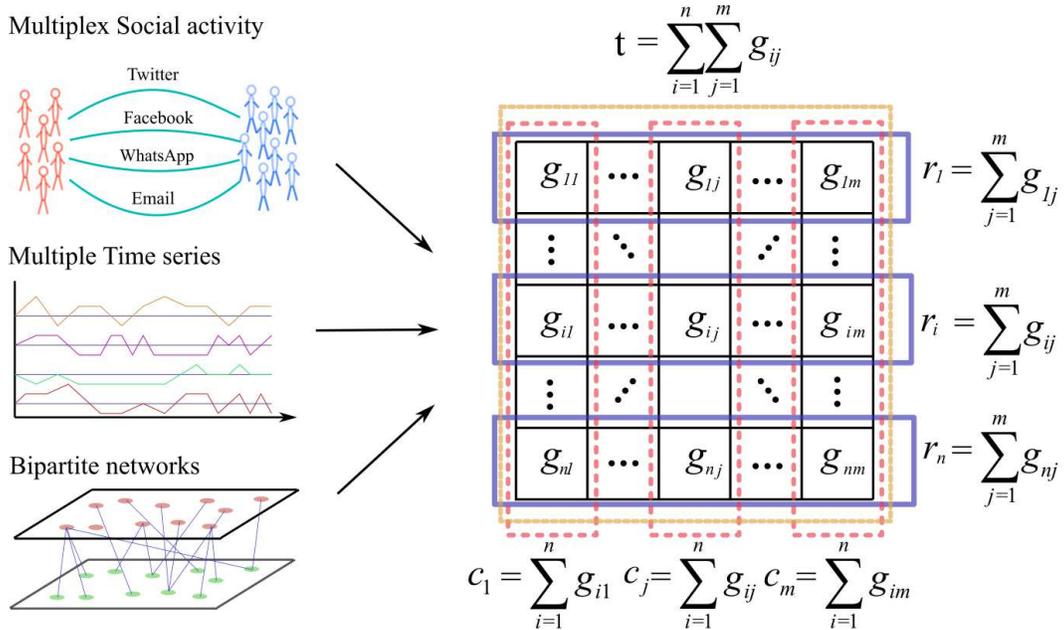}
	\caption{Schematic illustration of how the state of different real-world systems with $n$ elements and $m$ degrees of freedom can be represented in an $n\times m$ matrix $\mathbf{G}$.
The collection of all possible states of the system is the set of all such matrices. Typically, real-world systems have a strong heterogeneity or nonstationarity. This empirical fact implies that their possible states are not sufficiently characterized by the knowledge of a single \emph{global} constraint ($t$: solid orange box). More informative ensembles can be constructed by specifying one- ($\vec{r}$: solid blue boxes) or two-sided ($\vec{r},\vec{c}$: solid blue and dashed red boxes) \emph{local} constraints.}\label{examplex}
\end{figure*}

\subsection{Global and local constraints}
In each of the examples mentioned above, the `true' microscopic configuration (or \emph{microstate}) of the system can be uniquely represented by a specific `empirical' matrix $\mathbf{G}^*$ in the set $\mathcal{G}$ of all possible states.
A schematic illustration is shown in Fig.~\ref{examplex}.
As ordinary in statistical mechanics, when the size of the system is large one no longer focuses on the specific microstate $\mathbf{G}^*$ (which becomes not empirically accessible), but rather on the \emph{macrostate} defined by a collection of microscopic configurations compatible with the empirical value $\vec{C}^*\equiv\vec{C}(\mathbf{G}^*)$ of a certain observable quantify $\vec{C}(\mathbf{G})$ playing the role of a constraint.
The choice of $\vec{C}(\mathbf{G})$ determines the probability distribution $P(\mathbf{G})$ over $\mathcal{G}$ conditional on our knowledge of $\vec{C}^*$.
In other words, it determines how our estimate of the microstate of the system concentrates around the compatible configurations once we observe $\vec{C}^*$. Intuitively, before anything is obserbed, $P(\mathbf{G})$ is uniform on $\mathcal{G}$.

In ordinary statistical physics, the quantity $\vec{C}(\mathbf{G})$ is typically scalar (e.g. the total energy) or at most low-dimensional (e.g. a vector containing the total energy and the total number of particles), reflecting a few global conservation laws applying to a large homogeneous system at thermodynamic equilibrium.
However, in models of complex systems $\vec{C}(\mathbf{G})$ can be high-dimensional, as it may encode a large number of local constraints reflecting separate conservation laws imposed by spatial heterogeneity and/or temporal non-stationarity.
For instance, if $\mathbf{G}^*$ is the observed configuration of a complex network with $n$ nodes (i.e. the empirical $n\times n$ adjacency matrix), it is well known that the knowledge of purely global properties such as the overall number of links is insufficient in order to produce a statistical ensemble of networks with properties similar to those found in $\mathbf{G}^*$. Indeed, enforcing only the total number of links produces the popular Erd\H{o}s-R\'enyi random graph model, whose topological properties are way too homogenous as compared with those of real-world networks.
By contrast, if the number of links of each node is enforced separately (as in the so-called \emph{configuration model}), the resulting ensemble of graphs is found to successfully replicate many higher-order empirical topological properties~\cite{squartini2017maximum}.
As another example, if $\mathbf{G}^*$ represents a set of synchronous time series produced by the $n$ components of a non-stationary system observed over $m$ time steps, then the statistical properties of these time series will change over time. As a result, overall time-independent constraints will not be enough in order to produce ensembles of multivariate time series with properties close to those of $\mathbf{G}^*$, and time-dependent (i.e. local in time) constraints will in general be needed~\cite{almog2014binary}.

In our setting, we consider the general case where the distribution $P(\mathbf{G})$ defining the (binary or weighted) matrix ensemble is induced by a $K$-dimensional vector $\vec{C}(\mathbf{G})$ of \emph{constraints} imposed on the matrices.
We will assume that the $K$ constraints are all non-redundant, e.g. they are not trivial copies or linear combinations of each other~\cite{squartini2017reconnecting}.
We will consider both \emph{global} and \emph{local} constraints.
As global constraint we will consider the scalar quantity $t(\mathbf{G})$ defined as the total sum of all the entries of the matrix $\mathbf{G}$, i.e. $t(\mathbf{G})=\sum_{i=1}^{n}\sum_{j=1}^{m}g_{ij}$. The number of constraints in this case is $K=1$ and the `empirical' value of $t$ will be denoted as $t^*\equiv t(\mathbf{G}^*)$.
As local constraints, we will consider two possibilities: \emph{one-sided} local constraints and \emph{two-sided} local constraints.
A one-sided local constraint is the $n$-dimensional vector $\vec{r}(\mathbf{G})$ where the entry $r_i(\mathbf{G})=\sum_{j=1}^{m}g_{ij}$ $(i=1,n)$ represents the sum of the entries of the matrix $\mathbf{G}$ along its $i$-th row.
The number of constraints is in this case $K=n$ and the empirical value of $\vec{r}$ will be denoted as $\vec{r}^*\equiv\vec{r}(\mathbf{G}^*)$.
A two-sided local contraint is a pair of vectors $(\vec{r}(\mathbf{G}), \vec{c}(\mathbf{G}))$, where $\vec{r}(\mathbf{G})$ is still the $n$-dimensional vector representing the $n$ row sums of $\mathbf{G}$, while $\vec{c}(\mathbf{G})$ is the $m$-dimensional vector representing the $m$ column sums of $\mathbf{G}$, i.e. where each entry $c_j(\mathbf{G})=\sum_{i=1}^{n}g_{ij}$ ($j=1,m$) is the sum of the  entries of $\mathbf{G}$ along its $j$-th column.
The number of constraints is therefore $K=n+m$ and the empirical value of the pair $(\vec{r},\vec{c})$ will be denoted as $(\vec{r}^*,\vec{c}^*)\equiv(\vec{r}(\mathbf{G}^*),\vec{c}(\mathbf{G}^*))$.
A visual illustration of these constraints for possible data structures of practical interest is shown in Fig.~\ref{examplex}.

Purely global constraints lead to completely homogeneous expectations for the entries of the matrices in the ensemble. This result follows intuitively from symmetry arguments, and will be confirmed explicitly in the specific cases considered later. By contrast, local constraints lead to different expectations for entries in different rows and/or columns. Since, as we mentioned above, real-world complex systems are generally very heterogeneous in space and/or time, the only models that can capture the main features of such systems are those constructed from (one- or two-sided) local constraints. This is very important because, as we will show, it is precisely in presence of local constraints (of either type) that the property of EE breaks down.
This result implies that spatial heterogeneity and/or temporal non-stationarity might be natural origins for the breaking of EE.

\subsection{Soft constraints: canonical ensemble}
Any constraint, whether global or local, can be enforced either as a soft constraint (canonical ensemble) or as a hard constraint (microcanonical ensemble).
We start with the case of soft constraints, i.e. when one imposes that the ensemble average
\begin{equation}
\langle \vec{C}\rangle\equiv \sum_{\mathbf{G}\in\mathcal{G}}P(\mathbf{G}) \vec{C}(\mathbf{G})
\end{equation}
is fixed to a specific value $\vec{C}^*$.

The \emph{functional form} of the resulting canonical probability $P_{\textrm{can}}$ over $\mathcal{G}$ is found by maximizing Shannon's entropy functional
\begin{equation}
S_n[P]\equiv-\sum_{\mathbf{G}\in\mathcal{G}}P(\mathbf{G})\ln P(\mathbf{G})
\label{eq:shannon}
\end{equation}
(where the subscript $n$ indicates that the entropy is calculated for given $n$), subject to the condition $\langle\vec{C}\rangle=\vec{C}^*$.
The result~\cite{jaynes1957information} of this constrained maximization problem is the parametric solution
\begin{equation}\label{Can-prob}
P_{\textrm{can}}(\mathbf{G}|\vec{\theta})=\frac{e^{-H(\mathbf{G},\vec{\theta})}}{Z(\vec{\theta})},
\end{equation}
where $\vec{\theta}$ is a vector of Lagrange multipliers coupled to the constraint $\vec{C}$, the \emph{Hamiltonian} $H(\mathbf{G},\vec{\theta})=\vec{\theta}\cdot\vec{C}(\mathbf{G})$ is a linear combination of the constraints, and the \emph{partition function} $Z(\vec{\theta})=\sum_{\mathbf{G}\in\mathcal{G}}e^{-H(\mathbf{G},\vec{\theta})}$ is the normalization constant.

The \emph{numerical values} of the canonical probability are found by setting
\begin{equation}
P^*_{\textrm{can}}(\mathbf{G})\equiv P_{\textrm{can}}(\mathbf{G}|\vec{\theta}^*)
\label{eq:Pcan*}
\end{equation}
where $\vec{\theta}^*$ is the unique parameter value that realizes the `soft' constraint
\begin{equation}
 \langle\vec{C}\rangle_{\vec{\theta}^*}=\vec{C}^*
\label{eq:theta*}
\end{equation}
where the symbol $\langle\cdot\rangle_{\vec{\theta}}$ denotes an ensemble average with respect to $P_{{\textrm{can}}}(\mathbf{G}|{\vec{\theta}})$, i.e.
\begin{equation}
\langle\vec{C}\rangle_{\vec{\theta}}=\sum_{\mathbf{G}\in\mathcal{G}}P_{\textrm{can}}(\mathbf{G}|\vec{\theta})\, \vec{C}(\mathbf{G}).
\label{eq:angle}
\end{equation}
Equivalently, the unique value $\vec{\theta}^*$ is the one that maximizes the log-likelihood function
\begin{equation}
\mathcal{L}^*(\vec{\theta})\equiv\ln P_{\textrm{can}}(\mathbf{G}^*|\vec{\theta}),
\label{eq:likelihood}
\end{equation}
where $\mathbf{G}^*$ is the empirical configuration, or equivalently any configuration that realizes the empirical constraint exactly, i.e. such that $\vec{C}(\mathbf{G}^*)=\vec{C}^*$~\cite{jaynes1957information}.
The uniqueness of $\vec{\theta}^*$ follows whenever $\mathcal{L}^*(\vec{\theta})$ can be differentiated at least twice~\cite{squartini2017reconnecting}, as we confirm below for all the models considered in this paper.  

Inserting Eq.~\eqref{eq:Pcan*} into Eq.~\eqref{eq:shannon}, we obtain the value of the canonical entropy
\begin{equation}
S_{\textrm{can}}^*\equiv S_n[P^*_{\textrm{can}}] =-\mathcal{L}^*(\vec{\theta}^*)=-\ln P^*_{\textrm{can}}(\mathbf{G}^*)
\label{eq:S=-L}
\end{equation}
where we have omitted the dependence on $n$ to simplify the notation. The last equality is very useful, as it show that $S_{\textrm{can}}^*$ can be calculated by simply evaluating $P^*_{\textrm{can}}(\mathbf{G})$ on the single configuration $\mathbf{G}^*$~\cite{squartini2017reconnecting}.

\subsection{Hard constraints: microcanonical ensemble}
In the case of hard constraints, one requires that each individual configuration realizes the value $\vec{C}^*$.
This means that the `soft' constraint in Eq.~\eqref{eq:theta*} is replaced by the much stricter constraint
\begin{equation}
\vec{C}(\mathbf{G})=\vec{C}^*
\label{eq:hard}
\end{equation}
for each allowed configuration $\mathbf{G}$.
The microcanonical probability $P_{\textrm{mic}}$ is found by enforcing this stronger requirement, while still maximizing the entropy $S_n[P]$ defined in Eq.~\eqref{eq:shannon}. The result is the uniform distribution
\begin{equation}
P^*_{\textrm{mic}}(\mathbf{G})=\left\{\begin{array}{ll}\Omega_{\vec{C}^* }^{-1}&\vec{C}(\mathbf{G})=\vec{C}^*\\0&\vec{C}(\mathbf{G})\ne\vec{C}^*\end{array}\right.,
\label{eq:micro}
\end{equation}
where $\Omega_{\vec{C}^* }$ is the number of configurations in $\mathcal{G}$ realizing the `hard' constraint in Eq.~\eqref{eq:hard}.
The corresponding microcanonical entropy is obtained by inserting Eq.~\eqref{eq:micro} into Eq.~\eqref{eq:shannon}:
\begin{equation}
S^*_{\textrm{mic}}\equiv S_n[P_{\textrm{mic}}^*]=\ln \Omega_{\vec{C}^* },
\label{eq:Smic}
\end{equation}
which is also known as Boltzmann entropy.

Crucially, in order to define the microcanonical ensemble it is necessary that $\Omega_{\vec{C}^* }>0$, i.e. that there is at least one configuration realizing the constraint. 
In other words, the value of $\vec{C}^*$ should be realizable in (at least) one single configuration, and not only as an ensemble value.
This requirement is not strictly necessary for the canonical ensemble (even though our interpretation of $\vec{C}^*$ as the `empirical' value makes the requirement always natural). 
In any case, since in this paper we are going to study the (non)equivalence between the two ensembles, we need \emph{both} of them to be well defined in order to be compared, for a given value of $\vec{C}^*$.
Therefore we are going to assume that \emph{the value of $\vec{C}^*$, irrespective of the ensemble considered, is always realizable by at least one configuration, i.e. such that $\Omega_{\vec{C}^* }>0$.} 

Notably, calculating $\Omega_{\vec{C}^* }$ (especially in presence of many constraints and because of the discrete nature of the problem of interest for us) can be a complicated enumeration problem.
Therefore the microcanonical ensemble is typically much more difficult to deal with mathematically than the canonical ensemble.
For this reason, if the property of EE holds, one prefers to operate in the canonical ensemble and work out its asymptotics in the limit of large system size, trusting that the result would return the correct asymptotics for the microcanonical ensemble as well.
The above approach is at the core of many stardard calculations in statistical mechanics textbooks, where the property of EE is typically assumed to hold in general (at least in absence of phase transitions and long-range interactions).
However, when EE breaks down, this approach will lead to mathematically incorrect results.
We will study this problem in detail, for the ensembles considered, in the rest of the paper. To do so, we first need to define what we mean by \emph{thermodynamic limit}.

\subsection{The thermodynamic limit\label{sec:thermo}}
We will consider the thermodynamic limit defined as $n\to+\infty$, i.e. when the size of the system diverges.
However, the limit is not completely defined until we also specify how both $m$ and $\vec{C}^*$ behave as $n$ grows.

First of all, we consider two possibilities for the behaviour of $m$ as $n$ diverges:
\begin{itemize}
\item
$m$ remains finite as $n\to\infty$: in this case, we have $m=O(1)$ where $O(x)$ indicates a quantity that has a finite limit if divided by $x$ as $n\to+\infty$, i.e. $O(x)$ is asymptotically of the same leading order\footnote{Note that the `big-O' notation we use here is not always used with the same meaning throughout the literature: some authors prefer the `big-$\Theta$' notation $\Theta(x)$ to indicate a quantity that is of the same leading order as the argument $x$, and the `big-O' notation to indicate only an upper bound for it.} as $x$;
\item
$m$ diverges as $n\to\infty$: in this case, for simplicity and realism we assume that $m$ can diverge at most as fast as $n$, i.e. $m$ is at most $O(n)$; it is indeed difficult to imagine a physical situation where the number $m$ of state variables characterizing each of the $n$ units grows faster than the number $n$ of units themselves.
\end{itemize}
In simple words, the above assumptions mean that the number of state variables should be either (asymptotically) independent of the number $n$ of units being added to the system (as in the case of `intrinsic' observations, e.g. for multivariate time series) or at most proportional to $n$ (as in the case of `relational' observations, e.g. for networks).
We will show that these two situations lead to very different asymptotic results in terms of the strength of EN.
Importantly, the requirement that $m$ grows at most proportionally to $n$ implies that the number $K$ of both one-sided ($K=n$) and two-sided ($K=n+m$) local constraints is always \emph{extensive}, i.e. $K=O(n)$ which grows linearly in the size $n$ of the system, irrespective of the behaviour of $m$.

A separate, equally important consideration concerns the scaling of the value of the constraint $\vec{C}^*$ in the thermodynamic limit $n\to+\infty$. 
Also here, we distinguish between two situations that we denote as the \emph{sparse} and the \emph{dense} regimes.
\begin{itemize}
\item
We define \emph{sparse matrices} those for which each of the $m$ column sums (irrespective of whether such sums are chosen as constraints) is \emph{finite} in the thermodynamic limit, i.e. $c^*_j=O(1)$ ($j=1,\dots, m$). This implies that, in the canonical ensemble, the expected value of any entry $g_{ij}$ of the matrix $\mathbf{G}$ is on average $O(1/n)$; correspondingly, in the microcanonical ensemble the allowed matrices are dominated by zeroes (whence the name `sparse matrices'). 
Note that for the row sums one has $r^*_i=O(m/n)$ for all $i$. If $m$ grew slower than $n$, these row sums would vanish as $n\to+\infty$, which would imply that asympotically no microcanonical configuration would realize the local constraints. Since we require $\Omega_{\vec{C}^*}>0$ (see above), this means that in the sparse case we necessarily need $m=O(n)$. Consequently, $r^*_i=O(1)$, $r^*_i/m=O(1/n)$ ($i=1,\dots,n$), $t^*=O(n)$, and $t^*/mn=O(1/n)$.
\item
By contrast, we define \emph{dense matrices} those for which  each of the $m$ column sums (again, irrespective of whether they are chosen as constraints) \emph{diverges} proportionally to $n$ in the thermodynamic limit, i.e. $c^*_j=O(n)$ ($j=1,\dots, m$). In the canonical ensemble, the expected value of $g_{ij}$ is therefore $O(1)$, which makes the allowed matrices in the corresponding microcanonical ensemble `dense'. 
The row sums are now $r^*_i=O(m)$ and we have $r^*_i/m=O(1)$ (for all $i$) and $t^*/mn=O(1)$. 
In this case, we consider $m$ as either remaining finite, in which case we have $m=O(1)$, $r^*_i=O(1)$ ($i=1,\dots,n$) and $t^*=O(n)$, or diverging proportionally to $n$ (see above), in which case we have $m=O(n)$, $r^*_i=O(n)$ ($i=1,\dots,n$), and $t^*=O(n^2)$.
\item
Note that, in principle, in the weighted case we may even consider a sort of \emph{superdense} regime where some of the individual entries of the matrix diverge in the thermodynamic limit. This possibility is related to a Bose-Einstein condensation concentrating a finite fraction of the total weight $t^*$ of the matrix in a finite number of entries~\cite{zhang2020ensemble}.
However, we will not consider this extreme case here for simplicity, as it would not arise in most real-world applications.
\end{itemize}
Combined with the scaling of $\vec{C}^*$, the behaviour of $m$ as a function of $n$ in the thermodynamic limit can determine different asymptotic regimes, and in particular lead to the  weak or strong form of EN. The strong form, for the cases considered below, turns out to be possible in the regime where the matrices are \emph{dense} and $m$ is \emph{finite} as $n\to+\infty$.

\subsection{Ensemble (non)equivalence\label{sec:ensemble(non)equivalence}}
There are various ways to mathematically define the property of ensemble (non)equivalence. These include the notions of EE in the \emph{thermodynamic}, \emph{macrostate} and \emph{measure} sense which, under mild assumptions, can be proven to be equivalent~\cite{touchette2015equivalence}.
We will adopt the definition in the measure sense, which states that the ensembles are equivalent if the relative entropy
\begin{equation}\label{relative-entropy} S_n[P^*_{\textrm{mic}}||P^*_{\textrm{can}}]\equiv\sum_{\mathbf{G}\in\mathcal{G}}P^*_{\textrm{mic}}(\mathbf{G})\ln\frac{P^*_{\textrm{mic}}(\mathbf{G})}{P^*_{\textrm{can}}(\mathbf{G})}
\end{equation}
(which is the Kullback-Leibler divergence for given $n$ between the microcanonical and canonical entropies and is guaranteed to be non-negative~\cite{kullback1951information}), when rescaled by $n$, vanishes in the thermodynamic limit~\cite{touchette2015equivalence}, i.e. if the \emph{specific relative entropy} vanishes:
\begin{equation}
s[P^*_{\textrm{mic}}||P^*_{\textrm{can}}]\equiv\lim_{n\to+\infty}\frac{S_n[P^*_{\textrm{mic}}||P^*_{\textrm{can}}]}{n}=0
\label{eq:s}
\end{equation}
or equivalently
\begin{equation}
S_n[P^*_{\textrm{mic}}||P^*_{\textrm{can}}]=o(n),
\label{eq:KLon}
\end{equation}
where $o(x)$ indicates a quantity that goes to zero when divided by $x$ as $n\to+\infty$.

Importantly, it can be shown~\cite{squartini2015breaking,squartini2017reconnecting} that
\begin{equation}\label{relative-entropy=difference} S_n[P^*_{\textrm{mic}}||P^*_{\textrm{can}}]=\ln \frac{P^*_{\textrm{mic}}(\mathbf{G}^*)}{P^*_{\textrm{can}}(\mathbf{G}^*)}=S^*_{\textrm{can}}-S^*_{\textrm{mic}}.
\end{equation}
The inequality $S_n[P^*_{\textrm{mic}}||P^*_{\textrm{can}}]\ge0$, which is a general property of the relative entropy, implies therefore $S^*_{\textrm{can}}\ge S^*_{\textrm{mic}}$ and indicates the presence of an `extra entropy' in the canonical ensemble.
This extra entropy is due to the fact that, while in the microcanonical ensemble the constraint $\vec{C}$ is a deterministic quantity fixed to the value $\vec{C}^*$ through the hard constraint introduced in Eq.~\eqref{eq:hard}, in the canonical ensemble it is a random variable fluctuating around the expected value $\vec{C}^*$ as dictated by the soft constraint defined in Eq.~\eqref{eq:theta*}. With respect to the canonical ensemble, the hardness of the constraint in the microcanonical ensemble implies additional dependencies (i.e. smaller entropy) among the entries of $\mathbf{G}$.
The definition of EE in Eq.~\eqref{eq:s} states that, if the extra entropy $S_n[P^*_{\textrm{mic}}||P^*_{\textrm{can}}]$, once divided by $n$, vanishes in the thermodynamic limit, then the ensembles are equivalent. 

From Eq.~\eqref{relative-entropy=difference} it is clear that Eq.~\eqref{eq:s} is equivalent to the condition
\begin{equation}
\lim_{n\to+\infty}\frac{S_n[P_{\textrm{can}}^*]-S_n[P_{\textrm{mic}}^*]}{n}=0
\label{eq:assumption}
\end{equation}
or in other words to the asymptotic (for $n$ large) relation
\begin{equation}
S^*_{\textrm{mic}}=S^*_{\textrm{can}}-o(n).
\end{equation}
This implies
\begin{equation}
\Omega_{\vec{C}^* }=e^{S^*_{\textrm{can}}- o(n)},
\label{eq:approx}
\end{equation}
i.e. $\Omega_{\vec{C}^* }$ is approximated by $e^{S^*_{\textrm{can}}}$ up to a subexponential (in $n$) correction factor.
The above asymptotics is used in statistical mechanics textbooks whenever the property of EE is believed to hold, i.e. in absence of phase transitions or long-range interactions.
When EE does not hold, Eq.~\eqref{eq:approx} breaks down. In this case, the extra entropy in the canonical ensemble grows at least as fast as $n$.
Recent research has shown that this breakdown can happen even in complete absence of phase transitions, hence also in situations where EE was typically believed to hold. 
Here we are going to show that, additionally, the breakdown can occur with previously undocumented strength, i.e. the extra entropy can grow as fast as the entropy itself.

Combining Eqs.~\eqref{relative-entropy=difference} and~\eqref{eq:Smic}, one obtains the following exact generalization of Eq.~\eqref{eq:approx}, valid irrespective of whether EE holds:
\begin{equation}
\Omega_{\vec{C}^* }=e^{S^*_{\textrm{can}}- S_n[P^*_{\textrm{mic}}||P^*_{\textrm{can}}]}.
\label{eq:exact}
\end{equation}
Clearly, the above expression reduces to Eq.~\eqref{eq:approx} in case of EE, i.e. when Eq.~\eqref{eq:KLon} holds.
Although exact, Eq.~\eqref{eq:exact} is not very useful unless one can calculate $S_n[P^*_{\textrm{mic}}||P^*_{\textrm{can}}]$ explicitly.
An equivalent exact expression, which only requires the knowledge of $P^*_{\textrm{can}}$ and is again valid even when EE does not hold, has been derived~\cite{squartini2017reconnecting}:
\begin{eqnarray}
\Omega_{\vec{C}^* }&=&\sum_{\mathbf{G}\in\mathcal{G}}\int_{-\vec{\pi}}^{\vec{\pi}}\frac{{\mathrm{d}\vec{\psi}}}{{(2\pi)}^K}e^{\textrm{i}\vec{\psi}[\vec{C}^*-\vec{C}(\mathbf{G})]}\nonumber\\
&=&\int_{-\vec{\pi}}^{+\vec{\pi}}\frac{\mathrm{d}\vec{\psi}}{{(2\pi)}^K}P^{-1}_{\textrm{can}}(\mathbf{G}^*|\vec{\theta}^*+\textrm{i}\vec{\psi})
\label{Omega-hard-constraints}
\end{eqnarray}
(where $\int_{-\vec{\pi}}^{+\vec{\pi}}{\mathrm{d}\vec{\psi}}\equiv\prod_{k=1}^K\int_{-{\pi}}^{+{\pi}}{\mathrm{d}{\psi_k}}$).
We will confirm that the above expression provides the exact result in cases where the complex integral can be calculated explicitly and $\Omega_{\vec{C}^* }$ can be evaluated independently via combinatorial enumeration.
Indeed, Eq.~\eqref{Omega-hard-constraints} highlights a beautiful connection between canonical and microcanonical probabilities through an extension to complex numbers.

When the integral in Eq.~\eqref{Omega-hard-constraints} cannot be calculated directly, it is still possible to use a saddle-point technique leading to~\cite{squartini2017reconnecting}
\begin{eqnarray}
\Omega_{\vec{C}^* }&=&{\frac{e^{S^*_{\textrm{can}}}}{\sqrt{\det(2\pi\mathbf{\Sigma^*})}}\prod_{k=1}^{K}[1+O(1/\lambda^*_k)]}\nonumber\\
&=&e^{S^*_{\textrm{can}}}\prod_{k=1}^{K}\frac{1+O(1/\lambda^*_k)}{\sqrt{2\pi\lambda^*_k}}\label{Prob-micr-st-1}
\end{eqnarray}
where $\mathbf{\Sigma}^*$ is the covariance matrix of the $K$ constraints in the canonical ensemble, whose entries are defined as
\begin{equation}
\Sigma^*_{ij}=\left.\Sigma_{ij}\right|_{\vec{\theta}=\vec{\theta}^*}
\label{eq:whatstarmeans}
\end{equation}
with
\begin{eqnarray}
{\Sigma}_{ij}&\equiv&-\frac{\partial^2\mathcal{L}^*(\vec{\theta})}{\partial{\theta_i}\partial{\theta_j}}\nonumber\\
&=&\frac{\partial^2\ln Z(\vec{\theta})}{\partial{\theta_i}\partial{\theta_j}}\nonumber\\
&=&\langle C_i C_j\rangle_{\vec{\theta}}-\langle C_i\rangle_{\vec{\theta}}\langle C_j\rangle_{\vec{\theta}}\nonumber\\
&=&\textrm{Cov}_{\vec{\theta}}[C_i,C_j]
\label{Sigma_unit}
\end{eqnarray}
and $\{\lambda^*_k\}_{k=1}^K$ are the eigenvalues of $\mathbf{\Sigma}^*$.
We recall that covariance matrices are positive-semidefinite, so all their eigenvalues are non-negative.
If $\lambda_k^*$ is finite, then the quantity $O(1/\lambda^*_k)$ in Eq.~\eqref{Prob-micr-st-1} cannot in general be calculated explicitly, although it generates a correction that does not change the leading order of $\Omega_{\vec{C}^* }$ and $S^*_\textrm{mic}$. 
If $\lambda_k^*$ is infinite (i.e., if it diverges in the thermodynamic limit), then $O(1/\lambda^*_k)$ will vanish asymptotically and we have $1+O(1/\lambda^*_k)=1+o(1)$.
This implies that, if all the eigenvalues of $\mathbf{\Sigma}^*$ diverge, then Eq.~\eqref{Prob-micr-st-1}, when inserted into certain expressions, can lead to an exact result. This includes the case of local constraints, for which $K$ diverges in the thermodynamic limit.
We will therefore discuss the asymptotic behaviour of the eigenvalues of $\mathbf{\Sigma}^*$ in each of the examples considered later.

Equation~\eqref{Prob-micr-st-1} generalizes Eq.~\eqref{eq:approx} to the case where EE does not necessarily hold. Note that our initial assumption that the $K$ constraints are non-redundant implies that $\lambda^*_k>0$ for all $k$, i.e. $\mathbf{\Sigma}^*$ is positive-definite~\cite{squartini2017reconnecting}. Keeping this assumption also in the thermodynamic limit (as ensured by our choice of both global and local constraints defined above), we note two consequences. First, since Eq.~\eqref{Sigma_unit} shows that $\mathbf{\Sigma}^*$ is the Hessian matrix of second derivatives of $-\mathcal{L}^*(\vec{\theta})$, the fact that $\mathbf{\Sigma}^*$ is positive-definite implies that $\vec{\theta}^*$ is a unique global maximum for $\mathcal{L}^*(\vec{\theta})$~\cite{squartini2017reconnecting}, confirming what we had anticipated previously.
Second, the product in Eq.~\eqref{Prob-micr-st-1} is at most of the same order as the denominator. Therefore, in full generality, we can exploit Eq.~\eqref{Prob-micr-st-1} to rewrite Eq.~\eqref{eq:exact} as
\begin{equation}
\Omega_{\vec{C}^* }=e^{S^*_{\textrm{can}}-O(\alpha_n)},
\label{eq:approxalpha}
\end{equation}
where we have defined~\cite{squartini2017reconnecting}
\begin{equation}
\alpha_n\equiv\ln \sqrt{\det(2\pi\mathbf{\Sigma}^*)}=\frac{1}{2}\sum_{k=1}^{K}\ln({2\pi \lambda^*_k}).
\label{Relative-engenvale}
\end{equation}

We can now make three important considerations.
First, Eq.~\eqref{eq:approxalpha} means that
\begin{equation}
S_n[P^*_{\textrm{mic}}||P^*_{\textrm{can}}]=O(\alpha_n),
\label{eq:Salpha}
\end{equation}
showing that the speed of growth of $S_n[P^*_{\textrm{mic}}||P^*_{\textrm{can}}]$ with $n$ can be calculated explicitly through Eq.~\eqref{Relative-engenvale} using the knowledge of $\mathbf{\Sigma}^*$, which requires only the canonical ensemble. This is useful when microcanonical calculations are unfeasible.
Second, if $K$ is finite, or if $K$ diverges but all (except possibly a finite number of) the eigenvalues of $\mathbf{\Sigma}^*$ diverge, then the product inside Eq.~\eqref{Prob-micr-st-1} gives a subleading contribution to $S_n[P^*_{\textrm{mic}}||P^*_{\textrm{can}}]$, which therefore has the same asymptotic behaviour as $\alpha_n$:
\begin{equation}
S_n[P^*_{\textrm{mic}}||P^*_{\textrm{can}}]=\alpha_n[1+o(1)].
\label{eq:Salpha1}
\end{equation}
This result, which is stronger than Eq.~\eqref{eq:Salpha}, means that in such a case one can obtain exact estimates of quantities that depend on $S_n[P^*_{\textrm{mic}}||P^*_{\textrm{can}}]$, using only the knowledge of $\alpha_n$.
Third, Eq.~\eqref{eq:Salpha} shows that the definition of EE given by Eq.~\eqref{eq:KLon} coincides with
\begin{equation}
\alpha_n=o(n)
\label{eq:alphaon}
\end{equation}
which, again, can be ascertained by evaluating only $\mathbf{\Sigma}^*$ and avoiding any microcanonical calculation.
Indeed, Eq.~\eqref{eq:alphaon} can be formulated as an equivalent definition of EE in the measure sense~\cite{squartini2017reconnecting}.
If ${\alpha}_n$ grows faster than $o(n)$, then the system is under EN.

\section{Weak and strong ensemble nonequivalence}
In this section we illustrate the main results, i.e. we identify systems for which the breaking of EE occurs in a form that is at the same time `strong' and `unrestricted' and we calculate the relative entropy in various such systems.
To this end, we first make some general considerations leading to a rigorous definition of `strong' EN and subsequently study specific examples within our matrix ensembles.

\subsection{Relative entropy ratio\label{sec:ratio}}
Equation~\eqref{Prob-micr-st-1} reveals that the asymptotic behaviour of $\Omega_{\vec{C}^* }$ depends on that of $K$ and of the eigenvalues of the covariance matrix $\mathbf{\Sigma}^*$. 
We can indeed convince ourselves of this fact by looking at results of previous studies from a novel perspective.

Specifically, if $K=o(n)$ and if we exclude phase transitions, then Eq.~\eqref{Prob-micr-st-1} leads to Eq.~\eqref{eq:approx}, i.e. the ensembles are equivalent. This includes the traditional situation where one has a finite number of constraints, as well as more complicated cases where the number of constraints is subextensive (e.g. random graphs with constraints on a subextensive subset of node degrees~\cite{roccaverde2019is}).
In order to break EE in this case, one needs phase transitions corresponding to singularities of the partition function~\cite{touchette2015equivalence}. For instance, in the case of graphs with fixed numbers of edges and triangles (or wedges)~\cite{den2018ensemble}, there is a region in parameter space where one gets $S_n[P^*_{\textrm{mic}}||P^*_{\textrm{can}}]=O(n^2)$ and therefore $\Omega_{\vec{C}^* }=e^{S^*_{\textrm{can}}- O(n^2)}$. Since also $S^*_{\textrm{can}}$ and $S^*_{\textrm{mic}}$ are $O(n^2)$ in this case, it follows that
\begin{equation}
S_n[P^*_{\textrm{mic}}||P^*_{\textrm{can}}]=O(S^*_{\textrm{can}})
\label{eq:strong}
\end{equation}
(note that in general $S^*_{\textrm{can}}\ge S^*_{\textrm{mic}}$ due to the non-negativity of the Kullback-Leibler divergence and to Eq.~\eqref{relative-entropy=difference}, therefore $O(S^*_{\textrm{can}})$ is necessarily the leading order).
This is what we have previously referred to as a form of EN that is `restricted' (i.e. valid only in a certain region in parameter space arising from a phase transition and outside which EE is restored) but `strong' (i.e. where the relative entropy is of the same order as the entropy itself).

If $K=O(n)$, then Eq.~\eqref{eq:approx} is in general no longer valid.
For instance, in the case of \emph{sparse} graphs with fixed degrees ($K=n$), all the eigenvalues of $\mathbf{\Sigma}^*$ are finite in the thermodynamic limit~\cite{squartini2015breaking,garlaschelli2018covariance}; one indeed obtains $S_n[P^*_{\textrm{mic}}||P^*_{\textrm{can}}]=O(n)$~\cite{squartini2015breaking} and hence $\Omega_{\vec{C}^* }=e^{S^*_{\textrm{can}}- O(n)}$. Note that  in this case the product in Eq.~\eqref{Prob-micr-st-1} (which in general cannot be calculated exactly) is of the same order as the denominator and should be taken into account. 
In the case of \emph{dense} graphs with fixed degrees (again $K=n$), all the eigenvalues of $\mathbf{\Sigma}^*$ are instead $O(n)$~\cite{garlaschelli2018covariance}; one indeed obtains $S_n[P^*_{\textrm{mic}}||P^*_{\textrm{can}}]=O(n\ln n)$~\cite{garlaschelli2018covariance} and hence $\Omega_{\vec{C}^* }=e^{S^*_{\textrm{can}}-O(n\ln n)}$.
The product in Eq.~\eqref{Prob-micr-st-1} is in this case negliglible with respect to the denominator, which can be calculated exactly.
In any case, since $S^*_{\textrm{can}}$ and $S^*_{\textrm{mic}}$ are still $O(n^2)$ for both sparse and dense networks with fixed degrees, these situations correspond to
\begin{equation}
S_n[P^*_{\textrm{mic}}||P^*_{\textrm{can}}]=o(S^*_{\textrm{can}}),
\label{eq:weak}
\end{equation}
i.e. to what we have defined `weak' EN. On the other hand, this type of EN is not associated with phase transitions (which are indeed absent in the mentioned examples of graphs with fixed degrees) and is therefore `unrestricted', i.e. valid in the entire parameter space.

The above considerations suggest that, in order to rigorously define the strength of EN, we may define the ratio
\begin{equation}
R_n\equiv \frac{S_n[P^*_{\textrm{mic}}||P^*_{\textrm{can}}]}{S_n[P^*_{\textrm{can}}]}=1-\frac{S_n[P^*_{\textrm{mic}}]}{S_n[P^*_{\textrm{can}}]}
\label{eq:Rn}
\end{equation}
between the relative entropy and the  canonical entropy, calculated for fixed $n$, and consider its limit as $n\to+\infty$, i.e.
\begin{eqnarray} R_\infty&\equiv&\lim_{n\rightarrow\infty}R_n\nonumber\\
&=&\lim_{n\rightarrow\infty}\frac{S_n[P^*_{\textrm{mic}}||P^*_{\textrm{can}}]}{S_n[P^*_{\textrm{can}}]}\nonumber\\
&=&1-\lim_{n\rightarrow\infty}\frac{S_n[P^*_{\textrm{mic}}]}{S_n[P^*_{\textrm{can}}]}.
\label{limit-r-binary}
\end{eqnarray}
For brevity, we will call $R_n$ the \emph{relative entropy ratio} and $R_\infty$ the \emph{limiting relative entropy ratio}.
Note that the inequality $S^*_{\textrm{can}}\ge S^*_{\textrm{mic}}\ge 0$ implies $0\le R_n\le 1$ for all $n>0$.
The condition characterizing our notion of strong EN in Eq.~\eqref{eq:strong} coincides with $R_\infty$ being strictly positive.
The value of $R_\infty$ in that case quantifies exactly the asymptotic proportionality between $S_n[P^*_{\textrm{mic}}||P^*_{\textrm{can}}]$ and $S_n[P^*_{\textrm{can}}]$, which is otherwise left unquantified by Eq.~\eqref{eq:strong} alone. 
We will therefore adopt the strict inequality
\begin{equation}
R_\infty>0
\label{eq:strict}
\end{equation}
(which in turns implies the breakdown of Eq.~\eqref{eq:alphaon}, the converse being in general not true) as our definition of \emph{strong EN}. By contrast, the condition characterizing our notion of weak EN in Eq.~\eqref{eq:weak} can be rephrased as the equality $R_\infty=0$. Note  that one may have $R_\infty=0$ also in cases where the ensembles are equivalent.
We will therefore adopt the condition $R_\infty=0$, \emph{in conjunction with the violation of Eq.~\eqref{eq:alphaon}}, as our definition of \emph{weak EN}.
Note that our discussion following Eq.~\eqref{Prob-micr-st-1} implies that, if all but at most a finite number of the eigenvalues of $\mathbf{\Sigma}^*$ diverge, then the exact value of $R_\infty$ can be retrieved by replacing $S_n[P^*_{\textrm{mic}}||P^*_{\textrm{can}}]$ with $\alpha_n$ given by Eq.~\eqref{Relative-engenvale}, i.e. using only the canonical covariances between the constraints, without microcanonical calculations.

Note that Eq.~\eqref{eq:exact} implies
\begin{equation}
\Omega_{\vec{C}^* }=e^{S^*_{\textrm{can}}(1-R_n)}=O\left(\left(e^{S^*_{\textrm{can}}}\right)^{1-R_\infty}\right).
\label{eq:exactR}
\end{equation}
So, in presence of strong nonequivalence ($R_\infty>0$), $\Omega_{\vec{C}^* }$ is of strictly smaller order compared with the ordinary estimate in Eq.~\eqref{eq:approx}. 
This is actually due to the canonical ensemble having much bigger entropy than the microcanonical one: indeed, Eq.~\eqref{relative-entropy=difference} implies
\begin{equation}
S^*_{\textrm{mic}}=S^*_{\textrm{can}}(1-R_n)
\label{eq:exactRR}
\end{equation}
and, inverting,
\begin{equation}
S^*_{\textrm{can}}=\frac{1}{1-R_n}S^*_{\textrm{mic}}.
\label{eq:exactRRR}
\end{equation}
Note that the factor $1/(1-R_n)$ can be arbitrarily large since $R_n$ can be arbitrarily close to 1.

Given the above definitions of `weak' and `strong' EN in terms of the limiting relative entropy ratio, in what follows we will consider the specific ensembles of matrices introduced in the previous section, under both global and local constraints, and calculate the value of $\alpha_n$ and $R_\infty$ in each case.

\subsection{Global constraints}
As already discussed, ensembles of (binary or weighted) $n\times m$ matrices with a global constraint are defined by requiring that the single quantity $t(\mathbf{G})=\sum_{i=1}^n\sum_{j=1}^m g_{ij}$ takes, either `hardly' or `softly', a specific value $t^*\equiv t(\mathbf{G}^*)$.
For this simple choice of the constraint, both $S^*_{\textrm{can}}$  and $S^*_{\textrm{mic}}$ can be calculated exactly.
This allows us to check that the complex integral in Eq.~\eqref{Omega-hard-constraints} indeed provides the exact value of $\Omega_{t^*}$.
Moreover, we can confirm the correctness of the asymptotic formula in Eq.~\eqref{Prob-micr-st-1}.
All these approaches show that for both binary and weighted matrices with a global constraint the canonical and microcanonical ensembles are equivalent.

\subsubsection{Binary matrices under a global constraint}
Let us start with the case when the global constraint $t^*$ is imposed on binary matrix ensembles characterized by  $g_{ij}\in\{0,1\}$.
The calculation of the canonical entropy $S^*_{\textrm{can}}$ is straightforward (see Appendix) by first calculating the likelihood
\begin{equation}\label{can-prob-binary-glob-1}
P_{\textrm{can}}(\mathbf{G}^*|{\theta})=\frac{e^{-\theta\, t^*}}{(1+e^{-\theta})^{mn}}
\end{equation}
and then looking for the value $\theta^*$ that maximizes $P_{\textrm{can}}(\mathbf{G}^*|{\theta})$ or, equivalently, realizes the soft constraint $\langle t\rangle_{\theta^*}=t^*$. The result is
\begin{equation}\label{eq:thetaglobal}
	\theta^*=\ln\frac{mn-t^*}{t^*}.
\end{equation}
Using Eq.~\eqref{eq:S=-L}, we can then easily evaluate $S^*_{\textrm{can}}$ from Eqs.~\eqref{can-prob-binary-glob-1} and~\eqref{eq:thetaglobal} as
\begin{equation}\label{Canoni-entropy}
 S^*_{\textrm{can}}=-\ln P_{\textrm{can}}(\mathbf{G}^*|{\theta}^*)=\ln \frac{{(mn)}^{mn}}{{(t^*)}^{t^*}(mn-t^*)^{mn-t^*}}.
\end{equation}

The calculation of the microcanonical entropy $S^*_{\textrm{mic}}$ is in this case even simpler than that of the canonical one, since the number $\Omega_{t^*}$ of configurations realizing the hard constraint $t(\mathbf{G})=t^*$ is simply the number of ways in which $t^*$ `ones' can be placed in $mn$ available positions, i.e. the binomial coefficient $\Omega_{t^*}=\binom{mn}{t^*}$. This implies
\begin{equation}\label{Micro-canonical-entropy-binary}
S^*_{\textrm{mic}}=\ln\Omega_{t^*}=\ln\binom{mn}{t^*}.
\end{equation}
Importantly, it is possible to confirm that, upon extending the argument of the likelihood to the complex domain and calculating $P_{\textrm{can}}(\mathbf{G}^*|{\theta}^*+i\psi)$, the integral formula in Eq.~\eqref{Omega-hard-constraints}  returns a value of $\Omega_{t^*}$ that produces the exact value of the microcanonical entropy $S^*_{\textrm{mic}}$ given in Eq.~\eqref{Micro-canonical-entropy-binary}:
\begin{eqnarray}
S^*_{\textrm{mic}}&=&\ln\int_{-{\pi}}^{+{\pi}}\frac{\mathrm{d}{\psi}}{{2\pi}}P^{-1}_{\textrm{can}}(\mathbf{G}^*|{\theta}^*+\textrm{i}{\psi})\nonumber\\
&=&\ln\int_{-{\pi}}^{+{\pi}}\frac{\mathrm{d}{\psi}}{{2\pi}}
\frac{(1+e^{-\theta^*-\textrm{i}{\psi}})^{mn}}{e^{-(\theta^*+\textrm{i}{\psi})\, t^*}}\nonumber\\
&=&\ln\binom{mn}{t^*},
\label{Omega-hard-constraints-globalB}
\end{eqnarray}
where the (instructive) calculation justifying the last equality is reported in the Appendix.

Combining the expressions for $S^*_{\textrm{mic}}$ and $S^*_{\textrm{can}}$ into Eq.~\eqref{relative-entropy=difference}, we obtain the relative entropy between the two ensembles:
\begin{equation}
S_n[P^*_{\textrm{mic}}||P^*_{\textrm{can}}]=\ln\frac{{(mn)}^{mn}}{\displaystyle{\binom{mn}{t^*}}{(t^*)}^{t^*}(mn-t^*)^{mn-t^*}}.
\label{eq:relativeglobalbinary}
\end{equation}
In this simple example, the inequality $S_n[P^*_{\textrm{mic}}||P^*_{\textrm{can}}]>0$ clearly arises from the presence of dependencies among the entries of $\mathbf{G}$ in the microcanonical ensemble and the absence of such dependencies in the canonical one.
Indeed, while in the microcanonical ensemble the hard constraint $t(\mathbf{G})=t^*$ makes all the entries of $\mathbf{G}$ mutually dependent, in the canonical ensemble the soft constraint $\langle t\rangle_{\theta^*}=t^*$ leaves each entry $g_{ij}$ independent and identically (Bernoulli-)distributed with probability
\begin{equation}
p(g_{ij}|\theta^*)=\frac{e^{-\theta^* g_{ij}}}{1+e^{-\theta^*}},\quad g_{ij}\in\{0,1\}
\label{eq:bernoulli*}
\end{equation}
(see Appendix).
Consequently, while in the microcanonical ensemble the constraint $t(\mathbf{G})$ is a deterministic quantity fixed to the value $t^*$, in the canonical ensemble $t(\mathbf{G})$ is a random variable with expected value $t^*$ and variance
\begin{equation}
\Sigma^*=\textrm{Var}_{{\theta}^*}[t]=nm\frac{e^{-\theta^*}}{(1+e^{-\theta^*})^2}= t^*\left(1-\frac{t^*}{mn}\right)
\label{eq:varbinaryglo}
\end{equation}
(see Appendix), where $\Sigma^*$ is the only (recall that here $K=1$) of the covariance matrix $\mathbf{\Sigma}^*$ introduced in Eq.~\eqref{Sigma_unit}.

As discussed in Subsection~\ref{sec:ensemble(non)equivalence}, $S_n[P^*_{\textrm{mic}}||P^*_{\textrm{can}}]$ and $\Sigma^*$ are asymptotically related through Eq.~\eqref{eq:Salpha}, and the (non)equivalence of canonical and microcanonical ensembles is decided by the asymptotic behaviour of these two quantities. We will confirm both results in the particular case under consideration here. However, for compactness, we do this in conjunction with the weighted case, after introducing the latter below.

\subsubsection{Weighted matrices under a global constraint}
We now consider the case when the global constraint $t^*$ is enforced on weighted matrices where $g_{ij}$ is a non-negative integer.
As we show in the Appendix, in the canonical ensemble the likelihood can be calculated as
\begin{equation}\label{can-prob-weighted-glob-1}
P_{\textrm{can}}(\mathbf{G}^*|{\theta})={e^{-\theta\, t^*}}{(1-e^{-\theta})}^{mn}
\end{equation}
and is maximised by the parameter value
\begin{equation}\label{eq:thetaglobalw}
\theta^*=\ln\frac{mn+t^*}{t^*}
\end{equation}
(note the change of sign with respect to the binary case)
realizing the soft constraint $\langle t\rangle_{\theta^*}=t^*$.
The canonical entropy is therefore
\begin{equation}
 S^*_{\textrm{can}}=-\ln P_{\textrm{can}}(\mathbf{G}^*|{\theta}^*)=\ln\frac{{(mn+t^*)}^{mn+t^*}}{{(t^*)}^{t^*}{(mn)}^{mn}}.
\label{Canoni-entropyw}
\end{equation}

In the microcanonical ensemble, the number $\Omega_{t^*}$ of configurations realizing the hard constraint $t(\mathbf{G})=t^*$ coincides with the number of so-called \emph{weak compositions} of the positive integer $t^*$ into exactly $mn$ parts, i.e. the number of ways of writing the positive integer $t^*$ as the sum of an ordered sequence of $mn$ non-negative integers (note that two sequences that differ in the order of their terms represent different configurations).
This number is given by the \emph{negative binomial} coefficient $\Omega_{t^*}=\binom{mn+t^*-1}{t^*}$~\cite{heubach2009combinatorics}, whence
\begin{equation}\label{microcanonical-entropy-w-global}
S^*_{\textrm{mic}}=\ln \Omega_{t^*}=\ln\binom{mn+t^*-1}{t^*}.
\end{equation}
In this case as well, one can confirm that the integration of the complex quantity $P^{-1}_{\textrm{can}}(\mathbf{G}^*|{\theta}^*+i\psi)$ as specified in Eq.~\eqref{Omega-hard-constraints} produces precisely the same value of $\Omega_{t^*}$ used in Eq.~\eqref{microcanonical-entropy-w-global} (see Appendix), thus retrieving the exact entropy
\begin{eqnarray}
S^*_{\textrm{mic}}
&=&\ln\int_{-{\pi}}^{+{\pi}}\frac{\mathrm{d}{\psi}}{{2\pi}}P^{-1}_{\textrm{can}}(\mathbf{G}^*|{\theta}^*+\textrm{i}{\psi})\nonumber\\
&=&\ln\int_{-\pi}^{\pi}\frac{\mathrm{d}\psi}{2\pi}\frac{(1-e^{-\theta^*-\textrm{i}\psi})^{-mn}}{e^{-(\theta^*+\textrm{i}\psi)t^*}}\nonumber\\
&=&\ln\binom{mn+t^*-1}{t^*}.
\label{Omega-hard-constraints-globalW}
\end{eqnarray}

The relative entropy $S_n[P^*_{\textrm{mic}}||P^*_{\textrm{can}}]$, calculated using Eq.~\eqref{relative-entropy=difference}, equals
\begin{equation}
S_n[P^*_{\textrm{mic}}||P^*_{\textrm{can}}]=
\ln\frac{(mn+t^*)^{mn+t^*}}{\displaystyle{\binom{mn+t^*-1}{t^*}}{(t^*)}^{t^*}{{(mn)}^{mn}}}.
\label{eq:relativeglobalweighted}
\end{equation}
Again, the origin of a non-zero relative entropy lies in the presence of dependencies among all the entries of $\mathbf{G}$ in the microcanonical ensemble, where they are coupled by the hard constraint $t(\mathbf{G})=t^*$, and in the absence of such dependencies in the canonical ensemble, where each entry $g_{ij}$ is independent and now \emph{geometrically} (see Appendix) distributed with probability
\begin{equation}
p(g_{ij}|\theta^*)={e^{-\theta^* g_{ij}}}(1-e^{-\theta^*}),\quad g_{ij}\in\{0,1,2,\dots\}.
\label{eq:geometric*}
\end{equation}
As a consequence, while in the microcanonical ensemble the constraint $t(\mathbf{G})$ is fixed to the constant value $t^*$, in the canonical ensemble it is a random variable with expected value $t^*$ and variance
\begin{equation}
\Sigma^*=\textrm{Var}_{{\theta}^*}[t]=nm\frac{e^{-\theta^*}}{(1-e^{-\theta^*})^2}= t^*\left(1+\frac{t^*}{mn}\right)
\label{eq:varweightedglo}
\end{equation}
(see Appendix).

\subsubsection{Ensemble equivalence for matrices under a global constraint\label{sec:equivalence}}
We can now study, in a combined fashion, the (non)equivalence of the canonical and microcanonical ensembles of both binary and weighted matrices with a global constraint $t^*$.
To this end, we preliminarly notice that the reason why the quantity $\binom{k+l-1}{l}$ is called \emph{negative} binomial is the fact that it can be formally rewritten as the following binomial coefficient with negative signs:
\begin{equation}
\binom{k+l-1}{l}=(-1)^{l}\binom{-k}{l}.
\label{eq:negbinom}
\end{equation}
The above relation allows us to conveniently rewrite the relative entropy for the weighted case appearing in Eq.~\eqref{eq:relativeglobalweighted} as
\begin{equation}
S_n[P^*_{\textrm{mic}}||P^*_{\textrm{can}}]=\ln\frac{{(-mn)}^{-mn}}{\displaystyle{\binom{-mn}{t^*}}{(t^*)}^{t^*}(-mn-t^*)^{-mn-t^*}}.
\label{eq:relativeglobalweighted2}
\end{equation}
Upon comparison with the corresponding Eq.~\eqref{eq:relativeglobalbinary} valid in the binary case, we can express the relative entropy in general as
\begin{equation}
S^\pm_n[P^*_{\textrm{mic}}||P^*_{\textrm{can}}]=\ln\frac{{(\pm mn)}^{\pm mn}}{\displaystyle{\binom{\pm mn}{t^*}}{(t^*)}^{t^*}(\pm mn-t^*)^{\pm mn-t^*}},
\label{eq:relativeglobalboth}
\end{equation}
where the superscript ``$+$'' applies to binary matrices (note that $t^*\le mn$ in this case) and the superscript ``$-$'' applies to weighted matrices. Note that the expression for the weighted case can be formally retrieved by changing the sign of $m$ in the expression valid for the binary case.

As we discussed in Subsection~\ref{sec:ensemble(non)equivalence}, checking for (non)equivalence requires studying the asymptotic behaviour of the relative entropy. In this case, we can calculate the asymptotic behaviour of $S^\pm_n[P^*_{\textrm{mic}}||P^*_{\textrm{can}}]$ explicitly from the exact expression given by Eq.~\eqref{eq:relativeglobalboth}.
Note that, in both the sparse and dense case (see Subsection~\ref{sec:thermo}), $t^*$ and $mn$ diverge in the thermodynamic limit. 
We can therefore apply Stirling's formula
\begin{equation}
k!=\sqrt{2\pi k}\left(\frac{k}{e}\right)^k\left[1+o(1)\right]
\label{eq:stirling}
\end{equation}
to Eq.~\eqref{eq:relativeglobalboth}, which yields
\begin{equation}
S^\pm_n[P^*_{\textrm{mic}}||P^*_{\textrm{can}}]=\frac{1}{2}\ln\left[2\pi{ t^*\left(1-\frac{t^*}{\pm mn}\right)}\right]\left[1+o(1)\right].
\label{eq:relativeglobalbothapprox}
\end{equation}

For purely pedagogical reasons, we check that the asymptotic behaviour found above is consistent with the one we would retrieve by using the expansion in Eq.~\eqref{Prob-micr-st-1}, which leads to Eq.~\eqref{eq:Salpha} and reduces the problem of the calculation of $S^\pm_n[P^*_{\textrm{mic}}||P^*_{\textrm{can}}]$ to that of its leading order $\alpha_n$.
To this end, we note that in this case the matrix $\mathbf{\Sigma}^*$, being a $1\times 1$ matrix, coincides with its only eigenvalue 
\begin{equation}
(\lambda^*)^\pm=t^*\left(1-\frac{t^*}{\pm mn}\right),
\end{equation}
where we have used Eq.~\eqref{eq:varbinaryglo} for binary ($+$) and Eq.~\eqref{eq:varweightedglo} for weighted ($-$) matrices.
Therefore 
\begin{eqnarray}
\alpha^\pm_n&=&\ln\sqrt{2\pi(\lambda^*)^\pm}\nonumber\\
&=&\frac{1}{2}\ln{\left[2\pi t^*\left(1-\frac{t^*}{\pm mn}\right)\right]},
\label{eq:alpha+-global}
\end{eqnarray}
which has indeed the same leading order as Eq.~\eqref{eq:relativeglobalbothapprox}, thereby confirming the correctness of the saddle-point calculation. 
As an even stronger result, we are under the conditions for which Eq.~\eqref{eq:Salpha1} holds, a relationship that can be confirmed by comparing Eqs.~\eqref{eq:relativeglobalbothapprox} and~\eqref{eq:alpha+-global}.
It should also be noted that, since $t^*$ diverges in the thermodynamic limit, so does $(\lambda^*)^\pm$ and Eq.~\eqref{Prob-micr-st-1} leads to 
\begin{equation}
\Omega^{\pm}_{t^* }={\frac{e^{S_n^\pm[P^*_{\textrm{can}}]}}{\sqrt{2\pi t^*\left(1-\frac{t^*}{\pm mn}\right)}}[1+o(1)]},
\end{equation}
which is precisely what we get by applying Eq.~\eqref{eq:stirling} to the binomial and negative binomial coefficients appearing in the exact expression for $\Omega^{\pm}_{t^* }$ in the binary and weighted case respectively.

As stated in Eq.~\eqref{eq:alphaon}, checking whether the ensembles are equivalent boils down to checking whether $\alpha_n=o(n)$.
Note that the only effect of the asymptotic scaling of $t^*$ is that the quantity $1-t^*/(\pm mn)$ in Eq.~\eqref{eq:alpha+-global} converges to 1 in the sparse case $t^*/mn=O(1/n)$ and to a different, but still finite and positive constant in the dense case $t^*/mn=O(1)$ (see Subsection~\ref{sec:thermo}).
Therefore in both cases we have $\alpha_n=O(\ln t^*)=O(\ln mn)$. This implies $\alpha_n=o(n)$ independently of the asymptotic behaviour of $m$.
This result shows that, \emph{in presence of a global constraint, both binary and weighted matrices are under EE, irrespective of the scaling of $t^*$ and $m$.}
This finding confirms, in a generalized setting, the result obtained for networks with a given total number of links~\cite{squartini2015breaking}.
Since EE is preserved, we avoid the calculation of the limiting relative entropy ratio $R_\infty$ defined in Eq.~\eqref{limit-r-binary} in this case.

\subsection{One-sided local constraints}
We now consider ensembles of binary and weighted $n\times m$ matrices with one-sided local constraints, i.e. under the requirement that the $n$-dimensional vector $\vec{r}(\mathbf{G})$ with entries $r_i(\mathbf{G})=\sum_{j=1}^m g_{ij}$ ($i=1,n$) takes a specific value $\vec{r}^*\equiv\vec{r}(\mathbf{G}^*)$. 
Note that, unlike the case of global constraints, here the number of constraints is extensive.
As in the case with global constraints, it turns out that both $S^*_{\textrm{can}}$ and $S^*_{\textrm{mic}}$ can still be calculated exactly.
Therefore we can again confirm the correctness of both the exact integral formula in Eq.~\eqref{Omega-hard-constraints} and the asymptotic expansion in Eq.~\eqref{Prob-micr-st-1}.
Despite these extensions are mathematically straightforward, we find a deep physical difference with respect to the case with global constraints: the presence of an extensive number of local constraints implies the breaking of the equivalence of canonical and microcanonical ensembles for both binary and weighted matrices.
The calculation of the limiting relative entropy ratio $R_\infty$ allows us to quantify the strength of nonequivalence and also to identify the conditions leading to the `strong and unrestricted' form.

\subsubsection{Binary matrices under one-sided local constraints}
Let us first examine the case when the one-sided local constraints $\vec{r}^*$ are imposed on ensembles of binary matrices.
As we show in the Appendix, in the canonical ensemble the likelihood is
\begin{equation}\label{Can-prob-local-1main}
P_{\textrm{can}}(\mathbf{G}^*|\vec{\theta})=\frac{e^{-\vec{\theta}\cdot {\vec{r}}^*}}{\prod_{i=1}^{n}(1+e^{-\theta_i})^{m}}
\end{equation}
and reaches its maximum when the parameter $\vec{\theta}$ takes the value $\vec{\theta}^*$ with entries
\begin{equation}\label{eq:theta1sided}
	\theta^*_i=\ln\frac{m-r_i^*}{r_i^*}\qquad i=1,n,
\end{equation}
corresponding to the soft constraint $\langle \vec{r}\rangle_{\vec{\theta^*}}=\vec{r}^*$.
Substituting Eq.~\eqref{eq:theta1sided} into Eq.~\eqref{Can-prob-local-1main}, we obtain the canonical entropy as
\begin{eqnarray}
 S^*_{\textrm{can}}&=&-\ln P_{\textrm{can}}(\mathbf{G}^*|\vec{\theta}^*)\nonumber\\
&=&\sum_{i=1}^{n}\ln\frac{m^{m}}{{(r_i^*)}^{r^*_i}(m-r_i^*)^{m-r^*_i}}.\label{Canoni-entropy_bin1sided}
\end{eqnarray}

Let us now turn to the microcanonical ensemble.
Since the constraints are only one-sided, it is immediate to realize that the number $\Omega_{\vec{r}^*}$ of configurations realizing the hard constraint $\vec{r}(\mathbf{G})=\vec{r}^*$ is a product of row-specific binomial coefficients, so that the microcanonical entropy $S^*_{\textrm{mic}}$ can still be calculated exactly as the following simple generalization of Eq.~\eqref{Micro-canonical-entropy-binary}:
\begin{equation}\label{01-micro-states-local-z-redine-2}
S^*_{\textrm{mic}}=\ln\Omega_{\vec{r}^*}=\ln\prod_{i=1}^{n}\binom{m}{r^*_i}=\sum_{i=1}^{n}\ln\binom{m}{r^*_i}.
\end{equation}
For the same reason, $S^*_{\textrm{mic}}$ can also be exactly retrieved by explicitly integrating the complex-valued quantity $P_{\textrm{can}}(\mathbf{G}^*|\vec{\theta}^*+i\vec\psi)$ as prescribed  by Eq.~\eqref{Omega-hard-constraints} (see Appendix):
\begin{eqnarray}
S^*_{\textrm{mic}}&=&\ln\int_{-\vec{\pi}}^{+\vec{\pi}}\frac{\mathrm{d}\vec{\psi}}{{(2\pi)}^n}P^{-1}_{\textrm{can}}(\mathbf{G}^*|\vec{\theta}^*+\textrm{i}\vec{\psi})\nonumber\\
&=&\ln\prod_{i=1}^{n}\int_{-{\pi}}^{+{\pi}}\frac{\mathrm{d}{\psi_i}}{{2\pi}}\frac{(1+e^{-\theta_i^*-{\textrm{i}}\psi_i})^m}{e^{-(\theta_i^*+{\textrm{i}}\psi_i)r^*_i}}\nonumber\\
&=&\sum_{i=1}^{n}\ln\binom{m}{r^*_i}.
\end{eqnarray}

Combining the above results, we can calculate the relative entropy from Eq.~\eqref{relative-entropy=difference} as
\begin{equation}
{S_n[P^*_{\textrm{mic}}||P^*_{\textrm{can}}]=\sum_{i=1}^{n}\ln\frac{m^m}{\displaystyle{\binom{m}{r^*_i}}{(r_i^*)}^{r^*_i}(m-r^*_i)^{m-r^*_i}}}.
\label{Rebinaryoneside}
\end{equation}
The above quantity encodes the following difference between the two ensembles:
in the microcanonical ensemble, the hard constraint $\vec{r}(\mathbf{G})=\vec{r}^*$ makes all the entries in each row of $\mathbf{G}$ mutually dependent, while leaving different rows independent of each other;
on the other hand, in the canonical ensemble the soft constraint $\langle\vec{r}\rangle_{\vec{\theta}^*}=\vec{r}^*$ leaves \emph{all} entries of the matrix independent. As in the case with a global constraint, each
entry $g_{ij}$ is still Bernoulli-distributed, but now with row-specific probability
\begin{equation}
p(g_{ij}|\vec{\theta}^*)=\frac{e^{-\theta_i^* g_{ij}}}{1+e^{-\theta_i^*}},\quad g_{ij}\in\{0,1\},
\label{eq:bernoulli1sided*}
\end{equation}
as we show in the Appendix.
Correspondingly, in the microcanonical ensemble $\vec{r}$ is a deterministic vector fixed to the value $\vec{r}^*$, while in the canonical ensemble it is a random vector with expected value $\vec{r}^*$. The covariance matrix $\mathbf{\Sigma}^*$ between the entries of $\vec{r}$ (i.e. between the $n$ constraints) in the canonical ensemble is a diagonal matrix with entries
\begin{equation}
\Sigma_{ij}^*=\delta_{ij} r_i^*\left(1-\frac{r_i^*}{m}\right)
\label{eq:varbinary1sidedmain}
\end{equation}
where $\delta_{ij}=1$ if $i=j$ and $\delta_{ij}=0$ if $i\ne j$ (see Appendix).
This implies that the eigenvalues $\{\lambda^*_i\}_{i=1}^n$ of $\mathbf{\Sigma}^*$ are
\begin{equation}
\lambda^*_i= r_i^*\left(1-\frac{r_i^*}{m}\right)\qquad i=1,n.
\label{eq:lambda_bin}
\end{equation}
Again, we are going to discuss the (non)equivalence of the two ensembles together with the corresponding case of weighted matrices, after studying the latter below.

\subsubsection{Weighted matrices under one-sided local constraints}
We now move to the case when the one-sided local constraints $\vec{r}^*$ are imposed on ensembles of weighted matrices.
The canonical ensemble under such constraints is characterized by the likelihood
\begin{equation}\label{Can-prob-local-1mainW}
P_{\textrm{can}}(\mathbf{G}^*|\vec{\theta})=\frac{e^{-\vec{\theta}\cdot {\vec{r}}^*}}{\prod_{i=1}^{n}(1-e^{-\theta_i})^{-m}},
\end{equation}
which is maximized by the parameter value $\vec{\theta}^*$ with entries
\begin{equation}\label{eq:theta1sidedW}
	\theta^*_i=\ln\frac{m+r_i^*}{r_i^*}\qquad i=1,n
\end{equation}
realizing the soft constraint $\langle \vec{r}\rangle_{\vec{\theta^*}}=\vec{r}^*$ (see Appendix). If we insert Eq.~\eqref{eq:theta1sidedW} into Eq.~\eqref{Can-prob-local-1mainW}, we get
\begin{eqnarray}
 S^*_{\textrm{can}}&=&-\ln P_{\textrm{can}}(\mathbf{G}^*|\vec{\theta}^*)\nonumber\\
&=&\sum_{i=1}^{n}\ln\frac{{(m+r_i^*)}^{m+r_i^*}}{{(r_i^*)}^{r_i^*}{m}^{m}}.\label{Canoni-entropy_wei1sided}
\end{eqnarray}

The microcanonical entropy $S^*_{\textrm{mic}}$ is instead given by the following generalization of Eq.~\eqref{microcanonical-entropy-w-global}:
\begin{equation}\label{01-micro-states-local-z-redine-2W}
S^*_{\textrm{mic}}=\ln\Omega_{\vec{r}^*}
=\sum_{i=1}^{n}\ln\binom{m+r^*_i-1}{r^*_i},
\end{equation}
where we have expressed the number $\Omega_{\vec{r}^*}$ of configurations realizing the hard constraint $\vec{r}(\mathbf{G})=\vec{r}^*$ as a product of row-specific negative binomial coefficients.
Again, the microcanonical entropy can be obtained equivalently from Eq.~\eqref{Omega-hard-constraints} as follows (see Appendix):
\begin{eqnarray}
S^*_{\textrm{mic}}&=&\ln\int_{-\vec{\pi}}^{+\vec{\pi}}\frac{\mathrm{d}\vec{\psi}}{{(2\pi)}^n}P^{-1}_{\textrm{can}}(\mathbf{G}^*|\vec{\theta}^*+\textrm{i}\vec{\psi})\nonumber\\
&=&\ln\prod_{i=1}^{n}\int_{-{\pi}}^{+{\pi}}\frac{\mathrm{d}{\psi_i}}{{2\pi}}\frac{(1-e^{-\theta_i^*-{\textrm{i}}\psi_i})^{-m}}{e^{-(\theta_i^*+{\textrm{i}}\psi_i)r^*_i}}\nonumber\\
&=&\sum_{i=1}^{n}\ln\binom{m+r^*_i-1}{r^*_i}.
\label{micro:entropy:one:weighted}
\end{eqnarray}

The relative entropy, which can be obtained from Eq.~\eqref{relative-entropy=difference} as usual, equals
\begin{equation}
S_n[P^*_{\textrm{mic}}||P^*_{\textrm{can}}]=\sum_{i=1}^{n}\ln
\frac{{(m+r_i^*)}^{m+r_i^*}}{\displaystyle{\binom{m+r^*_i-1}{r^*_i}}{(r_i^*)}^{r_i^*}{m}^{m}}
\label{Relativeweightedoneside}
\end{equation}
and encodes the difference between the microcanonical ensemble, where the entries of each row of $\mathbf{G}$ are mutually coupled by the hard constraint $\vec{r}(\mathbf{G})=\vec{r}^*$ (while different rows are independent), and the canonical ensemble, where all entries of $\mathbf{G}$ are independent and \emph{geometrically} (see Appendix) distributed with row-dependent probability
\begin{equation}
p(g_{ij}|\vec{\theta}^*)={e^{-\theta_i^* g_{ij}}}(1-e^{-\theta_i^*}),\quad g_{ij}\in\{0,1,2,\dots\}.
\label{eq:geometric1sided**}
\end{equation}
As a consequence, while in the microcanonical ensemble the constraint $\vec{r}$ is fixed to the value $\vec{r}^*$, in the canonical ensemble it is a random vector fluctuating around $\vec{r}^*$ according to the diagonal covariance matrix $\mathbf{\Sigma}^*$ with entries
\begin{equation}
\Sigma_{ij}^*=\delta_{ij} r_i^*\left(1+\frac{r_i^*}{m}\right)
\label{eq:varweighted1sidedmain}
\end{equation}
(see Appendix) and eigenvalues
\begin{equation}
\lambda^*_i= r_i^*\left(1+\frac{r_i^*}{m}\right)\qquad i=1,n.
\label{eq:lambda_wei}
\end{equation}

\subsubsection{Ensemble nonequivalence for matrices under one-sided local constraints}
We can now compactly discuss the  (non)equivalence of canonical and microcanonical ensembles of both binary and weighted matrices under one-sided local constraints.
As for the case of global constraints discussed in Subsection~\ref{sec:equivalence}, we still have an exact knowledge of the canonical entropy, the microcanonical entropy, and the relative entropy.
Moreover, these quantities can all be written, using Eq.~\eqref{eq:negbinom}, in compact expressions formally valid for both binary ($+$) and weighted ($-$) matrices.
Indeed, the canonical entropy can be expressed by combining the expressions for $S_n[P^*_{\textrm{can}}]=S^*_{\textrm{can}}$ in Eqs.~\eqref{Canoni-entropy_bin1sided} and~\eqref{Canoni-entropy_wei1sided} into the unified formula
\begin{equation}
S^\pm_n[P^*_{\textrm{can}}]=\sum_{i=1}^{n}\ln\frac{m^{\pm m}}{{(r^*_i)}^{r^*_i}{(m\mp r^*_i)}^{\pm m-r^*_i}}
\label{Canentropy_combo}
\end{equation}
and, similarly, the microcanonical entropy can be obtained by formally combining Eqs.~\eqref{01-micro-states-local-z-redine-2} and~\eqref{01-micro-states-local-z-redine-2W} into 
\begin{equation}
S^\pm_n[P^*_{\textrm{mic}}]=\sum_{i=1}^{n}\ln\left[(\pm 1)^{r^*_i}\binom{\pm m}{r^*_i}\right].
\label{Micentropy_combo}
\end{equation}
The above expressions can be used to calculate the relative entropy as 
\begin{equation}
{S^\pm_n[P^*_{\textrm{mic}}||P^*_{\textrm{can}}]=\sum_{i=1}^{n}\ln\frac{(\pm m)^{\pm m}}{\displaystyle{\binom{\pm m}{r^*_i}}{(r_i^*)}^{r^*_i}(\pm m-r^*_i)^{\pm m-r^*_i}}}
\label{Rebothoneside}
\end{equation}
which indeed combines the expressions given in Eq.~\eqref{Rebinaryoneside} for binary  ($+$)  matrices and Eq.~\eqref{Relativeweightedoneside} for weighted ($-$) matrices.
Equation~\eqref{Rebothoneside} extends Eq.~\eqref{eq:relativeglobalboth} to the case of one-sided local constraints. We now consider different regimes.

\begin{itemize}
\item 
In the sparse case where $r^*_i=O(1)$ (for all $i$) and $m=O(n)$ (see Subsection~\ref{sec:thermo}), we can use Stirling's formula, given by Eq.~\eqref{eq:stirling}, to expand $m!$ (but not $r^*_i!$) appearing in the (negative) binomial coefficient to get 
\begin{equation}
\binom{\pm m}{r^*_i}\approx \frac{{(\pm m)}^{r_i^*}}{r_i^*!}
\label{eq:ammazzateo}
\end{equation}
and consequently
\begin{equation}
S^\pm_n[P^*_{\textrm{mic}}||P^*_{\textrm{can}}]\approx\sum_{i=1}^{n}\ln\frac{e^{r^*_i}r^*_i!}{{(r^*_i)}^{r^*_i}}=O(n).
\label{Rebothonesidesparse}
\end{equation}
\item 
In the dense case where both $r_i^*$ (for all $i$) and $m$ are $O(n)$, as discussed in Subsection~\ref{sec:thermo} (so that $r_i^*/m$ converges to a finite constant), we can use Stirling's formula to expand both $m!$ and $r_i^*!$ into Eq.~\eqref{Rebothoneside} to obtain
\begin{eqnarray}
S^{\pm}_n[P^*_{\textrm{mic}}||P^*_{\textrm{can}}]&=&\frac{1}{2}\sum_{i=1}^{n}\ln\left[2\pi{ r_i^*\left(1-\frac{r_i^*}{\pm m}\right)}\right]\left[1+o(1)\right]\nonumber\\
&=&O(n\ln n)
\label{eq:relativeonesidedbothapprox}
\end{eqnarray}
for binary ($+$) (in which case $r_i^*\le m$) and weighted ($-$) matrices. 
\item 
In the dense case where both $m$ and $r^*_i$ are finite, there is no asymptotic expansion that allows to simplify Eq.~\eqref{Rebothoneside} in general, so $S^\pm_n[P^*_{\textrm{mic}}||P^*_{\textrm{can}}]$ has to be evaluated explicitly (simple examples are provided below). The important general consideration is that, irrespective of the specific values of $m$ and $r^*_i$,
\begin{equation}
S^\pm_n[P^*_{\textrm{mic}}||P^*_{\textrm{can}}]=O(n).
\label{Rebothonesidedense}
\end{equation}
\end{itemize}

Again, we can confirm that the above asymptotic expressions are consistent with
what we would obtain from Eq.~\eqref{eq:Salpha}, which follows from the saddle-point approximation given by Eq.~\eqref{Prob-micr-st-1}.
To see this, noting that here $K=n$ and that Eqs.~\eqref{eq:varbinary1sidedmain} and~\eqref{eq:varweighted1sidedmain} indicate that $(\mathbf{\Sigma}^*)^{\pm}$ is a diagonal matrix with entries
\begin{equation}
(\Sigma_{ij}^*)^\pm=\delta_{ij} r_i^*\left(1-\frac{r_i^*}{\pm m}\right)
\label{eq:varboth1sidedmain}
\end{equation}
for binary ($+$) and weighted ($-$) one-sided constraints respectively, we can compactly express Eq.~\eqref{Relative-engenvale} as
\begin{eqnarray}
\alpha^{\pm
}_n&=&\ln\sqrt{\det\left[2\pi(\mathbf{\Sigma}^*)^\pm\right]}\nonumber\\
&=&\frac{1}{2}\sum_{i=1}^{n}\ln{\left[2\pi(\Sigma^*_{ii})^\pm\right]}\nonumber\\
&=&\frac{1}{2}\sum_{i=1}^{n}\ln{\left[2\pi r_i^*\left(1-\frac{r_i^*}{\pm m}\right)\right]}.\label{eq:alphaonesided}
\end{eqnarray}
In the dense regime with diverging $m$, the conditions guaranteeing the strong result in Eq.~\eqref{eq:Salpha1} (all but a finite number of diverging eigenvalues of $\mathbf{\Sigma}^*$) hold, as can be confirmed by comparing Eqs.~\eqref{eq:relativeonesidedbothapprox} and~\eqref{eq:alphaonesided}.
Moreover, noticing from Eq.~\eqref{eq:stirling} that ${e^{k}k!}/{{k}^{k}}=\sqrt{2\pi k}[1+o(1)]$, we see that Eq.~\eqref{eq:alphaonesided} confirms the asymptotic behaviour of the relative entropy obtained also in Eqs.~\eqref{Rebothonesidesparse} and~\eqref{Rebothonesidedense} for the other two regimes, under the respective assumptions on the scaling of $m$ and $k$. Coincidentally, we also see that the stronger result in Eq.~\eqref{eq:Salpha1} turns out to be a very good approximation for the relative entropy even in these two regimes where, technically, the required conditions are not met.
This means that, for the one-sided dense case with finite $m$, we can rewrite Eq.~\eqref{eq:Salpha} asymptotically (i.e. for large $n$) as
\begin{eqnarray}
S^\pm_n[P^*_{\textrm{mic}}||P^*_{\textrm{can}}]&=&C_1(m){\alpha}_n^\pm\nonumber\\
&=&C_1(m)\ln \sqrt{\det[2\pi(\mathbf{\Sigma}^*)^\pm]}
\label{mecojoni1}
\end{eqnarray}
where $C_1(m)$ is a finite and positive constant. Moreover, from the known inequality ${e^{k}k!}/{{k}^{k}}\ge\sqrt{2\pi k}$ for the factorial, we see that $C_1(m)\ge 1$ as implied by comparing Eqs.~\eqref{Rebothoneside} and~\eqref{eq:alphaonesided}. Finally, we also know from Stirling's approximation that $C_1(m)$ is not much bigger than $1$, i.e. $C_1(m)\gtrsim 1$, and that it rapidly approaches $1$: indeed when $m$ diverges Eq.~\eqref{eq:Salpha1} holds exactly, which implies 
\begin{equation}
\lim_{m\to\infty}C_1(m)=1.
\label{eq:apijanderculo}
\end{equation}

The fact that, in all regimes, $S^\pm_n[P^*_{\textrm{mic}}||P^*_{\textrm{can}}]$ (or equivalently $\alpha_n$) is at least of order $O(n)$ shows that Eq.~\eqref{eq:alphaon} is violated and that \emph{EE breaks down for both binary and weighted matrices under one-sided local constraints}, irrespective of the density and of the behaviour of $m$.
This important finding generalizes the result, documented so far only for ensembles of binary graphs with given degree sequence~\cite{squartini2015breaking,garlaschelli2018covariance,roccaverde2019breaking} (and possibly modular structure~\cite{garlaschelli2016ensemble}) and weighted graphs with given strength sequence~\cite{zhang2020ensemble}, that EE breaks down in the presence of an extensive (i.e. growing like $n$) number of local constraints.
Here, this result is extended to more general ensembles of matrices, i.e. asymmetric, rectangular matrices describing e.g. bipartite graphs, multivariate time series, multiplex social activity, multi-cast communication systems and multi-cell gene expression profiles with variable $m$.
More importantly, this generalized setting allows for a qualitatively new phenomenon to emerge, namely the onset of `strong' EN, as we now show.

Indeed, we can investigate the `strength' of nonequivalence by comparing the asymptotic behaviour of the relative entropy with that of the canonical entropy given by Eq.~\eqref{Canentropy_combo}.
This expression can be evaluated in the usual three regimes as follows.

In the sparse case with $r^*_i=O(1)$ and $m=O(n)$, noticing that asymptotically (for large $n$) we have ${(m\mp r^*_i)}^{\pm m-r^*_i}\approx m^{\pm m-r^*_i}e^{\mp r^*_i}$, Eq.~\eqref{Canentropy_combo} reduces to
\begin{equation}\label{Canentropy_comboasympto}
S^\pm_n[P^*_{\textrm{can}}]\approx\sum_{i=1}^{n}r^*_i\ln\frac{e^{\pm 1}m}{r^*_i}=O(n\ln n),
\end{equation}
which dominates over the order $O(n)$ of the corresponding relative entropy $S^\pm_n[P^*_{\textrm{mic}}||P^*_{\textrm{can}}]$ calculated previously in Eq.~\eqref{Rebothonesidesparse} for the sparse case.
This implies that the limiting relative entropy ratio defined in Eq.~\eqref{limit-r-binary} is $R^{\pm}_\infty=0$ for both binary ($+$) and weighted ($-$) constraints, meaning that in this case the breaking of EE is still `weak' as in the case of graphs with local constraints.

In the dense case with $r^*_i=O(n)$ and $m=O(n)$, Eq.~\eqref{Canentropy_combo} can be evaluated as
\begin{eqnarray}
S^\pm_n[P^*_{\textrm{can}}]&=&\sum_{i=1}^{n}\left[\pm m\ln\frac{m}{m\mp r^*_i}+r^*_i\ln\frac{m\mp r^*_i}{r^*_i}\right]\nonumber\\
&=&O(n^2)
\label{Canentropy_comboweak}
\end{eqnarray}
which, again, dominates over the order $O(n\ln n)$ of the corresponding relative entropy calculated in Eq.~\eqref{eq:relativeonesidedbothapprox}. Therefore we still have $R^{\pm}_\infty=0$ (weak nonequivalence).

Finally, the dense case where both $m$ and $r^*_i$ remain finite as $n\to\infty$ is the subject of the rest of this Section. Equation~\eqref{Canentropy_combo} implies that 
\begin{equation}
S^\pm_n[P^*_{\textrm{can}}]=O(n)
\label{Canentropy_combostrong}
\end{equation}
which, upon comparison with Eq.~\eqref{Rebothonesidedense}, shows that now the relative entropy grows as fast as the canonical entropy, signalling the `strong' form of EN. 
Using the combined expressions given in Eqs.~\eqref{Canentropy_combo} and~\eqref{Micentropy_combo}, we can explicitly calculate the relative entropy ratio introduced in Eq.~\eqref{eq:Rn} as follows:
\begin{equation}
R^\pm_n=1-\frac{\sum_{i=1}^{n}\ln\left[(\pm 1)^{r^*_i}\binom{\pm m}{r^*_i}\right]}{\sum_{i=1}^{n}\ln\frac{m^{\pm m}}{{(r^*_i)}^{r^*_i}{(m\mp r^*_i)}^{\pm m-r^*_i}}}>0.
\label{Rbothn}
\end{equation}
Using Eqs.~\eqref{eq:alphaonesided} and~\eqref{mecojoni1}, we obtain the alternative asymptotic (for large $n$) expression
\begin{eqnarray}
R^\pm_n&=&\frac{S^\pm_n[P^*_{\textrm{mic}}||P^*_{\textrm{can}}]}{S^\pm_n[P^*_{\textrm{can}}]}\nonumber\\
&=&C_1(m)\frac{\alpha^{\pm
}_n}{\sum_{i=1}^{n}\ln\frac{m^{\pm m}}{{(r^*_i)}^{r^*_i}{(m\mp r^*_i)}^{\pm m-r^*_i}}}\nonumber\\
&=&\frac{C_1(m)}{2}\frac{\sum_{i=1}^{n}\ln{\left[2\pi r_i^*\left(1-\frac{r_i^*}{\pm m}\right)\right]}}{\sum_{i=1}^{n}\ln\frac{m^{\pm m}}{{(r^*_i)}^{r^*_i}{(m\mp r^*_i)}^{\pm m-r^*_i}}}.
\label{Rbothnasympto}
\end{eqnarray}
Comparing Eqs.~\eqref{Rbothn} and~\eqref{Rbothnasympto} confirms that, as noticed above, $C_1(m)\approx 1$ also for finite $m$.

\begin{figure*}[t]
	\centering
	\includegraphics[height=7cm]{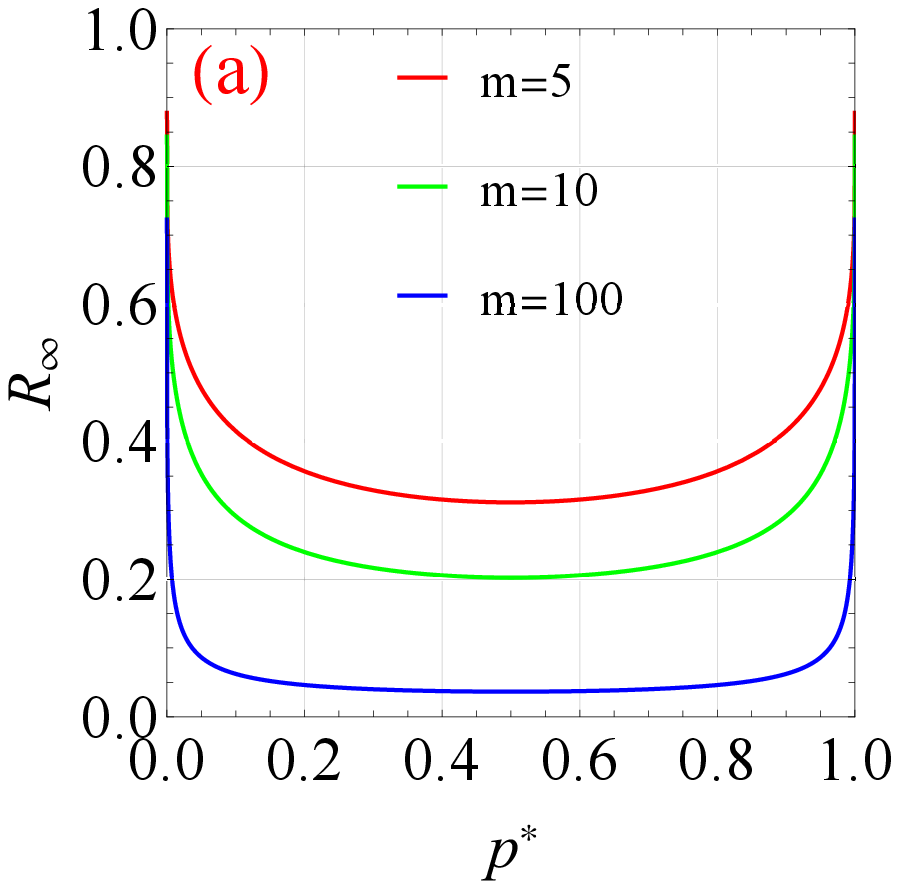}
	\hspace{1cm}
	\includegraphics[height=7cm]{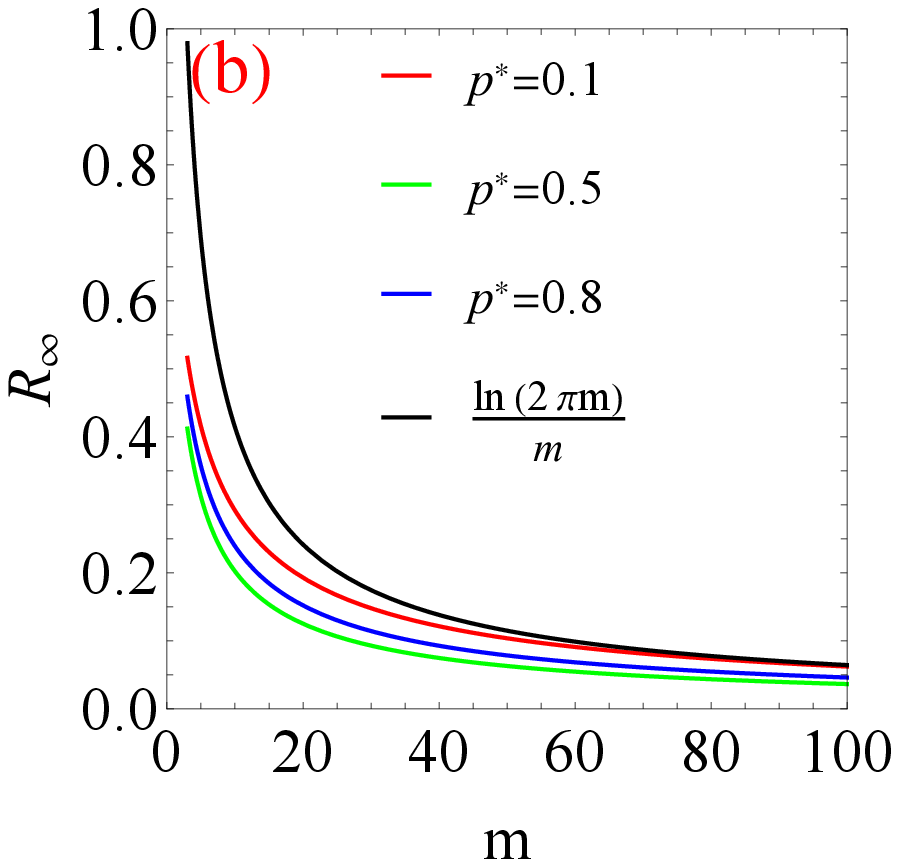}
	\caption{Strong ensemble nonequivalence, signalled by a positive limiting entropy ratio $R^+_\infty>0$, for binary matrices under homogeneous one-sided local constraints ($r^*_i=r^*$ $\forall i$) in the dense case with finite $m$ and $r^*$. (a) $R^+_\infty$ as a function of $p^*= r^*/m$ for various values of $m$. Note that $R^+_\infty$ is larger for smaller $m$ and for values of $p^*$ more distant from the uniform case ($p^*=1/2$). (b) $R^+_\infty$ as a function of $m$ for various values of $p^*$. Note that, as $m$ grows, $R^+_\infty$ decays like ${\ln (2\pi m)}/{m}$.}\label{BinarylS}
\end{figure*}

In general, taking the thermodynamic limit $n\rightarrow\infty$ in Eq.~\eqref{Rbothn} or~\eqref{Rbothnasympto} requires the specification of the value of $r^*_i$ for all $i$.
For the sake of illustration, we can consider the simplest case where each constraint has the same value $r^*_i=r^*$ ($i=1,n$). Note that the resulting canonical entropy of matrices with constant one-sided constraint $r^*$, given by Eq.~\eqref{Canentropy_combo}, coincides with the canonical entropy of matrices with the implied global constraint $t^*=n r^*$, given by Eqs.~\eqref{Canoni-entropy} and~\eqref{Canoni-entropyw} in the binary and weighted case respectively. 
However, the microcanonical entropy in the one-sided case, given by Eq.~\eqref{Micentropy_combo}, is strictly smaller than the corresponding one for matrices with the implied global constraint $t^*=n r^*$, given by Eqs.~\eqref{Micro-canonical-entropy-binary} and~\eqref{microcanonical-entropy-w-global} in the binary and weighted case respectively. 
From Eq.~\eqref{Rbothn} we immediately find
\begin{equation}
R^\pm_\infty=\lim_{n\to\infty}R^\pm_n=1-\frac{\ln\left[(\pm 1)^{r^*}\binom{\pm m}{r^*}\right]}{\ln\frac{m^{\pm m}}{{(r^*)}^{r^*}{(m\mp r^*)}^{\pm m-r^*}}}>0,
\label{Ratio:oneside:bothequal}
\end{equation}
confirming strong nonequivalence as defined in Eq.~\eqref{eq:strict}.
To gain numerical and visual insight about the behaviour of $R^\pm_\infty$ in Eq.~\eqref{Ratio:oneside:bothequal}, let us consider the binary and weighted cases separately.

In the binary case, Eq.~\eqref{eq:theta1sided} implies that, if $r^*_i=r^*$ for all $i$, then the Lagrange multipliers $\theta^*_i$ are all equal to
\begin{equation}\label{eq:theta1sidedbinall}
\theta^*_+\equiv\ln\frac{m- r^*}{r^*}.
\end{equation}
Then, writing $p^*\equiv r^*/m=e^{-\theta_+^*}/(1+e^{-\theta_+^*})\in(0,1)$, from Eq.~\eqref{Ratio:oneside:bothequal} we obtain
\begin{eqnarray}
R^+_\infty&=&1-\frac{{\ln\binom{m}{r^*}}}{{\ln\frac{m^{m}}{{(r^*)}^{r^*}(m-r^*)^{m-r^*}}}}\label{eq:Rinf1sidedbinary}\\
&=&1-\frac{{\ln\binom{m}{p^*m}}}{{\ln\frac{{m^m}}{{{(p^*m)}^{p^*m}(m-p^*m)^{m-p^*m}}}}}>0.\nonumber
\end{eqnarray}
Using the above expression, in Fig.~\ref{BinarylS} we plot $R^+_\infty$ as a function of either $p^*$ (for fixed $m$) or $m$ (for fixed $p^*$). We see that, for a wide range of values of $p^*$, $R^+_\infty$ remains appreciably large for values of $m$ up to one hundred. Moreover, values of $p^*$ closer to $0$ or $1$ than to $1/2$ make $R^+_\infty$ larger. So, for empirical applications where the level of `multiplexity' is moderate (i.e. small $m$), and especially away from the uniform case ($p^*=1/2$), there is a significant entropy reduction from the canonical to the microcanonical ensemble.
By contrast, as $m$ increases while $p^*$ remains fixed, $R^+_\infty$ decreases like $\frac{\ln2\pi m}{m}$, as can be easily realized by applying Stirling's formula to Eq.~\eqref{eq:Rinf1sidedbinary}. This coincides with the system progressively moving to the different regime where both $m$ and $r^*_i$ grow as $n$ grows, which results in weak EN and $R^+_\infty=0$ as previously noticed. Similarly, if $r^*_i$ remains finite while $m$ grows, we enter the sparse regime for which $R^+_\infty=0$ as previously noticed.

\begin{figure*}[htbp]
	\centering
	\includegraphics[height=7cm]{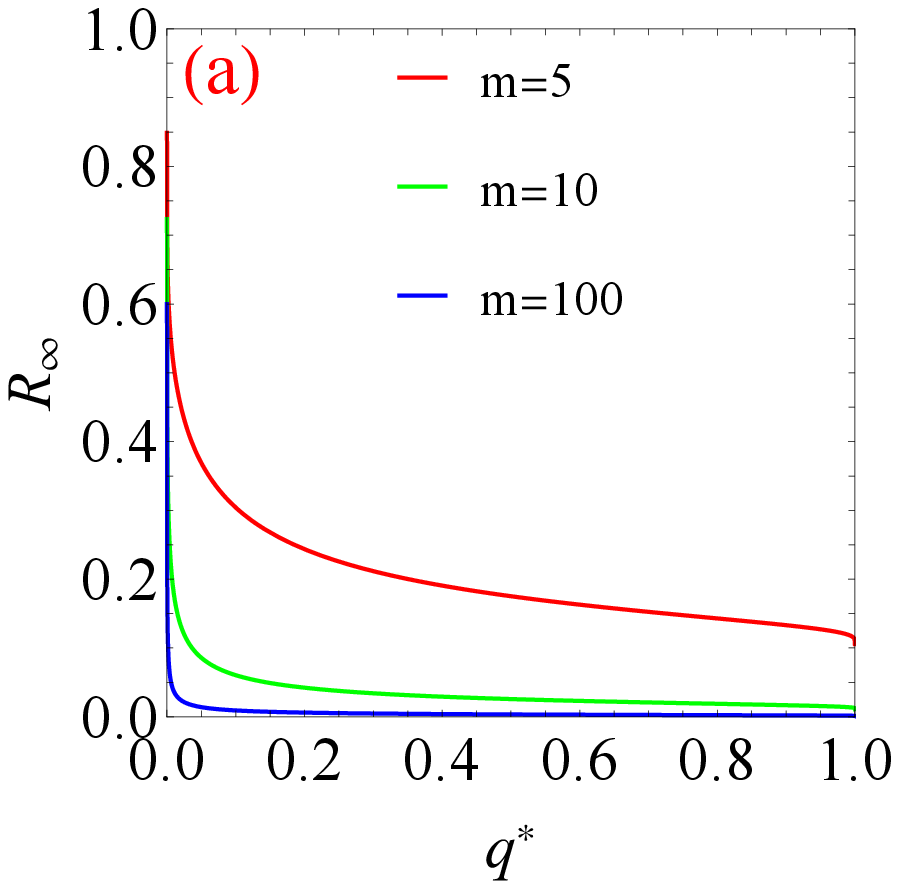}
	\hspace{1cm}
	\includegraphics[height=7cm]{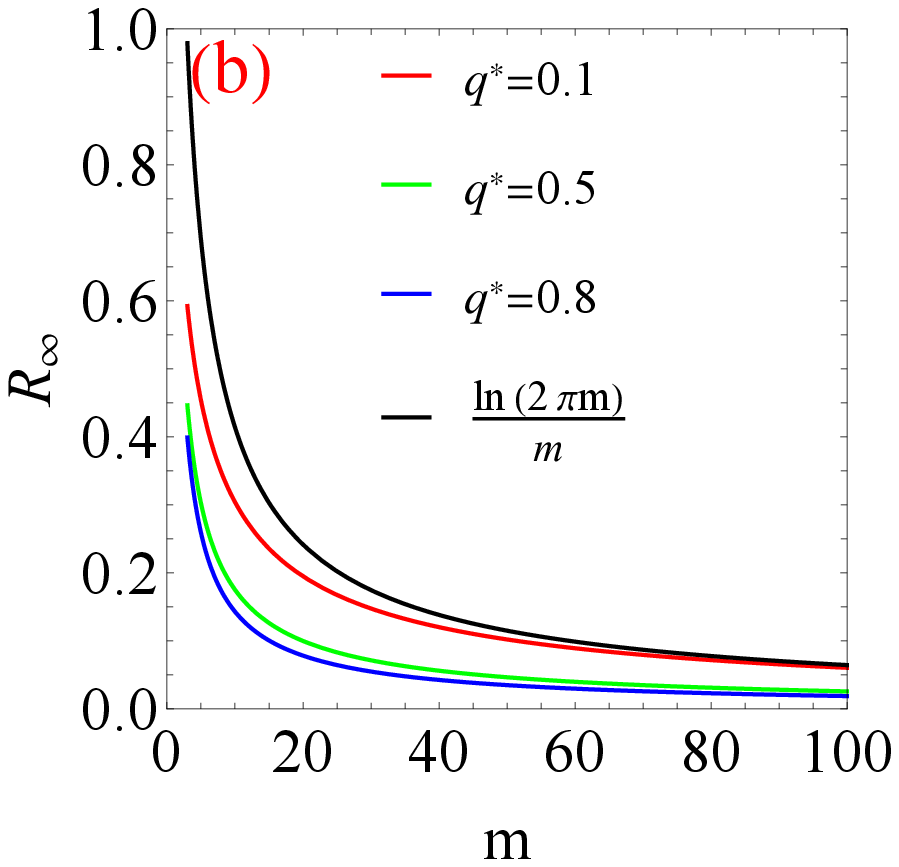}
	\caption{Strong ensemble nonequivalence, signalled by a positive limiting entropy ratio $R^-_\infty>0$, for weighted matrices under homogeneous one-sided local constraints ($r^*_i=r^*$ $\forall i$) in the dense case with finite $m$ and $r^*$. (a) $R^-_\infty$ as a function of $q^*=\frac{r^*}{m+r^*}$ for various values of $m$. Note that $R^-_\infty$ is larger for smaller $m$ and $q^*$. (b) $R^-_\infty$ as a function of $m$ for various values of $q^*$. As in the binary case, $R^-_\infty$ decays like ${\ln (2\pi m)}/{m}$ as $m$ grows.}\label{WeightedlS}
\end{figure*}

In the weighted case, Eq.~\eqref{eq:theta1sidedW} implies that if $r^*_i=r^*$ for all $i$, then $\theta^*_i=\theta^*_-$ for all $i$ with
\begin{equation}\label{eq:theta1sidedweiall}
\theta^*_-\equiv\ln\frac{m+ r^*}{r^*}.
\end{equation}
Then, writing $q^*\equiv e^{-\theta^*_-}=\frac{r^*}{m+r^*}\in(0,1)$, from Eq.~\eqref{Ratio:oneside:bothequal} we obtain 
\begin{eqnarray}
R^-_\infty&=&1-\frac{\ln\binom{m+r^*-1}{r^*}}{\ln\frac{{(m+r^*)^{m+r^*}}}{{m^m{(r^*)}^{r^*}}}}\label{R-weighted}\\
&=&1-\frac{\ln\binom{{(m-1+q^*)}/{(1-q^*)}}{{mq^*}/{(1-q^*)}}}{{\frac{m}{1-q^*}}\ln{\frac{m}{1-q^*}}-m\ln m-{\frac{mq^*}{1-q^*}}\ln{\frac{mq^*}{1-q^*}}}>0.\nonumber
\end{eqnarray}
Figure~\ref{WeightedlS} shows the behaviour of $R^-_\infty$ either as a function of $q^*$ (with $m$ fixed) or as a function of $m$ (with $q^*$ fixed). Considerations similar to the binary case apply. The main difference is that, while $R^+_\infty$ has a symmetric behaviour around the value $p^*=1/2$ (arising from the fundamental symmetry of exchanging $p^*$ with $1-p^*$ and $g_{ij}=1$ with $g_{ij}=0$ in the binary case), $R^-_\infty$ decreases monotonically as a function of $q^*$ (due to the lack of any symmetry of that sort in the weighted case).
So now $R^-_\infty$ is larger for smaller $m$ and $q^*$.

The above results illustrate what we had anticipated previously, i.e. that if $m$ is finite (and the matrices are necessarily dense) then the ensembles feature a strong form of EN. 
Here, this form of EN is also `unrestricted', as it holds irrespective of the value of $\vec{C}^*$ or $\vec{\theta}^*$, i.e. throughout the parameter space.
To the best of our knowledge, this is the first evidence of a situation for which EN occurs in a simultaneously `strong and unrestricted' form, i.e. the most robust manifestation of the breaking of EE documented so far. Its ultimate origin is the presence of an extensive number of local constraints, and not of phase transitions.

\subsection{Two-sided local constraints}
We now discuss binary and weighted matrices under two-sided local constraints $(\vec{r}(\mathbf{G}), \vec{c}(\mathbf{G}))$, where $\vec{r}(\mathbf{G})$ is still the $n$-dimensional vector of row sums while $\vec{c}(\mathbf{G})$ is the $m$-dimensional vector of column sums, with entries $c_j(\mathbf{G})=\sum_{i=1}^{n}g_{ij}$ $(j=1,\dots,m)$. 
We constrain both vectors to a given value $(\vec{r}^*, \vec{c}^*)\equiv (\vec{r}(\mathbf{G}^*), \vec{c}(\mathbf{G}^*))$.
Note that the number of constraints is still extensive.
Unlike the case of one-sided constraints, for two-sided constraints it is not possible to calculate the exact number $\Omega_{\vec{r}^*, \vec{c}^*}$ of configurations in the microcanonical ensemble.
By contrast, all canonical calculations can still be carried out analytically (although the value of the Lagrange multipliers can be determined only via implicit expressions).
Therefore, as we show below, the matrix $\mathbf{\Sigma}^*$ of canonical covariances between the constraints becomes a crucial tool to calculate the asymptotic behaviour of the relevant microcanonical quantities.

\subsubsection{Binary matrices under two-sided local constraints}
As usual, we start from the binary case. 
As shown in the Appendix, in the canonical ensemble the likelihood is  
\begin{equation}\label{Likeli:can:binary:2}
  P_{\textrm{can}}(\mathbf{G}^*|\vec{\alpha}, \vec{\beta})=\frac{e^{-\vec{\alpha}\cdot\vec{r}^*-\vec{\beta}\cdot\vec{c}^*}}{\prod_{i=1}^{n}\prod_{j=1}^{m}[1+e^{-(\alpha_i+\beta_j)}]}
\end{equation}
and is maximized by the parameter values $(\vec{\alpha}^*, \vec{\beta}^*)$ defined implicitly by the following set of $n+m$ coupled nonlinear equations:
\begin{eqnarray}
 r_i^*&=&\sum_{j=1}^{m}\frac{e^{-(\alpha^*_i+\beta^*_j)}}{1+e^{-(\alpha^*_i+\beta^*_j)}},\quad i=1,n,\label{eq:2side1b}\\
c_j^*&=&\sum_{i=1}^{n}\frac{e^{-(\alpha^*_i+\beta^*_j)}}{1+e^{-(\alpha^*_i+\beta^*_j)}},\quad j=1,m.\label{eq:2side2b}
\end{eqnarray}
Unfortunately, in general these equations cannot be solved analytically to express $(\vec{\alpha}^*, \vec{\beta}^*)$ as an explicit function of $(\vec{r}^*, \vec{c}^*)$. This is due to the fact that the presence of both row and column constraints couples all parameters. However, the equations can be solved numerically and the unique solution $(\vec{\alpha}^*, \vec{\beta}^*)$ can then be inserted into $P_{\textrm{can}}(\mathbf{G}|\vec{\alpha}, \vec{\beta})$. This gives complete analytical control over the canonical ensemble.
In particular, the canonical entropy is 
\begin{eqnarray}
  S^*_{\textrm{can}}&=&-\ln P_{\textrm{can}}(\mathbf{G}^*|\vec{\alpha}^*,\vec{\beta}^*)\label{Canonical:entropy:binary:two}
\\
&=&\vec{\alpha}^*\cdot\vec{r}^*+\vec{\beta}^*\cdot\vec{c}^*
+\sum_{i=1}^{n}\sum_{j=1}^{m}\ln[1+e^{-(\alpha^*_i+\beta^*_j)}].\nonumber
\end{eqnarray}
Note that, since this model has additional constraints with respect to the one-sided case with the same row sums $\vec{r}^*$, the canonical entropy above cannot be larger than the corresponding one-sided canonical entropy given by Eq.~\eqref{Canoni-entropy_bin1sided}, i.e. we have the following upper bound:
\begin{equation}
S^*_{\textrm{can}}\le
\sum_{i=1}^{n}\ln\frac{m^{m}}{{(r_i^*)}^{r^*_i}(m-r_i^*)^{m-r^*_i}}.
\label{eq:boundcan}
\end{equation}

On the other hand, the microcanonical entropy $S^*_{\textrm{mic}}$ cannot be computed analytically, although the asymptotic formulas based on Eqs.~\eqref{eq:approxalpha} and~\eqref{Relative-engenvale} can be used to estimate it from the canonical covariance matrix $\mathbf{\Sigma}^*$.
As we show later, this leads to an asymptotic estimate of the relative entropy based on Eq.~\eqref{eq:Salpha}.
As for the canonical entropy, the microcanonical one cannot be larger than the corresponding one given by Eq.~\eqref{01-micro-states-local-z-redine-2} in the one-sided case with the same row sums, with the only difference that now the resulting upper bound is tight:
\begin{equation}\label{eq:boundmic}
S^*_{\textrm{mic}}<\sum_{i=1}^{n}\ln\binom{m}{r^*_i}.
\end{equation}
Indeed, the configurations matching both the row and the column constraints in the two-sided case form a proper subset of the configurations matching only the row constraints in the one-sided case.

It should be noted that in the microcanonical ensemble both $\vec{r}$ and $\vec{c}$ are deterministic vectors fixed to the values $\vec{r}^*$ and $\vec{c}^*$ respectively, while in the canonical ensemble they are random vectors with expected values $\vec{r}^*$ and $\vec{c}^*$.
In the microcanonical ensemble, the hard constraints $(\vec{r}(\mathbf{G}),\vec{c}(\mathbf{G}))=(\vec{r}^*,\vec{c}^*)$ create mutual dependencies among all the entries of $\mathbf{G}$. In the canonical ensemble, the soft constraint $(\langle\vec{r}\rangle_{\vec{\alpha}^*},\langle\vec{c}\rangle_{\vec{\beta}^*})=(\vec{r}^*,\vec{c}^*)$ leaves all entries of $\mathbf{G}$ independent. As in all other canonical binary ensembles considered above, each
entry $g_{ij}$ is Bernoulli-distributed, but now with its specific parameters:
\begin{equation}
p(g_{ij}|\vec{\alpha}^*,\vec{\beta}^*)=\frac{e^{-(\alpha_i^*+\beta^*_i) g_{ij}}}{1+e^{-(\alpha_i^*,\beta^*_i)}},\quad g_{ij}\in\{0,1\},
\label{eq:bernoulli2sided*}
\end{equation}
as shown in the Appendix.

For illustration, we consider the special case where the column sums are all equal to each other, i.e. $c^*_j=c^*$ for all $j$. In this case, since the corresponding Lagrange multipliers must also be all equal to each other ($\beta_j^*=\beta^*$ for all $j$), it is indeed possible to solve for the parameters explicitly. Indeed, Eqs.~\eqref{eq:2side1b} and~\eqref{eq:2side2b} reduce to the $n+1$ independent equations
\begin{eqnarray}
r_i^*&=&{m}\frac{e^{-(\alpha^*_i+\beta^*)}}{1+e^{-(\alpha^*_i+\beta^*)}},\quad i=1,n,\label{eq:2side1bex}\\
c^*&=&\sum_{i=1}^{n}\frac{e^{-(\alpha^*_i+\beta^*)}}{1+e^{-(\alpha^*_i+\beta^*)}},\label{eq:2side2bex}
\end{eqnarray}
where the second equation is simply the consistency condition $c^*=\sum_{i=1}^{n}r^*_i/m$ implied by the first one. This means that the parameter $\beta^*$ is actually redundant, as it could in principle be reabsorbed into a shift of all the $\alpha^*_i$'s.
In any case, the combination $\alpha^*_i+\beta^*$ is found explicitly by inverting Eq.~\eqref{eq:2side1bex}:
\begin{equation}\label{Param:localrow:binary:two}
\alpha^*_i+\beta^*=\ln \frac{m-r^*_i}{r^*_i}.
\end{equation}
Note that, inserting this value into the expression for $S^*_{\textrm{can}}$ in Eq.~\eqref{Canonical:entropy:binary:two}, we obtain exactly the canonical entropy found previously in Eq.~\eqref{Canoni-entropy_bin1sided} for the binary ensemble with one-sided (row) constraints specified by the same vector $\vec{r}^*$, i.e. the bound in Eq.~\eqref{eq:boundcan} is fully saturated:
\begin{eqnarray}
 S^*_{\textrm{can}}&=&\vec{\alpha}^*\cdot\vec{r}^*+m\beta^*c^*
+m\sum_{i=1}^{n}\ln[1+e^{-(\alpha^*_i+\beta^*)}]\nonumber\\
&=&\sum_{i=1}^{n}\ln\frac{m^{m}}{{(r_i^*)}^{r^*_i}(m-r_i^*)^{m-r^*_i}}.\label{Canonical:entropy:binary:twoc}
\end{eqnarray}
Indeed, the two canonical ensembles are indistinguishable and all their properties are the same. However, the corresponding microcanonical ensembles remain very different, because in the two-sided case each of the $m$ column sums has to match the exact value $c^*$ separately, while in the one-sided case only the total sum $m c^*$ of all the $m$ column sums (which is necessarily implied by the row constraints) has to be matched exactly. Indeed Eq.~\eqref{eq:boundmic} is a tight bound that cannot be saturated.
Similarly, the covariance matrix $\mathbf{\Sigma}^*$ is now a $(n+m)\times(n+m)$ matrix (calculated later) and its determinant is different from the one obtained in the one-sided case, where the matrix is $n\times n$. 

Again, we are going to discuss the (non)equivalence of the two ensembles together with the corresponding case of weighted matrices, after studying the latter below.

\subsubsection{Weighted matrices under two-sided local constraints}
We now discuss EN in weighted matrices with two-sided local constraints. 
The likelihood (see Appendix) is now  
\begin{equation}\label{Weighted:two:can:like}
  P_{\textrm{can}}(\mathbf{G}^*|\vec{\alpha}, \vec{\beta})=\frac{e^{-\vec{\alpha}\cdot\vec{r}^*-\vec{\beta}\cdot\vec{c}^*}}{\prod_{i=1}^{n}\prod_{j=1}^{m}[1-e^{-(\alpha_i+\beta_j)}]^{-1}}
\end{equation}
and is maximized by the unique parameter values $(\vec{\alpha}^*, \vec{\beta}^*)$ defined implicitly through the $n+m$ coupled nonlinear equations
\begin{eqnarray}
 r_i^*&=&\sum_{j=1}^{m}\frac{e^{-(\alpha^*_i+\beta^*_j)}}{1-e^{-(\alpha^*_i+\beta^*_j)}},\quad i=1,n,\label{eq:2side1w}\\
c_j^*&=&\sum_{i=1}^{n}\frac{e^{-(\alpha^*_i+\beta^*_j)}}{1-e^{-(\alpha^*_i+\beta^*_j)}},\quad j=1,m.\label{eq:2side2w}
\end{eqnarray}
that can be solved numerically. The solution $(\vec{\alpha}^*, \vec{\beta}^*)$, when inserted into $P_{\textrm{can}}(\mathbf{G}|\vec{\alpha}, \vec{\beta})$, completely characterizes the canonical ensemble.
The resulting canonical entropy is 
\begin{eqnarray}
  S^*_{\textrm{can}}&=&-\ln P_{\textrm{can}}(\mathbf{G}^*|\vec{\alpha}^*,\vec{\beta}^*)\label{Canonical:entropy:weighted:two}\\
&=&\vec{\alpha}^*\cdot\vec{r}^*+\vec{\beta}^*\cdot\vec{c}^*-\sum_{i=1}^{n}\sum_{j=1}^{m}\ln[1-e^{-(\alpha^*_i+\beta^*_j)}]\nonumber
\end{eqnarray}
and an upper bound is provided by the canonical entropy given in Eq.~\eqref{Canoni-entropy_wei1sided} for the one-sided case with the same row constraints $\vec{r}^*$:
\begin{equation}
 S^*_{\textrm{can}}\le\sum_{i=1}^{n}\ln\frac{{(m+r_i^*)}^{m+r_i^*}}{{(r_i^*)}^{r_i^*}{m}^{m}}.\label{eq:boundcanw}
\end{equation}

As in the binary two-sided case, the microcanonical entropy $S^*_{\textrm{mic}}$ cannot be computed explicitly, but it can still be evaluated asymptotically from the determinant of the canonical covariance matrix $\mathbf{\Sigma}^*$ using Eqs.~\eqref{eq:approxalpha} and~\eqref{Relative-engenvale}.
Correspondingly, the relative entropy can be computed using Eq.~\eqref{eq:Salpha}.
The microcanonical entropy given by Eq.~\eqref{01-micro-states-local-z-redine-2W} for the corresponding one-sided case is still a strict upper bound for the two-sided entropy:
\begin{equation}\label{eq:boundmicw}
S^*_{\textrm{mic}}<\sum_{i=1}^{n}\ln\binom{m+r^*_i-1}{r^*_i}.
\end{equation}

As in the corresponding binary case, in the microcanonical ensemble both $\vec{r}$ and $\vec{c}$ are deterministic and fixed to the values $\vec{r}^*$ and $\vec{c}^*$, while in the canonical ensemble they are random with expected values $\vec{r}^*$ and $\vec{c}^*$.
The coupled hard constraints $(\vec{r}(\mathbf{G}),\vec{c}(\mathbf{G}))=(\vec{r}^*,\vec{c}^*)$ create mutual dependencies among all the entries of $\mathbf{G}$ in the microcanonical ensemble. By contrast, the soft constraint $(\langle\vec{r}\rangle_{\vec{\alpha}^*},\langle\vec{c}\rangle_{\vec{\beta}^*})=(\vec{r}^*,\vec{c}^*)$ leaves all entries of $\mathbf{G}$ independent in the canonical ensemble. 
In the latter, as for all weighted matrices discussed so far, each entry $g_{ij}$ is geometrically distributed, but now with its specific parameters:
\begin{equation}
p(g_{ij}|\vec{\alpha}^*,\vec{\beta}^*)={e^{-(\alpha_i^*+\beta^*_i) g_{ij}}}\left[1-e^{-(\alpha_i^*+\beta^*_i) }\right]
\label{eq:geometric2sided**}
\end{equation}
for $g_{ij}\in\{0,1,2,\dots\}$, as we show in the Appendix.

Here as well, the special case where the column sums are all equal to each other ($c^*_j=c^*$ for all $j$) provides a nice example. The corresponding Lagrange multipliers are in this case all equal to each other ($\beta_j^*=\beta^*$ for all $j$) and this allows us to solve for all parameters explicitly. In particular, Eqs.~\eqref{eq:2side1w} and~\eqref{eq:2side2w} reduce to the $n+1$ independent equations
\begin{eqnarray}
r_i^*&=&{m}\frac{e^{-(\alpha^*_i+\beta^*)}}{1-e^{-(\alpha^*_i+\beta^*)}},\quad i=1,n,\label{eq:2side1wex}\\
c^*&=&\sum_{i=1}^{n}\frac{e^{-(\alpha^*_i+\beta^*)}}{1-e^{-(\alpha^*_i+\beta^*)}},\label{eq:2side2wex}
\end{eqnarray}
where, again, the second equation is equivalent to the consistency condition $c^*=\sum_{i=1}^{n}r^*_i/m$. Inverting Eq.~\eqref{eq:2side1wex}, we obtain explicitly
\begin{equation}\label{Param:localrow:weighted:two}
\alpha^*_i+\beta^*=\ln \frac{m+r^*_i}{r^*_i}
\end{equation}
which, if inserted into the expression for $S^*_{\textrm{can}}$ in Eq.~\eqref{Canonical:entropy:weighted:two}, produces exactly the canonical entropy found previously in Eq.~\eqref{Canoni-entropy_wei1sided} for the weighted ensemble with one-sided (row) constraints specified by the same vector $\vec{r}^*$:
\begin{eqnarray}
  S^*_{\textrm{can}}&=&\vec{\alpha}^*\cdot\vec{r}^*+m{\beta}^*{c}^*-m\sum_{i=1}^{n}\ln[1-e^{-(\alpha^*_i+\beta^*)}]\nonumber\\
&=&\sum_{i=1}^{n}\ln\frac{{(m+r_i^*)}^{m+r_i^*}}{{(r_i^*)}^{r_i^*}{m}^{m}}.\label{Canonical:entropy:weighted:twoc}
\end{eqnarray}
The upper bound in Eq.~\eqref{eq:boundcanw} is therefore fully saturated.
Again, while the canonical ensembles are identical for the two cases, the microcanonical ensembles remain very different and the microcanonical entropy under two-sided constraints is strictly smaller than the one under one-sided constraints: the upper bound in Eq.~\eqref{eq:boundmicw} cannot be saturated.
Similarly, the determinant of the covariance matrix $\mathbf{\Sigma}^*$, which here is a $(n+m)\times(n+m)$ matrix (that we calculate later on), is different from the one obtained in the one-sided case.

The (non)equivalence of the two ensembles is discussed below, in conjunction with the case of two-sided binary matrices.

\subsubsection{Ensemble nonequivalence for matrices under two-sided local constraints}
To investigate EN in the two-sided case, it is convenient to preliminary combine the results obtained so far in the binary ($+$) and weighted ($-$) cases as follows.

The canonical entropy $S^\pm_n[P^*_{\textrm{can}}]$ can be evaluated by combining Eqs.~\eqref{Canonical:entropy:binary:two} and~\eqref{Canonical:entropy:weighted:two}, as well as the corresponding upper bounds given by Eqs.~\eqref{eq:boundcan} and~\eqref{eq:boundcanw}, into 
\begin{eqnarray}
S^\pm_n[P^*_{\textrm{can}}]&=&\vec{\alpha}^*\cdot\vec{r}^*+\vec{\beta}^*\cdot\vec{c}^*
\pm\sum_{i=1}^{n}\sum_{j=1}^{m}\ln[1\pm e^{-(\alpha^*_i+\beta^*_j)}]\nonumber\\
&\le& \sum_{i=1}^{n}\ln\frac{m^{\pm m}}{{(r^*_i)}^{r^*_i}{(m\mp r^*_i)}^{\pm m-r^*_i}}\label{Canonical:entropy:both:two}
\end{eqnarray}
(see Eq.~\eqref{Canentropy_combo} for a comparison).
It is easy to check that, in all the three regimes considered (sparse, dense with diverging $m$, dense with finite $m$), the above canonical entropy has the same qualitative behaviour as the corresponding quantity obtained previously in Eq.~\eqref{Canentropy_combo} for the one-sided case.

Unlike the one-sided case, the microcanonical entropy cannot be evaluated exactly, neither through a direct combinatorial formula nor via the complex integral approach, and we only have strict upper bounds given by Eqs.~\eqref{eq:boundmic} and~\eqref{eq:boundmicw} in the binary and weighted case respectively, which we can combine as follows:
\begin{equation}
S^\pm_n[P^*_{\textrm{mic}}]< \sum_{i=1}^{n}\ln\left[(\pm 1)^{r^*_i}\binom{\pm m}{r^*_i}\right]
\label{micentropy:both:bounds}
\end{equation}
(see Eq.~\eqref{Micentropy_combo} for a comparison).

We can now discuss EN in a combined fashion for binary and weighted matrices.
While we cannot calculate the relative entropy exactly, we can correctly evaluate its asymptotic scaling via Eq.~\eqref{eq:Salpha}, because the canonical covariance matrix $(\mathbf{\Sigma}^*)^\pm$ between the constraints can still be calculated analytically as a function of the parameters $(\vec{\alpha}^*,\vec{\beta}^*)$, in both the binary and weighted cases. 
In particular, it is easy to see that the entries $({\Sigma}^*_{ij})^\pm$ are arranged into a block structure, with a square $n\times n$ diagonal block ($i,j\in[1,n]$) representing the covariance matrix between pairs of row sums, a square $m\times m$ diagonal block ($i,j\in[n+1,n+m]$) representing the covariance matrix between pairs of column sums, and two rectangular ($n\times m$ and $m\times n$) off-diagonal blocks representing the covariances between row and column sums ($i\in[1,n],\,j\in[n+1,n+m]$ and $i\in[n+1,n+m],\,j\in[1,n]$).
As we show in the Appendix, these entries are
\begin{widetext}
\begin{equation}\label{eq:curlymain1}
	{(\Sigma^*_{ij})}^\pm
=\left\{\begin{array}{rl}\delta_{ij}\sum_{k=1}^m\frac{\displaystyle e^{-(\alpha^*_i+\beta^*_k)}}{\displaystyle\left[1\pm e^{-(\alpha^*_i+\beta^*_k)}\right]^2}&i,j\in[1,n],\\
\frac{\displaystyle e^{-(\alpha^*_i+\beta^*_{j-n})}}{\displaystyle\left[1\pm e^{-(\alpha^*_i+\beta^*_{j-n})}\right]^2}&i\in[1,n],\,j\in[n+1,n+m]\\
\frac{\displaystyle e^{-(\alpha^*_j+\beta^*_{i-n})}}{\displaystyle\left[1\pm e^{-(\alpha^*_j+\beta^*_{i-n})}\right]^2}&i\in[n+1,n+m],\,j\in[1,n]\\
\delta_{ij}\sum_{k=1}^n\frac{\displaystyle e^{-(\alpha^*_k+\beta^*_{j-n})}}{\displaystyle\left[1\pm e^{-(\alpha^*_k+\beta^*_{j-n})}\right]^2}&i,j\in[n+1,n+m]
\end{array}\right..
\end{equation}
\end{widetext}
The above expression is the generalization of Eq.~\eqref{eq:varboth1sidedmain} to the case of two-sided constraints.
Once the values of $\vec{r}^*$ and $\vec{c}^*$ are specified, one can calculate the determinant of the above matrix and, through Eq.~\eqref{eq:Salpha}, the leading order of the relative entropy $S^\pm_n[P^*_{\textrm{mic}}||P^*_{\textrm{can}}]$. 
As we show in the Appendix, the order of $\alpha^\pm_n$ confirms the same scalings for the relative entropy found previously in Eqs.~\eqref{Rebothonesidesparse},~\eqref{eq:relativeonesidedbothapprox} and~\eqref{Rebothonesidedense} for the one-sided case: namely, $\alpha^\pm_n =O(n)$ in the sparse regime, $\alpha^\pm_n=O(n\ln n)$ in the dense regime with $m=O(n)$, and $\alpha^\pm_n=O(n)$ in the dense regime with finite $m$.

In practice, unlike the one-sided case, calculating the values of $S^\pm_n[P^*_{\textrm{mic}}||P^*_{\textrm{can}}]$ and $R^\pm_\infty$ (or bounds for them) as explicit functions of the constraints is not easy in general. 
It is however possible, and instructive, to consider a special case where  $S^\pm_n[P^*_{\textrm{mic}}||P^*_{\textrm{can}}]$ and $R^\pm_\infty$ in this two-sided case (given the vectors $\vec{r}^*$ and $\vec{c}^*$) can be related to the corresponding values obtained in the one-sided case with the same vector $\vec{r}^*$ (but without a constraint on $\vec{c}^*$).
Indeed, if we consider again the special case with constant column constraints ($c_j^*=c^*$, $j=1,m$) then from our previous results in Eqs.~\eqref{Canonical:entropy:binary:twoc} and~\eqref{Canonical:entropy:weighted:twoc} we recall that, for any given value of $n$, the two-sided canonical entropy $S^\pm_n[P^*_{\textrm{can}}]$ is exactly equal to the one-sided canonical entropy given in Eq.~\eqref{Canentropy_combo} corresponding to the same vector $\vec{r}^*$, while of course the two-sided microcanonical entropy $S^\pm_n[P^*_{\textrm{mic}}]$ is strictly smaller than the one-sided one given in Eq.~\eqref{Micentropy_combo}.
This automatically implies that $S^\pm_n[P^*_{\textrm{mic}}||P^*_{\textrm{can}}]$ in the two-sided case is strictly larger than the corresponding one-sided relative entropy given in Eq.~\eqref{Rebothoneside}. This proves that the scaling of the relative entropy is always at least $O(n)$, irrespective of the density and of the value of $m$: in all regimes, EE breaks down for binary and weighted matrices under two-sided local constraints, as found in the one-sided case. The presence of the extra column constraints is not changing the qualitative behaviour of the relative entropy, but only its numerical value. Since the assumption of constant column sums only changes the values, but not the order, of the relative entropy, we expect that the scalings remain unchanged in the general case as well.

Moreover, EN has again the strong form ($R^\pm_\infty>0$) in the sparse regime with finite $m$, because the value of $R^\pm_ n=1-{S^\pm_n[P^*_{\textrm{mic}}]}/{S^\pm_n[P^*_{\textrm{can}}]}$ in the two-sided case is strictly larger than the corresponding one calculated previously for the one-sided case.
In particular, we can use Eq.~\eqref{Rbothn} to establish the following lower bound in the two-sided case with constant column constraints and finite $m$:
\begin{equation}
R^\pm_n>1-\frac{\sum_{i=1}^{n}\ln\left[(\pm 1)^{r^*_i}\binom{\pm m}{r^*_i}\right]}{\sum_{i=1}^{n}\ln\frac{m^{\pm m}}{{(r^*_i)}^{r^*_i}{(m\mp r^*_i)}^{\pm m-r^*_i}}}>0.
\label{Rbothn2side}
\end{equation}
The above inequality proves strong EN in this case as well. Again, we expect that relaxing the assumption of constant column sums will change only the value of $R^\pm_n$, but not its strict positivity.

Finally, we can also establish an upper bound for $R^\pm_n$ by rewriting Eq.~\eqref{eq:Salpha} asymptotically for large $n$, in analogy with Eq.~\eqref{mecojoni1}, as
\begin{equation}
S^\pm_n[P^*_{\textrm{mic}}||P^*_{\textrm{can}}]=C_2(m){\alpha}_n^\pm
\label{mecojoni2}
\end{equation}
where $C_2(m)$ is a finite positive constant and noticing that, since the covariance matrix $(\mathbf{\Sigma}^*)^\pm$ is positive-definite, we can use Hadamard's inequality stating that the determinant of a positive-definite matrix is less than or equal to the product of the diagonal entries of the matrix. This means
\begin{equation}
{\alpha}_n^\pm=\ln\sqrt{\det\left[2\pi(\mathbf{\Sigma}^*)^\pm\right]}\le\tilde{\alpha}_n^{\pm}
\label{eq:sobonitutti}
\end{equation}
where, using Eq.~\eqref{eq:curlymain1}, we have introduced
\begin{eqnarray}
\tilde{\alpha}_n^{\pm}&=&\frac{1}{2}\sum_{i=1}^{n}\ln{\left[2\pi(\Sigma^*_{ii})^\pm\right]}+\frac{1}{2}\sum_{j=1}^{m}\ln{\left[2\pi(\Sigma^*_{jj})^\pm\right]}.\quad
\label{eq:dettwosided}
\end{eqnarray}
Now, using Eqs.~\eqref{Param:localrow:binary:two} and~\eqref{Param:localrow:weighted:two} in the binary and weighted case respectively, it is easy to show that, in the two-sided case with constant column sums, the first $n$ diagonal entries of $(\mathbf{\Sigma}^*)^\pm$ are identical to the $n$ diagonal entries of the covariance matrix in the corresponding one-sided case given by Eq.~\eqref{eq:varboth1sidedmain}, i.e.
\begin{equation}
(\Sigma_{ii}^*)^\pm=r_i^*\left(1-\frac{r_i^*}{\pm m}\right),\quad i=1,n.
\end{equation}
Inserting the above expression into Eq.~\eqref{eq:dettwosided}, and noticing that the last sum in the latter is strictly positive, we can write
\begin{eqnarray}
\tilde{\alpha}_n^{\pm}<\frac{1}{2}\sum_{i=1}^{n}\ln{\left[2\pi r_i^*\left(1-\frac{r_i^*}{\pm m}\right)\right]}.
\label{apritutto}
\end{eqnarray}
Combining Eqs.~\eqref{mecojoni2},~\eqref{eq:sobonitutti} and~\eqref{apritutto} we obtain the upper bound (for large $n$)
\begin{eqnarray}
R^\pm_n&=&C_2(m)\frac{{\alpha}_n^\pm}{S^\pm_n[P^*_{\textrm{can}}]}\label{semprepeggio}\\
&\le&C_2(m)\frac{\tilde{\alpha}_n^\pm}{S^\pm_n[P^*_{\textrm{can}}]}\nonumber\\
&<&\frac{C_2(m)}{2}\frac{\sum_{i=1}^{n}\ln{\left[2\pi r_i^*\left(1-\frac{r_i^*}{\pm m}\right)\right]}}{\sum_{i=1}^{n}\ln\frac{m^{\pm m}}{{(r^*_i)}^{r^*_i}{(m\mp r^*_i)}^{\pm m-r^*_i}}}\nonumber\\
&=&\frac{C_2(m)}{C_1(m)}\left[1-\frac{\sum_{i=1}^{n}\ln\left[(\pm 1)^{r^*_i}\binom{\pm m}{r^*_i}\right]}{\sum_{i=1}^{n}\ln\frac{m^{\pm m}}{{(r^*_i)}^{r^*_i}{(m\mp r^*_i)}^{\pm m-r^*_i}}}\right],\nonumber
\end{eqnarray}
where we have used Eqs.~\eqref{Rbothn} and~\eqref{Rbothnasympto} established in the one-sided case.
Comparing Eq.~\eqref{semprepeggio} with Eq.~\eqref{Rbothn2side} we see that we must have $C_2(m)/C_1(m)>1$. We conjecture that, in analogy with $C_1(m)$ in the one-sided case, $C_2(m)\gtrsim 1$. Moreover, here as well we know that $\lim_{m\to\infty}C_2(m)=1$ as in Eq.~\eqref{eq:apijanderculo}. This means that we expect that, for $n$ large, $C_2(m)/C_1(m)\approx1$ so that the upper bound in Eq.~\eqref{semprepeggio} approaches the lower bound in Eq.~\eqref{Rbothn2side}, which is therefore a very good estimate of the actual value of $R^\pm_n$ in the two-sided case with constant column constraints:
\begin{equation}
R^\pm_n\approx 1-\frac{\sum_{i=1}^{n}\ln\left[(\pm 1)^{r^*_i}\binom{\pm m}{r^*_i}\right]}{\sum_{i=1}^{n}\ln\frac{m^{\pm m}}{{(r^*_i)}^{r^*_i}{(m\mp r^*_i)}^{\pm m-r^*_i}}}.
\label{Rbothn2sideguess}
\end{equation}
Upon comparison with Eq.~\eqref{Rbothn}, we see that $R^\pm_n$ remains practically unchanged with respect to the one-sided case with the same value of $\vec{r}^*$. 

Note that the above result means that the decrease $\Delta_\textrm{mic}$ in microcanonical entropy introduced by the extra column constraints is subleading with respect to the canonical entropy. Indeed, denoting with $\{\cdot\}_{h}$ a quantity evaluated in the $h$-sided case (where $h=1,2$), and exploiting again the identity of the canonical entropies $\{S^\pm_n[P^*_{\textrm{can}}]\}_1=\{S^\pm_n[P^*_{\textrm{can}}]\}_2$ and our conjecture $C_2(m)/C_1(m)\approx1$, we can use Eqs.~\eqref{relative-entropy=difference} and the results obtained so far to express the decrease in microcanonical entropy as  
\begin{eqnarray}
\Delta_\textrm{mic}&=&\{S^\pm_n[P^*_{\textrm{mic}}]\}_1 -\{S^\pm_n[P^*_{\textrm{mic}}]\}_2\nonumber\\
&=&\{S^\pm_n[P^*_{\textrm{mic}}||P^*_{\textrm{can}}]\}_2
-\{S^\pm_n[P^*_{\textrm{mic}}||P^*_{\textrm{can}}]\}_1\nonumber\\
&=&C_2(m)\{{\alpha}_n^\pm\}_2-C_1(m)\{{\alpha}_n^\pm\}_1\nonumber\\
&<&C_2(m)\{\tilde{\alpha}_n^\pm\}_2-C_1(m)\{{\alpha}_n^\pm\}_1\nonumber\\
&\approx&\frac{C_1(m)}{2}\sum_{j=1}^{m}\ln{\left[2\pi(\Sigma^*_{jj})^\pm\right]}\nonumber\\
&=&C_1(m)\frac{m}{2}\ln\sum_{k=1}^n\frac{ 2\pi\, e^{-(\alpha^*_k+\beta^*)}}{\left[1\pm e^{-(\alpha^*_k+\beta^*)}\right]^2}\nonumber\\
&=&C_1(m)\frac{m}{2}\ln\sum_{k=1}^n{\left[2\pi \frac{r_i^*}{m}\left(1-\frac{r_i^*}{\pm m}\right)\right]},
\end{eqnarray}
which is of order $O(\ln n)$, while the canonical entropy is of order $O(n)$ in the dense case with finite $m$ considered here.

Clearly, if we additionally consider constant row constraints, i.e. $r^*_i=r^*$ for $i=1,n$ (where necessarily $r^*=c^*m/n$), then in analogy with Eq.~\eqref{Ratio:oneside:bothequal} we can establish the following explicit lower bound for the value of $R^\pm_\infty$ in the two-sided case with constant row and column constraints:
\begin{equation}
R^\pm_\infty>1-\frac{\ln\left[(\pm 1)^{r^*}\binom{\pm m}{r^*}\right]}{\ln\frac{m^{\pm m}}{{(r^*)}^{r^*}{(m\mp r^*)}^{\pm m-r^*}}}>0.
\label{Ratio:twoside:bothequal}
\end{equation}
Our expectation in Eq.~\eqref{Rbothn2sideguess} suggests that the above lower bound is a very good approximation for the actual value of $R^\pm_\infty$:
\begin{equation}
R^\pm_\infty\approx1-\frac{\ln\left[(\pm 1)^{r^*}\binom{\pm m}{r^*}\right]}{\ln\frac{m^{\pm m}}{{(r^*)}^{r^*}{(m\mp r^*)}^{\pm m-r^*}}},
\label{Ratio:twoside:bothequalguess}
\end{equation}
leading to the same result as in Eq.~\eqref{Ratio:oneside:bothequal} for the one-sided case.

The above results generalize the finding of strong EN to the two-sided case, again in the dense regime with finite $m$. The results do not change qualitatively, and apparently only slightly quantitatively, with respect to the one-sided case. This result points again at the fact that it is the extensivity of the constraints that plays the key role for EN: adding a finite number $m$ of (column) constraints does not relevantly change the picture already obtained in the one-sided case.

\section{Discussion and conclusions}
We have studied the problem of EN in the general context of $n\times m$ matrices with given constraints. 
Such matrices can represent high-dimensional data such as multivariate time series, expression profiles, multiplex social activity, and other relational or structured data encountered in many settings. Their entries can either be binary (Boolean) or weighted (non-negative integers).
The constraints imposed on these matrices represent sums over either all the entries of the matrix (single global constraint) or over individual rows (local one-sided constraints) and possibly also columns (local two-sided constraints). These constraints take the form of linear terms into the Hamiltonian at the exponent of the maximum-entropy probability distribution characterizing the matrix ensemble.

Global constraints do not account for the heterogeneity (either spatial or temporal, i.e. nonstationarity) in the physical data-generating process, as they lead to probability distributions with identical parameters for all the entries of the matrix.
By contrast, local constraints produce probability distributions with different local (row- and possibly column-specific) parameters.
Most modern data structures are heterogeneous and/or nonstationary, and are therefore characterized by (at least) the type of local constraints considered here.
Indeed, maximum-entropy ensembles with local constraints are being increasingly used, either as null models for pattern detection or even as generative models and inference methods whenever there is only partial, local information available about the system~\cite{squartini2017maximum,cimini2019statistical}.

We have shown that local constraints break the asymptotic (i.e. for large $n$) equivalence of canonical and microcanonical ensembles, where the constraints are enforced in a soft and hard manner respectively. By contrast, global constraints preserve EE. 
Mathematically, EE is encountered when the relative entropy between the canonical and microcanonical probability distributions is $o(n)$. 
Importantly, the breakdown of EE observed here under local constraints occurs without phase transitions, which would require nonlinear constraints in the Hamiltonian and are therefore deliberately excluded from the cases we considered. 
The form of EN we observe under local constraints is also `unrestricted', i.e. it holds for any value of the model parameters (here, for any graphical value of the constraints), while the mechanism for EN based on phase transitions requires specific parameters or phases.
Our results hold in all regimes of density and for all values of $m$, and therefore generalize a recently discovered, alternative mechanism for the breakdown of EE observed so far in ensembles of binary graphs with given degree sequence~\cite{squartini2015breaking,garlaschelli2016ensemble,garlaschelli2018covariance,roccaverde2019breaking} and weighted graphs with given strength sequence~\cite{zhang2020ensemble}.

At the same time, our results highlight a qualitatively new finding. While the systems with local constraints studied in the past exhibited a `weak' degree of EN (where the relative entropy is of smaller order compared with the canonical entropy, while still growing at least linearly in $n$), here we identified a regime for which EN is as `strong' as in presence of phase transitions (i.e. with the relative entropy being of the same order as the canonical entropy).
This regime is obtained when both $m$ and the expected value of each entry of the matrix are finite, i.e. $O(1)$.
In practice, this means that the data structure is one where $n$ grows as the size of the system grows, while $m$ remains finite. This circumstance is naturally encountered e.g. when $n$ represents a large number of timesteps during which a small number $m$ of synchronous time series are observed (e.g. for EEG signals), or when $n$ represents a large number of genes for which expression levels are observed in a small number $m$ of cells at the same time, or when $n$ represents a large number of users whose activities or preferences are recorded for a small number $m$ of platforms, items, or other dimensions.

The simultaneously `strong' and `unrestricted' form of EN discussed here has never been documented so far, to the best of our knowledge. Indeed, in all the settings that had been studied previously, $m$ was necessarily equal to $n$ since the matrices represented the special case of square adjacency matrices of graphs, therefore the regime leading to strong EN could not be observed.

EN has important practical consequences. 
A traditional expectation in statistical physics is that, in absence of phase transitions or long-range interactions, ensembles are equivalent and it is therefore legitimate to freely choose the ensemble to work with, e.g. based purely on mathematical or computational convenience. 
For instance, if the ensemble is used as a null model for a real system, one may either want to randomize the data numerically by keeping certain quantities fixed (in which case the microcanonical ensemble is the most efficient choice) or prefer an exact mathematical characterization of the probability of each configuration in the ensemble (in which case the canonical ensemble is the easiest to work with). 
This view has been challenged by the recent discovery of EN under local constraints. Nonequivalence imposes a principled choice of the ensemble, that can no longer be based on practical convenience. 
For instance, if one has reasons to believe that the hypothesis underlying the null model, or the partial information available about the system, should be treated as a hard constraint, then one is forced to choose the microcanonical ensemble. By contrast, if one believes the constraints should be treated as soft (for instance to account for possible measurement errors leading to noisy values of the constraints in the data), then one should take the canonical route. 

Our observation of strong EN shows that the quantitative differences between the two descriptions of the same system are much bigger than previously encountered in the case of weak EN.
These big quantitative differences are exemplified by Eqs.~\eqref{eq:exactR},~\eqref{eq:exactRR} and~\eqref{eq:exactRRR}.
A `wrong' choice of the ensemble can therefore lead to major errors in the estimation of the probability distribution characterizing the ensemble, of the resulting entropy, of the expected values of higher-order properties that are nonlinear functions of the constraints, etc. Conclusions of statistical analyses can therefore be highly biased.

However, besides this warning, the findings presented here are intended to offer also a constructive solution.
The fact that it is possible to rigorously quantify the differences between the two ensembles via the explicit calculation of the relative entropy ratio $R_n$ and its limiting value $R_\infty$ implies that one can still make a convenient choice of the ensemble, while at the same time being able to retrieve the desired results for the other ensemble via the calculated value of $R_\infty$.
In other words, besides being a warning signal for strong EN, $R_\infty$ is also a concrete tool allowing researchers to switch more easily between alternative descriptions of the same system, by compensating for their irreducible differences.
The calculations carried out here can hopefully serve as useful references for future quantitative research in a variety of domains.

\section*{Acknowledgments}
This work is supported by the Chinese Scholarship Council (No. 201606990001), the Dutch Econophysics Foundation (Stichting Econophysics, Leiden, the Netherlands), the Netherlands Organization for Scientific Research (NWO/OCW) and the EU project {SoBigData++}, contract no. 871042.

\bibliography{qi_refrence}

\appendix
\section{Global constraints}
Here we derive the main mathematical expressions for the case of global constraints on our ensembles of binary and weighted matrices. This corresponds to the case where the constraint $\vec{C}(\mathbf{G})$ is a simple scalar quantity ${C}(\mathbf{G})$ defined as the total value ${C}(\mathbf{G})\equiv t(\mathbf{G})\equiv\sum_{i=1}^n\sum_{j=1}^mg_{ij}$.
There is only one scalar Lagrange multiplier $\theta$ entering the definition of the Hamiltonian
\begin{equation}\label{Hamil-global-binary-1}
H(\mathbf{G},\theta)=\theta\, t(\mathbf{G})=\theta\sum_{i=1}^n\sum_{j=1}^mg_{ij}.
\end{equation}
The above Hamiltonian is the same for both binary and weighted matrices under a global constraint. However, the calculation of the partition function (hence of all the other properties) is different in the two cases.

\subsection{Binary matrices under a global constraint}
Let us consider binary matrices first ($g_{ij}=0,1$).
The partition function can be calculated as follows:
\begin{eqnarray}	Z(\theta)
&=&\sum_{\mathbf{G\in\mathcal{G}}}e^{-H(\mathbf{G},{\theta})}\nonumber\\
&=&\sum_{\mathbf{G}\in\mathcal{G}}e^{-\theta\sum_{i=1}^n\sum_{j=1}^m g_{ij}}\nonumber\\
&=&\sum_{\mathbf{G}\in\mathcal{G}}\prod_{i=1}^n\prod_{j=1}^me^{-\theta g_{ij}}\nonumber\\
&=&\prod_{i=1}^n\prod_{j=1}^m\sum_{g_{ij}=0,1}e^{-\theta g_{ij}}\nonumber\\
&=&\prod_{i=1}^n\prod_{j=1}^m(1+e^{-\theta})\nonumber\\
&=&(1+e^{-\theta})^{mn}.\label{Par-fun-1}
\end{eqnarray}
This leads to
\begin{eqnarray}
P_{\textrm{can}}(\mathbf{G}|{\theta})&=&\frac{e^{-H(\mathbf{G},{\theta})}}{Z(\theta)}\nonumber\\
&=&\frac{e^{-\theta\, t(\mathbf{G})}}{(1+e^{-\theta})^{mn}}\nonumber\\
&=&\prod_{i=1}^n\prod_{j=1}^m\frac{e^{-\theta g_{ij}}}{1+e^{-\theta}}
\label{eq:probglobal}
\end{eqnarray}
and to Eq.~\eqref{can-prob-binary-glob-1} in the main text.
Notice that Eq.~\eqref{eq:probglobal} reveals that each entry $g_{ij}$ of the matrix $\mathbf{G}$ is a Bernoulli-distributed random variable taking the value $g_{ij}=1$ with probability $p(1|\theta)={e^{-\theta}}/{(1+e^{-\theta})}$ and the value $g_{ij}=0$ with probability $p(0|\theta)={1}/{(1+e^{-\theta})}$, i.e.
\begin{equation}
p(g_{ij}|\theta)=\frac{e^{-\theta g_{ij}}}{1+e^{-\theta}}.
\label{eq:bernoulli}
\end{equation}
(the entries of $\mathbf{G}$ are therefore \emph{i.i.d.}).
The expected value of $g_{ij}$ is
\begin{equation}
\langle g_{ij}\rangle_\theta\equiv\sum_{g_{ij}=0,1}g_{ij}p(g_{ij}|\theta)
=\frac{e^{-\theta}}{1+e^{-\theta}},
\label{eq:expbinary}
\end{equation}
while its variance is
\begin{equation}
\textrm{Var}_\theta[g_{ij}]\equiv\langle g_{ij}^2\rangle_\theta-\langle g_{ij}\rangle_\theta^2=\frac{e^{-\theta}}{(1+e^{-\theta})^2}.
\label{eq:varbinary}
\end{equation}
The resulting expected value and variance of the constraint $t(\mathbf{G})$ are
\begin{eqnarray}
\langle t\rangle_\theta&=&\sum_{i=1}^n\sum_{j=1}^m\langle g_{ij}\rangle_\theta=nm\frac{e^{-\theta}}{1+e^{-\theta}},
\label{eq:exptbinary}\\
\textrm{Var}_\theta[t]&=&\sum_{i=1}^n\sum_{j=1}^m\textrm{Var}_\theta[g_{ij}]=nm\frac{e^{-\theta}}{(1+e^{-\theta})^2},
\label{eq:vartbinary}
\end{eqnarray}
the latter identity following from the fact that, since all the entries of $\mathbf{G}$ are mutually independent, the variance of the constraint $t(\mathbf{G})$ is the sum of all variances.

Now, we have to find the parameter value $\theta^*$ that solves Eq.~\eqref{eq:theta*} or equivalently maximizes the log-likelihood $\ln P_{\textrm{can}}(\mathbf{G}^*|\theta)$.
This can be done by setting the expected value $\langle t\rangle_{\theta^*}$ equal to the desired value $t^*$, which leads to
\begin{equation}\label{relationship-global-binary-1}
	e^{-\theta^*}=\frac{t^*}{mn-t^*},
\end{equation}
and to Eq.~\eqref{eq:thetaglobal} in the main text.
Inserting Eq.~\eqref{relationship-global-binary-1} into the expressions for $P_{\textrm{can}}(\mathbf{G}^*|{\theta})$ and $\textrm{Var}_{\theta}[t]$ leads to the values of $S^*_{\textrm{can}}$ and $\textrm{Var}_{\theta^*}[t]$ given in Eqs.~\eqref{Canoni-entropy} and~\eqref{eq:varbinaryglo} in the main text.
One can easily confirm that $\textrm{Var}_{\theta^*}[t]$ coincides with the only ($K=1$) entry of the $1\times 1$ covariance matrix $\mathbf{\Sigma}^*$ obtained through Eq.~\eqref{Sigma_unit}, i.e.
\begin{eqnarray}
{\Sigma}^*&=&\left.\frac{\partial^2\ln Z(\theta)}{\partial{\theta}^2}\right|_{{\theta}={\theta}^*}\nonumber\\
&=&t^*\left(1-\frac{t^*}{mn}\right)\nonumber\\
&=&\textrm{Var}_{\theta^*}[t]
\label{Sigma_unit1}
\end{eqnarray}
and, trivially,
\begin{equation}
 \det(\mathbf{\Sigma^*})=t^*\left(1-\frac{t^*}{mn}\right).
\end{equation}

The calculation of the microcanonical entropy $S^*_{\textrm{mic}}$ is in this case trivial and given by Eq.~\eqref{Micro-canonical-entropy-binary} in the main text.
Given the simplicity of this example, it is instructive to show explicitly that the integral formula in Eq.~\eqref{Omega-hard-constraints}, which can be calculated exactly in this case, gives the correct value of $\Omega_{t^*}$.
To do this, we use Eq.~\eqref{eq:probglobal} to obtain the complex-valued quantity
\begin{equation}
P_{\textrm{can}}(\mathbf{G}^*|{\theta}^*+\textrm{i}\psi)=\frac{e^{-(\theta^*+\textrm{i}\psi)t^*}}{[1+e^{-(\theta^*+\textrm{i}\psi)}]^{mn}}
\label{eq:complex}
\end{equation}
and use Eq.~\eqref{Omega-hard-constraints} to calculate $\Omega_{t^*}$ as
\begin{equation}
\Omega_{t^*}=\frac{1}{{2\pi}}\int_{-{\pi}}^{{\pi}}{[1+e^{-(\theta^*+\textrm{i}\psi)}]^{mn}}{e^{{(\theta^*+\textrm{i}\psi)t^*}}}\mathrm{d}\psi.
\end{equation}
To calculate the above integral, we change variable from $\psi$ to $z\equiv e^{-(\theta^*+\textrm{i}\psi)}$, so that
$\mathrm{d}z=\mathrm{d}e^{-(\theta^*+\textrm{i}\psi)}=-\textrm{i}z\mathrm{d}\psi$ and
$\mathrm{d}\psi=\textrm{i}\,\mathrm{d}z/z$.
Then the integral becomes
\begin{equation}\label{states-z1-redefin-1}
\Omega_{t^*}=\frac{\textrm{i}}{{2\pi}}\int_{e^{-(\theta^*-\textrm{i}\pi)}}^{e^{-(\theta^*+\textrm{i}\pi)}}(1+z)^{mn}z^{-(t^*+1)}\mathrm{d}z.
\end{equation}
and, using the binomial formula
\begin{equation}
(1+x)^{l}=\sum_{k=0}^{l}\binom{l}{k}x^{l-k},
\label{eq:bino}
\end{equation}
we obtain
\begin{equation}\label{states-z1-redefin-2}
\Omega_{t^*}=\frac{\textrm{i}}{2\pi}\sum_{k=1}^{mn}\binom{mn}{k}\int_{e^{-(\theta^*-\textrm{i}\pi)}}^{e^{-(\theta^*+\textrm{i}\pi)}}z^{mn-k-t^*-1}\mathrm{d}z.
\end{equation}
Now, each integral in the above sum can be calculated using Cauchy's residue theorem, from which we know that the integral is non-zero only when the exponent of $z$ is $-1$, in which case it equals $-2\pi\textrm{i}$. This selects the only value $k=mn-t^*$ in the sum, so that
\begin{equation}\label{OMGa-1}
\Omega_{t^*}=\frac{\textrm{i}}{2\pi}\binom{mn}{mn-t^*}(-2\pi\textrm{i})=\binom{mn}{t^*},
\end{equation}
which coincides with the binomial coefficient used in Eq.~\eqref{Micro-canonical-entropy-binary}.

\subsection{Weighted matrices under a global constraint}

We now consider the case of weighted matrices ($g_{ij}=0,1,2,\dots,+\infty$) with a global constraint $t^*$.
The Hamiltonian is still given by Eq.~\eqref{Hamil-global-binary-1}, while the partition function is now calculated differently as follows:
\begin{eqnarray}	Z(\theta)
&=&\sum_{\mathbf{G\in\mathcal{G}}}e^{-H(\mathbf{G},{\theta})}\nonumber\\
&=&\sum_{\mathbf{G}\in\mathcal{G}}e^{-\theta\sum_{i=1}^n\sum_{j=1}^m g_{ij}}\nonumber\\
&=&\sum_{\mathbf{G}\in\mathcal{G}}\prod_{i=1}^n\prod_{j=1}^me^{-\theta g_{ij}}\nonumber\\
&=&\prod_{i=1}^n\prod_{j=1}^m\sum_{g_{ij}=0}^{+\infty}e^{-\theta g_{ij}}\nonumber\\
&=&\prod_{i=1}^n\prod_{j=1}^m\frac{1}{1-e^{-\theta}}\nonumber\\
&=&\frac{1}{{(1-e^{-\theta})}^{mn}}.
\label{eq:partitionfunctionglobalW}
\end{eqnarray}
The canonical probability is therefore
\begin{eqnarray}
P_{\textrm{can}}(\mathbf{G}|{\theta})&=&\frac{e^{-H(\mathbf{G},{\theta})}}{Z(\theta)}\nonumber\\
&=&{e^{-\theta\, t(\mathbf{G})}}{(1-e^{-\theta})}^{mn}\nonumber\\
&=&\prod_{i=1}^n\prod_{j=1}^m e^{-\theta g_{ij}}(1-e^{-\theta}),\label{eq:probglobalw}
\end{eqnarray}
which leads to Eq.~\eqref{can-prob-weighted-glob-1} in the main text.
Equation~\eqref{eq:probglobalw} shows that all the entries of $\mathbf{G}$ are \emph{i.i.d.} random variables, in this case distributed according to a geometric distribution with success probability $e^{-\theta}$:
\begin{equation}
p(g_{ij}|\theta)=e^{-\theta g_{ij}}(1-e^{-\theta}).
\label{eq:geometric}
\end{equation}
The expected value of $g_{ij}$ is now
\begin{equation}
\langle g_{ij}\rangle_\theta=\sum_{g_{ij}=0}^{+\infty}g_{ij}p(g_{ij}|\theta)=\frac{e^{-\theta}}{1-e^{-\theta}}
\label{eq:expweighted}
\end{equation}
and its variance is
\begin{equation}
\textrm{Var}_\theta[g_{ij}]\equiv\langle g_{ij}^2\rangle_\theta-\langle g_{ij}\rangle_\theta^2=\frac{e^{-\theta}}{(1-e^{-\theta})^2}
\label{eq:varweighted}
\end{equation}
(note the change of sign at the denominator with respect to Eq.~\eqref{eq:varbinary}), from which we calculate the expected value and variance of the constraint $t(\mathbf{G})$ as
\begin{eqnarray}
\langle t\rangle_\theta&=&\sum_{i=1}^n\sum_{j=1}^m\langle g_{ij}\rangle_\theta=nm\frac{e^{-\theta}}{1-e^{-\theta}},
\label{eq:exptweighted}\\
\textrm{Var}_\theta[t]&=&\sum_{i=1}^n\sum_{j=1}^m\textrm{Var}_\theta[g_{ij}]=nm\frac{e^{-\theta}}{(1-e^{-\theta})^2}.
\label{eq:vartweighted}
\end{eqnarray}

The maximum-likelihood parameter value $\theta^*$ is found by setting the expected value $\langle t\rangle_{\theta^*}$ equal to $t^*$, resulting in
\begin{equation}
	e^{-\theta^*}=\frac{t^*}{mn+t^*}
\label{relationship-global-weighted-1}
\end{equation}
(notice again the change of sign with respect to the binary case) and to Eq.~\eqref{eq:thetaglobalw} in the main text.
Substituting Eq.~\eqref{relationship-global-weighted-1} into Eqs.~\eqref{eq:probglobalw} and~\eqref{eq:vartweighted} produces the expressions for $S^*_{\textrm{can}}$ and $\textrm{Var}_{\theta^*}[t]$ shown in Eqs.~\eqref{Canoni-entropyw} and~\eqref{eq:varweightedglo} in the main text.
As for the binary case, one can easily confirm that $\textrm{Var}_{\theta^*}[t]$ coincides with
\begin{eqnarray}
{\Sigma}^*&=&\left.\frac{\partial^2\ln Z(\theta)}{\partial{\theta}^2}\right|_{{\theta}={\theta}^*}\nonumber\\
&=&t^*\left(1+\frac{t^*}{mn}\right)\nonumber\\
&=&\textrm{Var}_{\theta^*}[t]
\label{Sigma_unit2}
\end{eqnarray}
so that
\begin{equation}
 \det(\mathbf{\Sigma^*})=t^*\left(1+\frac{t^*}{mn}\right).
\end{equation}

Again, it is instructive to show that the complex integral in Eq.~\eqref{Omega-hard-constraints} gives the exact result corresponding to the microcanonical entropy reported in Eq.~\eqref{microcanonical-entropy-w-global}.
Calculating the quantity
\begin{equation} P_{\textrm{can}}(\mathbf{G}^*|\theta^*+\textrm{i}\psi)=\frac{e^{-(\theta^*+\textrm{i}\psi)t^*}}{[1-e^{-(\theta^*+\textrm{i}\psi)}]^{-mn}}
\end{equation}
and inserting it into Eq.~\eqref{Omega-hard-constraints} yields
\begin{equation}
		\Omega_{t^*}= \frac{1}{2\pi}\int_{-\pi}^{\pi}\frac{e^{(\theta^*+\textrm{i}\psi)t^*}}{[1-e^{-(\theta^*+\textrm{i}\psi)}]^{mn}}\mathrm{d}\psi.
\end{equation}
We first perform the change of variable $y\equiv e^{-(\theta^*+\textrm{i}\psi)}$, $\mathrm{d}\psi=\textrm{i}\mathrm{d}y/y$ and rearrange the integral as
\begin{eqnarray}
\Omega_{t^*} &=&\frac{\textrm{i}}{2\pi }\int_{e^{-(\theta^*-\textrm{i}\pi)}}^{e^{-(\theta^*+\textrm{i}\pi)}} {y^{-(t^*+1)}}{\left(\frac{1}{1-y}\right)}^{mn}\mathrm{d}y\label{wm-ms-01-psi-global-2}\\
&=&\frac{\textrm{i}}{2\pi}\int_{e^{-(\theta^*-\textrm{i}\pi)}}^{e^{-(\theta^*+\textrm{i}\pi)}} {y^{-(t^*+1)}}{\left(1+\frac{y}{1-y}\right)}^{mn}\mathrm{d}y.\nonumber
\end{eqnarray}

Then we perform a second change of variable $z\equiv y/{(1-y)}$, $\mathrm{d}y=\mathrm{d}z/{(z+1)^2}$ and apply the binomial formula in Eq.~\eqref{eq:bino} twice to obtain
\begin{eqnarray}
\Omega_{t^*} & =&\frac{\textrm{i}}{2\pi }\sum_{k=0}^{mn}\binom{mn}{k} \int_{z_-}^{z_+}
{\left({\frac{z}{z+1}}\right)^{-(t^*+1)}}\frac{{z}^{k}}{(z+1)^2}\mathrm{d}z\nonumber\\
& =&\frac{\textrm{i}}{2\pi } \sum_{k=0}^{mn}\binom{mn}{k}\int_{z_-}^{z_+}   z^{k-t^*-1}(z+1)^{t^*-1}\mathrm{d}z\nonumber\\
& =&\frac{\textrm{i}}{2\pi} \sum_{k=0}^{mn}\binom{mn}{k}
\sum^{t^*-1}_{h=0}\binom{t^*-1}{h}
\int_{z_-}^{z_+}z^{k+h-t^*-1}\mathrm{d}z\nonumber
\end{eqnarray}
where we have defined
\begin{equation}
z_\pm\equiv{\frac{e^{-(\theta^*\pm\textrm{i}\pi)}}{1-e^{-(\theta^*\pm\textrm{i}\pi)}}}.
\end{equation}
Using again the residue theorem, the only non-zero integral is obtained for $h=t^*-k$, which selects the value
\begin{eqnarray}
\Omega_{t^*} & =&\frac{1}{2\pi\textrm{i}}\sum_{k=0}^{mn}\binom{mn}{k}\binom{t^*-1}{k-1} (-2\pi \textrm{i})\nonumber\\
& =&\sum_{k=0}^{mn}\binom{mn}{k}\binom{t^*-1}{k-1}\nonumber\\
& =&\sum_{k=0}^{mn}\binom{mn}{k}\binom{t^*-1}{t^*-k}\nonumber\\
& =&\binom{t^*+mn-1}{t^*}\label{value-omega-global-wm-1}
\end{eqnarray}
(where we have used the generalized Vandermonde's identity).
The above calculation retrieves exactly the negative binomial coefficient used in Eq.~\eqref{microcanonical-entropy-w-global}.

\section{One-sided local constraints}
We now consider the case of one-sided local constraints on ensembles of binary and weighted $n\times m$ matrices.
The constraint $\vec{C}(\mathbf{G})$ is now an $n$-dimensional ($K=n$) vector $\vec{r}(\mathbf{G})$ where the entry $r_i(\mathbf{G})=\sum_{j=1}^mg_{ij}$ is the $i$-th row sum of the matrix $\mathbf{G}$.
Correspondingly, there is an $n$-dimensional vector $\vec{\theta}$ of Lagrange multipliers and the Hamiltonian is
\begin{equation}
H(\mathbf{G},\vec{\theta})=\vec{\theta}\cdot \vec{r}(\mathbf{G})=\sum_{i=1}^n\theta_ir_i(\mathbf{G})=\sum_{i=1}^n\theta_i\sum_{j=1}^mg_{ij}
\label{Hamil-local-colum-1}
\end{equation}
for both binary and weighted matrices.
The calculation of the resulting properties of binary and weighted ensembles is discussed separately below.

\subsection{Binary matrices under one-sided local constraints}
In the binary case, the partition function $Z(\vec{\theta})$ can be calculated from Eq.~\eqref{Hamil-local-colum-1} according to the following generalization of Eq.~\eqref{Par-fun-1}:
\begin{eqnarray}	Z(\vec{\theta})
&=&\sum_{\mathbf{G\in\mathcal{G}}}e^{-H(\mathbf{G},\vec{\theta})}\nonumber\\
&=&\sum_{\mathbf{G}\in\mathcal{G}}e^{-\sum_{i=1}^n\theta_i\sum_{j=1}^m g_{ij}}\nonumber\\
&=&\sum_{\mathbf{G}\in\mathcal{G}}\prod_{i=1}^n\prod_{j=1}^me^{-\theta_i g_{ij}}\nonumber\\
&=&\prod_{i=1}^n\prod_{j=1}^m\sum_{g_{ij}=0,1}e^{-\theta_i g_{ij}}\nonumber\\
&=&\prod_{i=1}^n\prod_{j=1}^m(1+e^{-\theta_i})\nonumber\\
&=&\prod_{i=1}^{n}(1+e^{-\theta_i})^{m}.\label{Parti-fun-local-1}
\end{eqnarray}
The resulting canonical probability is
\begin{eqnarray}
P_{\textrm{can}}(\mathbf{G}|\vec{\theta})&=&\frac{e^{-H(\mathbf{G},\vec{\theta})}}{Z(\vec{\theta})}\nonumber\\
&=&\frac{e^{-\vec{\theta}\cdot \vec{r}(\mathbf{G})}}{\prod_{i=1}^{n}(1+e^{-\theta_i})^{m}}\nonumber\\
&=&\prod_{i=1}^{n}\prod_{j=1}^{m}\frac{e^{-\theta_ig_{ij}}}{1+e^{-\theta_i}},\label{Can-prob-local-1}
\end{eqnarray}
which leads to Eq.~\eqref{Can-prob-local-1main} in the main text.
As in the case of binary matrices under a global constraint, each entry $g_{ij}$ of the matrix $\mathbf{G}$ is a Bernoulli-distributed random variable. However, while all these entries are still independent, the parameter of the distribution depends on the row being considered:
\begin{equation}
p(g_{ij}|\vec{\theta})=\frac{e^{-\theta_i g_{ij}}}{1+e^{-\theta_i}}.
\label{eq:bernoulli1sided}
\end{equation}
Consequently, Eqs.~\eqref{eq:expbinary} and~\eqref{eq:varbinary} generalize to
\begin{eqnarray}
\langle g_{ij}\rangle_{\vec{\theta}}&\equiv&\sum_{g_{ij}=0,1}g_{ij}p(g_{ij}|\vec{\theta})
=\frac{e^{-\theta_i}}{1+e^{-\theta_i}},
\label{eq:expbinary1sided}\\
\textrm{Var}_{\vec{\theta}}[g_{ij}]&\equiv&\langle g_{ij}^2\rangle_{\vec{\theta}}-\langle g_{ij}\rangle_{\vec{\theta}}^2=\frac{e^{-\theta_i}}{(1+e^{-\theta_i})^2}.
\label{eq:varbinary1sided}
\end{eqnarray}
We can therefore calculate the expected value of each constraint $r_i(\mathbf{G})$ as
\begin{equation}
\langle r_i\rangle_{\vec{\theta}}=\sum_{j=1}^m\langle g_{ij}\rangle_{\vec{\theta}}=m\frac{e^{-\theta_i}}{1+e^{-\theta_i}},\quad i=1,n.
\label{eq:exptbinary1sided}
\end{equation}
Similarly, the variance of $r_i$ is
\begin{equation}
\textrm{Var}_{\vec{\theta}}[r_i]=\sum_{j=1}^m\textrm{Var}_{\vec{\theta}}[g_{ij}]=m\frac{e^{-\theta_i}}{(1+e^{-\theta_i})^2},\quad i=1,n
\label{eq:vartbinary1sided}
\end{equation}
while all covariances between different constraints are zero, because of the independence of distinct entries of $\mathbf{G}$:
\begin{equation}
\textrm{Cov}_{\vec{\theta}}[r_i,r_j]=\sum_{k=1}^m\sum_{l=1}^m\textrm{Cov}_{\vec{\theta}}[g_{ik},g_{jl}]=0,\quad i\ne j.
\label{eq:covtbinary1sided}
\end{equation}
We can combine Eqs.~\eqref{eq:vartbinary1sided} and~\eqref{eq:covtbinary1sided} as follows:
\begin{equation}
\textrm{Cov}_{\vec{\theta}}[r_i,r_j]=\delta_{ij}m\frac{e^{-\theta_i}}{(1+e^{-\theta_i})^2},
\label{eq:covvtbinary1sided}
\end{equation}
where $\delta_{ij}=1$ if $i=j$ and $\delta_{ij}=0$ if $i\ne j$.

Now, the parameter value $\vec{\theta}^*$ that maximizes the log-likelihood is found by equating the expected value $\langle \vec{r}\rangle_{\vec{\theta}^*}$ with the desired value $\vec{r}^*$. Inverting Eq.~\eqref{eq:exptbinary1sided}, this leads to
\begin{equation}\label{Relationship-theta-c-1}
	e^{-\theta^*_i}=\frac{r_i^*}{m-r_i^*}\qquad i=1,n
\end{equation}
or equivalently to Eq.~\eqref{eq:theta1sided} in the main text.
The expression for the canonical entropy $S^*_{\textrm{can}}$ given in Eq.~\eqref{Canoni-entropy_bin1sided} in the main text follows from substituting Eqs.\eqref{Relationship-theta-c-1} into Eq.~\eqref{Can-prob-local-1}.
Similarly, the expression for the entries of the $n\times n$ covariance matrix $\mathbf{\Sigma}^*$ given in Eq.~\eqref{eq:varbinary1sidedmain} in the main text follows from combining Eqs.~\eqref{eq:covvtbinary1sided} and~\eqref{Relationship-theta-c-1}.
Note that Eq.~\eqref{eq:varbinary1sidedmain} can also be obtained by differentiating the logarithm of Eq.~\eqref{Parti-fun-local-1} as prescribed by Eq.~\eqref{Sigma_unit}:
\begin{eqnarray}
{\Sigma}^*_{ij}&=&\left.\frac{\partial^2\ln Z(\vec{\theta})}{\partial{\theta_i}\partial{\theta_j}}\right|_{\vec{\theta}=\vec{\theta}^*}\nonumber\\
&=&\delta_{ij}r_i^*\left(1-\frac{r_i^*}{m}\right)\nonumber\\
&=&\textrm{Cov}_{\vec{\theta}^*}[r_i,r_j].
\label{Sigma_1sided}
\end{eqnarray}
These results imply
\begin{equation}
\det(\mathbf{\Sigma^*})=\prod_{i=1}^n r_i^*\left(1-\frac{r_i^*}{m}\right).
\end{equation}

The microcanonical entropy $S^*_{\textrm{mic}}$ can be directly calculated as Eq.~\eqref{01-micro-states-local-z-redine-2} in the main text.
We can still confirm that its value is exactly retrieved by using the integral formula in Eq.~\eqref{Omega-hard-constraints}.
From Eq.~\eqref{Can-prob-local-1} we obtain
\begin{equation}
	P_{\textrm{can}}(\mathbf{G}^*|\vec{\theta}^*+{\textrm{i}}\vec{\psi})=\prod_{i=1}^{n}\frac{e^{-(\theta_i^*+{\textrm{i}}\psi_i)r^*_i}}{[1+e^{-(\theta_i^*+{\textrm{i}}\psi_i)}]^m}.
\end{equation}
Using Eq.~\eqref{Omega-hard-constraints}, we can  calculate $\Omega_{\vec{r}^*}$ by exploiting again the binomial theorem, a change of variables ($z_i\equiv e^{-(\theta_i^*+\textrm{i}\psi_i)}$,
$\mathrm{d}z_i=-\textrm{i}z_i\mathrm{d}\psi_i$) and the residue theorem as in Eqs.~\eqref{states-z1-redefin-1},~\eqref{states-z1-redefin-2} and~\eqref{OMGa-1}:
\begin{eqnarray}
		\Omega_{\vec{r}^*}
&=&\int_{-\vec{\pi}}^{+\vec{\pi}}\frac{\mathrm{d}\vec{\psi}}{{(2\pi)}^n}\prod_{i=1}^{n}\frac{[1+e^{-(\theta_i^*+{\textrm{i}}\psi_i)}]^m}{e^{-(\theta_i^*+{\textrm{i}}\psi_i)r^*_i}}\nonumber\\
 &=&\prod_{i=1}^{n}\int_{-{\pi}}^{+{\pi}}\frac{\mathrm{d}{\psi_i}}{{2\pi}}\frac{[1+e^{-(\theta_i^*+{\textrm{i}}\psi_i)}]^m}{e^{-(\theta_i^*+{\textrm{i}}\psi_i)r^*_i}} \nonumber\\
&=&\prod_{i=1}^{n}\int_{-{\pi}}^{+{\pi}}\frac{\mathrm{d}{\psi_i}}{{2\pi}}\sum_{k=0}^{m}\binom{m}{k}e^{-(\theta_i^*+{\textrm{i}}\psi_i)(k-r^*_i)}\nonumber\\
&=&\prod_{i=1}^{n}\int_{\theta_i^*-\mathrm{i}{\pi}}^{\theta_i^*+\mathrm{i}{\pi}}\frac{\mathrm{d}{z_i}}{(-2\pi\textrm{i})}\sum_{k=0}^{m}{\binom{m}{k}}{z_i}^{k-r^*_i-1}\nonumber\\
&=&\prod_{i=1}^{n}\frac{(-2\pi\textrm{i})}{(-2\pi\textrm{i})}\binom{m}{r^*_i}\nonumber\\
&=&\prod_{i=1}^{n}\binom{m}{r^*_i}
\end{eqnarray}
which coincides with Eq.~\eqref{01-micro-states-local-z-redine-2}.

\subsection{Weighted matrices under one-sided local constraints}
In the weighted case, the partition function is given by the following generalization of Eq.~\eqref{eq:partitionfunctionglobalW}:
\begin{eqnarray}	Z(\vec{\theta})
&=&\sum_{\mathbf{G\in\mathcal{G}}}e^{-H(\mathbf{G},\vec{\theta})}\nonumber\\
&=&\sum_{\mathbf{G}\in\mathcal{G}}e^{-\sum_{i=1}^n\theta_i\sum_{j=1}^m g_{ij}}\nonumber\\
&=&\sum_{\mathbf{G}\in\mathcal{G}}\prod_{i=1}^n\prod_{j=1}^me^{-\theta_i g_{ij}}\nonumber\\
&=&\prod_{i=1}^n\prod_{j=1}^m\sum_{g_{ij}=0}^{+\infty}e^{-\theta_i g_{ij}}\nonumber\\
&=&\prod_{i=1}^n\frac{1}{(1-e^{-\theta_i})^m}.
\label{eq:partitionfunction1localW}
\end{eqnarray}
The resulting canonical probability is
\begin{eqnarray}
P_{\textrm{can}}(\mathbf{G}|\vec{\theta})&=&\frac{e^{-H(\mathbf{G},\vec{\theta})}}{Z(\vec{\theta})}\nonumber\\
&=&\frac{e^{-\vec{\theta}\cdot \vec{r}(\mathbf{G})}}{\prod_{i=1}^{n}(1-e^{-\theta_i})^{-m}}\nonumber\\
&=&\prod_{i=1}^{n}\prod_{j=1}^{m}\frac{e^{-\theta_ig_{ij}}}{(1-e^{-\theta_i})^{-1}},\label{Can-prob-local-1W}
\end{eqnarray}
leading to Eq.~\eqref{Can-prob-local-1mainW} in the main text.
As in the case of weighted matrices under a global constraint, each entry $g_{ij}$ of the matrix $\mathbf{G}$ is an independent and geometrically distributed random variable. On the other hand, as in the case of binary matrices under local constraints, the parameter of the distribution depends on the row being considered:
\begin{equation}
p(g_{ij}|\vec{\theta})={e^{-\theta_i g_{ij}}}{(1-e^{-\theta_i})}.
\label{eq:geometric1sided}
\end{equation}
The resulting expected value and variance of $g_{ij}$ are given by the following generalizations of Eqs.~\eqref{eq:expweighted} and~\eqref{eq:varweighted}:
\begin{eqnarray}
\langle g_{ij}\rangle_{\vec{\theta}}&\equiv&\sum_{g_{ij}=0,1}g_{ij}p(g_{ij}|\vec{\theta})
=\frac{e^{-\theta_i}}{1-e^{-\theta_i}},
\label{eq:expweighted1sided}\\
\textrm{Var}_{\vec{\theta}}[g_{ij}]&\equiv&\langle g_{ij}^2\rangle_{\vec{\theta}}-\langle g_{ij}\rangle_{\vec{\theta}}^2=\frac{e^{-\theta_i}}{(1-e^{-\theta_i})^2}.
\label{eq:varweighted1sided}
\end{eqnarray}
The expected value of each constraint $r_i(\mathbf{G})$ is therefore
\begin{equation}
\langle r_i\rangle_{\vec{\theta}}=\sum_{j=1}^m\langle g_{ij}\rangle_{\vec{\theta}}=m\frac{e^{-\theta_i}}{1-e^{-\theta_i}},\quad i=1,n,
\label{eq:exptweighted1sided}
\end{equation}
while the covariances between different constraints are
\begin{equation}
\textrm{Cov}_{\vec{\theta}}[r_i,r_j]=\delta_{ij}m\frac{e^{-\theta_i}}{(1-e^{-\theta_i})^2}.
\label{eq:covvtweighted1sided}
\end{equation}
As usual, note the change of sign at the denominator of Eqs.~\eqref{eq:exptweighted1sided} and~\eqref{eq:covvtweighted1sided} with respect to the corresponding Eqs.~\eqref{eq:exptbinary1sided} and~\eqref{eq:covvtbinary1sided} valid in the binary case.

Using Eq.~\eqref{eq:exptweighted1sided}, we set $\langle \vec{r}\rangle_{\vec{\theta}^*}=\vec{r}^*$ and solve for $\vec{\theta}^*$, finding
\begin{equation}\label{Relationship-theta-c-1W}
	e^{-\theta^*_i}=\frac{r_i^*}{m+r_i^*}\qquad i=1,n
\end{equation}
as the parameter value  that maximizes the log-likelihood. From the above expression, we get Eq.~\eqref{eq:theta1sidedW} and, using Eq.~\eqref{Can-prob-local-1W}, Eq.~\eqref{Canoni-entropy_wei1sided} in the main text.

Similarly, the expression for the entries of the $n\times n$ covariance matrix $\mathbf{\Sigma}^*$ given in Eq.~\eqref{eq:varbinary1sidedmain} in the main text follows from combining Eqs.~\eqref{eq:vartbinary1sided},~\eqref{eq:covtbinary1sided} and~\eqref{Relationship-theta-c-1}.
Note that Eq.~\eqref{eq:varbinary1sidedmain} can also be obtained by differentiating the logarithm of Eq.~\eqref{Parti-fun-local-1} as prescribed by Eq.~\eqref{Sigma_unit}:
\begin{eqnarray}
{\Sigma}^*_{ij}&=&\left.\frac{\partial^2\ln Z(\vec{\theta})}{\partial{\theta_i}\partial{\theta_j}}\right|_{\vec{\theta}=\vec{\theta}^*}\nonumber\\
&=&\delta_{ij}r_i^*\left(1+\frac{r_i^*}{m}\right)\nonumber\\
&=&\textrm{Cov}_{\vec{\theta}^*}[r_i,r_j].
\label{Sigma_1sidedW}
\end{eqnarray}
So, in analogy with the binary case,
\begin{equation}
\det(\mathbf{\Sigma^*})=\prod_{i=1}^n r_i^*\left(1+\frac{r_i^*}{m}\right).
\end{equation}

The microcanonical entropy $S^*_{\textrm{mic}}$ can be directly calculated as Eq.~\eqref{01-micro-states-local-z-redine-2W} in the main text.
We can still confirm that its value is correctly retrieved by using the integral formula in Eq.~\eqref{Omega-hard-constraints}.
From Eq.~\eqref{Can-prob-local-1W} we obtain
\begin{equation} P_{\textrm{can}}(\mathbf{G}^*|\vec{\theta}^*+{\textrm{i}}\vec{\psi})=\prod_{i=1}^{n}\frac{e^{-(\theta_i^*+{\textrm{i}}\psi_i)r^*_i}}{[1-e^{-(\theta_i^*+{\textrm{i}}\psi_i)}]^{-m}}.
\end{equation}
Using Eq.~\eqref{Omega-hard-constraints}, we can  calculate $\Omega_{\vec{r}^*}$ by exploiting again the binomial theorem as
\begin{equation}
\Omega_{\vec{r}^*}=\int_{-\vec{\pi}}^{\vec{\pi}}\frac{\mathrm{d}\vec{\psi}}{(2\pi)^{n}}\prod_{i=1}^{n}\frac{{e^{(\beta_i^*+\textrm{i}\psi_i)(r^*_i)}}}{[1-{e^{-(\beta_i^*+\textrm{i}\psi_i)}}]^m}.
\end{equation}
We can use the change of variables $y_i\equiv{e^{-(\beta_i^*+\textrm{i}\psi_i)}}$, $\mathrm{d}\psi_i=-{\textrm{i}}\mathrm{d}y_i/y_i$ and the relation $(1-y_i)^{-m}=(1+\frac{y_i}{1-y_i})^m$ to calculate $\Omega_{\vec{r}^*}$ as
\begin{eqnarray}
		\Omega_{\vec{r}^*} & =&\prod_{i=1}^{n}\int_{{\beta}_i^*-\textrm{i}{\pi}}^{{\beta}_i^*+\textrm{i}{\pi}}\frac{\mathrm{d}y_i}{2\pi\textrm{i}} \left(1+\frac{
			y_i}{1-y_i}\right)^m y_i^{-r^*_i-1}\\
		                      & =&\prod_{i=1}^{n}\int_{{\beta}_i^*-\textrm{i}{\pi}}^{{\beta}_i^*+\textrm{i}{\pi}}\frac{\mathrm{d}y_i}{2\pi\textrm{i}}\sum_{k=0}^{m}\binom{m}{k}y_i^{-r^*_i-1}\left(\frac{y_i}{1-y_i}\right)^k.\nonumber
\end{eqnarray}
Using another change of variables $z_i={y_i}/{(1-y_i)}$, $y_i={z_i}/{(z_i+1)}$, $\mathrm{d}y_i={\mathrm{d}z_i}/{(z_i+1)^2}$, and denoting $u^*_i={\frac{e^{-({\beta}_i^*+\textrm{i}{\pi})}}{1-e^{-({\beta}_i^*+\textrm{i}{\pi})}}}$, we find
\begin{eqnarray}
		\Omega_{\vec{r}^*} &=& \prod_{i=1}^{n}\int_{-{u}_i^*}^{{u}_i^*}\frac{\mathrm{d}z_i}{2\pi\textrm{i}} \sum_{k=0}^{m}\binom{m}{k}
		                       {\left(\frac{z_i}{z_i+1}\right)}^{-r^*_i-1}\frac{(z_i)^k}{(z_i+1)^2}\nonumber\\
 & =&\prod_{i=1}^{n}\int_{-{u}_i^*}^{{u}_i^*}\frac{\mathrm{d}z_i}{2\pi \textrm{i}} \sum_{k=0}^{m}\binom{m}{k}
		                       {(z_i+1)}^{r^*_i-1}
		(z_i)^{k-r_i^*-1}\nonumber\\
&=&\prod_{i=1}^{n}\int_{-{u}_i^*}^{{u}_i^*}\frac{\mathrm{d}z_i}{2\pi\textrm{i}}
\sum_{k=0}^{m}\binom{m}{k}
		                       \sum_{l=0}^{r^*_i-1}\binom{r^*_i-1}{l}z_i^{l+k-r^*_i-1}.\nonumber
\end{eqnarray}
Now, according to Cauchy's residue theorem, only when $l+k=r^*_i$ we get a non-zero value. This allows us to further write
\begin{equation} \Omega_{\vec{r}^*}=\prod_{i=1}^{n}\sum_{k=1}^{m}\binom{m}{k}\binom{r^*_i-1}{k-1}=\prod_{i=1}^{n}\binom{m+r_i^*-1}{r_i^*},\nonumber
\end{equation}
which coincides with Eq.~\eqref{micro:entropy:one:weighted} in the main text.

\section{Two-sided local constraints}
We now discuss ensembles of binary and weighted $n\times m$ matrices with two-sided local constraints. In this case $\vec{C}(\mathbf{G})$ is $(n+m)$-dimensional $(K=n+m)$ and specified by the two vectors $(\vec{r}(\mathbf{G}), \vec{c}(\mathbf{G}))$, where $\vec{r}(\mathbf{G})$ is still the $n$-dimensional vector of row sums of the matrix $\mathbf{G}$ (as in the one-sided case) and, additionally, $\vec{c}(\mathbf{G})$ is the $m$-dimensional vector of column sums of $\mathbf{G}$, with entries $c_j(\mathbf{G})=\sum_{i=1}^{n}g_{ij}$ ($j=1,m$). The corresponding Lagrange multipliers take the form $(\vec{\alpha}, \vec{\beta})$ where $\vec{\alpha}$ is $n$-dimensional and coupled to $\vec{r}(\mathbf{G})$, while $\vec{\beta}$ is $m$-dimensional and coupled to $\vec{c}(\mathbf{G})$. The corresponding Hamiltonian is
\begin{eqnarray}
H(\mathbf{G},\vec{\alpha}, \vec{\beta})&=&\sum_{i=1}^{n}\alpha_i r_i(\mathbf{G})+\sum_{j=1}^{m}\beta_jc_j(\mathbf{G})\nonumber\\
&=&\sum_{i=1}^{n}\sum_{j=1}^{m}(\alpha_i+\beta_j)g_{ij}.
\label{eq:H2}
\end{eqnarray}
As usual, the binary and weighted cases are discussed separately below.

\subsection{Binary matrices under two-sided local constraints}
Starting from the Hamiltonian in Eq.~\eqref{eq:H2}, the partition function of the canonical binary matrix ensemble can be still calculated exactly as a simple generalization of Eq.~\eqref{Parti-fun-local-1}:
\begin{eqnarray}
Z(\vec{\alpha},\vec{\beta})
&=&\sum_{\mathbf{G\in\mathcal{G}}}e^{-H(\mathbf{G},\vec{\alpha}, \vec{\beta})}\nonumber\\
&=&\sum_{\mathbf{G\in\mathcal{G}}}e^{-\sum_{i=1}^{n}\sum_{j=1}^{m}(\alpha_i+\beta_j)g_{ij}}\nonumber\\
&=&\sum_{\mathbf{G\in\mathcal{G}}}\prod_{i=1}^{n}\prod_{j=1}^{m}e^{-(\alpha_i+\beta_j)g_{ij}}\nonumber\\
&=&\prod_{i=1}^{n}\prod_{j=1}^{m}\sum_{g_{ij}=0,1}e^{-(\alpha_i+\beta_j)g_{ij}}\nonumber\\
&=&\prod_{i=1}^{n}\prod_{j=1}^{m}[1+e^{-(\alpha_i+\beta_j)}].
\label{eq:Z+}
\end{eqnarray}
The resulting probability is
\begin{eqnarray} 
P_{\textrm{can}}(\mathbf{G}|\vec{\alpha}, \vec{\beta})
&=&\frac{e^{-H(\mathbf{G},\vec{\alpha}, \vec{\beta})}}{Z(\vec{\alpha},\vec{\beta})}\nonumber\\
&=&\frac{e^{-\vec{\alpha}\cdot\vec{r}(\mathbf{G})-\vec{\beta}\cdot\vec{c}(\mathbf{G})}}{\prod_{i=1}^{n}\prod_{j=1}^{m}[1+e^{-(\alpha_i+\beta_j)}]}
\nonumber\\
&=&\prod_{i=1}^{n}\prod_{j=1}^{m}\frac{e^{-(\alpha_i+\beta_j)g_{ij}}}{1+e^{-(\alpha_i+\beta_j)}}.\label{eq:bibbo}
\end{eqnarray}
As in the case of binary matrices under global and one-sided local constraints, each entry $g_{ij}$ of the matrix $\mathbf{G}$ is still an independent and Bernoulli-distributed random variable, now controlled by the two parameters $\alpha_i$ and $\beta_j$. We can write the probability of $g_{ij}$ as
\begin{equation}\label{prob:entry:binary:two}
  p(g_{ij}|\vec{\alpha}, \vec{\beta})=\frac{e^{-(\alpha_i+\beta_j)g_{ij}}}{1+e^{-(\alpha_i+\beta_j)}}.
\end{equation}
The expected value of $g_{ij}$ is now
\begin{equation}\label{Covariance:binary;two}
\langle g_{ij}\rangle_{\vec{\alpha}, \vec{\beta}}\equiv\sum_{g_{ij}=0,1}g_{ij}\,p(g_{ij}|\vec{\alpha}, \vec{\beta})=\frac{e^{-(\alpha_i+\beta_j)}}{1+e^{-(\alpha_i+\beta_j)}}
\end{equation}
and the variance is
\begin{eqnarray}
  \textrm{Var}_{\vec{\alpha}, \vec{\beta}}[g_{ij}]&\equiv&\langle g_{ij}^2\rangle_{\vec{\alpha}, \vec{\beta}}-\langle g_{ij}\rangle_{\vec{\alpha}, \vec{\beta}}^2\nonumber\\
&=&\frac{e^{-(\alpha_i+\beta_j)}}{[1+e^{-(\alpha_i+\beta_j)}]^2}.\label{Variance:binary:two}
\end{eqnarray}
The resulting expected values of the constraints are
\begin{eqnarray}
\langle r_i\rangle_{\vec{\alpha}, \vec{\beta}}&=&\sum_{j=1}^{m}\frac{e^{-(\alpha_i+\beta_j)}}{1+e^{-(\alpha_i+\beta_j)}},\quad i=1,n,\\
\langle c_j\rangle_{\vec{\alpha}, \vec{\beta}}&=&\sum_{i=1}^{n}\frac{e^{-(\alpha_i+\beta_j)}}{1+e^{-(\alpha_i+\beta_j)}},\quad j=1,m.
\end{eqnarray}
The unique parameter values $(\vec{\alpha}^*, \vec{\beta}^*)$ that maximize the likelihood are found as usual by imposing that the expected values $(\langle\vec{r}\rangle_{\vec{\alpha}^*, \vec{\beta}^*},\langle\vec{c}\rangle_{\vec{\alpha}^*, \vec{\beta}^*})$ match the desired values $(\vec{r}^*,\vec{c}^*)$.
Unfortunately, in this case the values $(\vec{\alpha}^*, \vec{\beta}^*)$ cannot be determined analytically as a function of $(\vec{r}^*,\vec{c}^*)$, but they are defined implicitly by imposing 
\begin{equation}
(\vec{r}^*,\vec{c}^*)=(\langle\vec{r}\rangle_{\vec{\alpha}^*, \vec{\beta}^*},\langle\vec{c}\rangle_{\vec{\alpha}^*, \vec{\beta}^*}),
\label{eq:2sidedbapp}
\end{equation}
which leads to Eqs.~\eqref{eq:2side1b} and~\eqref{eq:2side2b} in the main text.

\subsection{Weighted matrices under two-sided local constraints}
In the canonical ensemble of weighted matrices under two-sided local constraints, the partition function is the following generalization of Eq.~\eqref{eq:partitionfunction1localW}:
\begin{eqnarray}
Z(\vec{\alpha},\vec{\beta})
&=&\sum_{\mathbf{G\in\mathcal{G}}}e^{-H(\mathbf{G},\vec{\alpha}, \vec{\beta})}\nonumber\\ &=&\sum_{\mathbf{G}\in\mathcal{G}}e^{-\sum_{i=1}^{n}\sum_{j=1}^{m}(\alpha_i+\beta_j)g_{ij}}\nonumber\\
&=&\sum_{\mathbf{G}\in\mathcal{G}}\prod_{i=1}^{n}\prod_{j=1}^{m}e^{-\sum_{i=1}^{n}\sum_{j=1}^{m}(\alpha_i+\beta_j)g_{ij}}\nonumber\\
&=&\prod_{i=1}^{n}\prod_{j=1}^{m}\sum_{g_{ij}=0}^{+\infty} e^{-(\alpha_i+\beta_j)g_{ij}}\nonumber\\
& =&\prod_{i=1}^{n}\prod_{j=1}^{m}\frac{1}{1-e^{-(\alpha_i+\beta_j)}}.
\label{eq:Z-}
\end{eqnarray}
The resulting canonical probability is
\begin{eqnarray} P_{\textrm{can}}(\mathbf{G}|\vec{\alpha},\vec{\beta})&=&\frac{e^{-H(\mathbf{G},\vec{\alpha}, \vec{\beta})}}{Z(\vec{\alpha},\vec{\beta})}\nonumber\\
&=&\frac{e^{-\vec{\alpha}\cdot\vec{r}(\mathbf{G})-\vec{\beta}\cdot\vec{c}(\mathbf{G})}}{\prod_{i=1}^{n}\prod_{j=1}^{m}{[1-e^{-(\alpha_i+\beta_j)}]}^{-1}}\nonumber\\
&=&\prod_{i=1}^{n}\prod_{j=1}^{m}\frac{e^{-(\alpha_i+\beta_j)g_{ij}}}{{[1-e^{-(\alpha_i+\beta_j)}]}^{-1}}.
\label{eq:bobbo}
\end{eqnarray}
As in the case of weighted matrices under global and one-sided local constraints, each entry $g_{ij}$ of the matrix $\mathbf{G}$ is an independent geometrically distributed random variable defined by the probability 
\begin{equation}\label{prob:entry:weighted:two}
  p(g_{ij}|\vec{\alpha}, \vec{\beta})=e^{-(\alpha_i+\beta_j)g_{ij}}[1-e^{-(\alpha_i+\beta_j)}],
\end{equation}
which is now controlled by the entry-specific pair of parameters $\alpha_i,\beta_j$. 
The expected value of $g_{ij}$ is
\begin{equation}\label{Covariance:weighted;two}
\langle g_{ij}\rangle_{\vec{\alpha}, \vec{\beta}}\equiv\sum_{g_{ij}=0}^{+\infty}g_{ij}p(g_{ij}|\vec{\alpha}, \vec{\beta})=\frac{e^{-(\alpha_i+\beta_j)}}{{1-e^{-(\alpha_i+\beta_j)}}}
\end{equation}
and the variance is
\begin{eqnarray}
  \textrm{Var}_{\vec{\alpha}, \vec{\beta}}[g_{ij}]&\equiv&\langle g_{ij}^2\rangle_{\vec{\alpha}, \vec{\beta}}-\langle g_{ij}\rangle_{\vec{\alpha}, \vec{\beta}}^2\nonumber\\
&=&\frac{e^{-(\alpha_i+\beta_j)}}{[1-e^{-(\alpha_i+\beta_j)}]^2}.\label{Variance:weighted:two}
\end{eqnarray}
The expected values of the constraints are
\begin{eqnarray}
\langle r_i\rangle_{\vec{\alpha}, \vec{\beta}}&=&\sum_{j=1}^{m}\frac{e^{-(\alpha_i+\beta_j)}}{1-e^{-(\alpha_i+\beta_j)}},\quad i=1,n,\\
\langle c_j\rangle_{\vec{\alpha}, \vec{\beta}}&=&\sum_{i=1}^{n}\frac{e^{-(\alpha_i+\beta_j)}}{1-e^{-(\alpha_i+\beta_j)}},\quad j=1,m.
\end{eqnarray}
As in the two-sided binary case, the values $(\vec{\alpha}^*, \vec{\beta}^*)$ maximizing the likelihood cannot be determined analytically as a function of the empirical values $(\vec{r}^*,\vec{c}^*)$, but they are defined implicitly by imposing the equality 
\begin{equation}
(\vec{r}^*,\vec{c}^*)=(\langle\vec{r}\rangle_{\vec{\alpha}^*, \vec{\beta}^*},\langle\vec{c}\rangle_{\vec{\alpha}^*, \vec{\beta}^*})
\end{equation}
between the empirical and the expected values of the constraints. This equality leads to Eqs.~\eqref{eq:2side1w} and~\eqref{eq:2side2w} in the main text.

\subsection{Determinant of the covariance matrix for two-sided local constraints\label{sec:determinant}}
The covariance matrix $(\mathbf{\Sigma}^*)^\pm$ in binary ($+$) and weighted ($-$) ensembles of matrices under two-sided local constraints is an $(n+m)\times(n+m)$ matrix. 
It contains all covariances among the $n$ row sums, all covariances among the $m$ column sums, and all the covariances between row and column sums.
If we order the constraints by considering first the $n$ row sums $\vec{r}^*$ and then the $m$ column sums $\vec{c}^*$ into the $(n+m)$-dimensional vector $\vec{C}^*=(\vec{r}^*,\vec{c}^*)$, and combine Eqs.~\eqref{eq:Z+} and~\eqref{eq:Z-} into the general partition function
\begin{equation}
Z^\pm(\vec{\alpha},\vec{\beta})
=\prod_{i=1}^{n}\prod_{j=1}^{m}[1\pm e^{-(\alpha_i+\beta_j)}]^{\pm 1}
\end{equation} 
valid for binary ($+$) and weighted ($-$) matrices, we can determine the entries of $\mathbf{\Sigma}^\pm$ by applying the definition in Eq.~\eqref{Sigma_unit}.
This yields
\begin{eqnarray}\label{eq:curlyapp1}
	{\Sigma}^\pm_{ij}&=&\left\{\begin{array}{ll}\displaystyle{\frac{\partial^2\ln Z^\pm(\vec{\alpha},\vec{\beta})}{{\partial{\alpha}_i}{\partial{\alpha}_j}}}&i,j\in[1,n],\\
\displaystyle{\frac{\partial^2\ln Z^\pm(\vec{\alpha},\vec{\beta})}{{\partial{\alpha}_i}{\partial{\beta}_{j-n}}}}&i\in[1,n],\,j\in[n+1,n+m]\\
\displaystyle{\frac{\partial^2\ln Z^\pm(\vec{\alpha},\vec{\beta})}{{\partial{\alpha}_{i-n}}{\partial{\beta}_{j}}}}&i\in[n+1,n+m],\,j\in[1,n]\\
\displaystyle{\frac{\partial^2\ln Z^\pm(\vec{\alpha},\vec{\beta})}{{\partial{\beta}_{i-n}}{\partial{\beta}_{j-n}}}}&i,j\in[n+1,n+m]
\end{array}\right.\nonumber\\
&=&\left\{\begin{array}{ll}\displaystyle{\textrm{Cov}^\pm_{\vec{\alpha},\vec{\beta}}[r_i,r_j]}&i,j\in[1,n],\\
\displaystyle{\textrm{Cov}^\pm_{\vec{\alpha},\vec{\beta}}[r_i,c_{j-n}]}&i\in[1,n],\,j\in[n+1,n+m]\\
\displaystyle{\textrm{Cov}^\pm_{\vec{\alpha},\vec{\beta}}[c_{i-n},r_j]}&i\in[n+1,n+m],\,j\in[1,n]\\
\displaystyle{\textrm{Cov}^\pm_{\vec{\alpha},\vec{\beta}}[c_{i-n},c_{j-n}]}&i,j\in[n+1,n+m]
\end{array}\right.\nonumber
\end{eqnarray}
and, following Eq.~\eqref{eq:whatstarmeans},
\begin{equation}\label{eq:whatstarmeans2side}
	(\Sigma^*_{ij})^\pm=\left.(\Sigma_{ij})^\pm\right|_{(\vec{\alpha}, \vec{\beta})=(\vec{\alpha}^*, \vec{\beta}^*)}.
\end{equation}

It is easy to see that $(\mathbf{\Sigma}^*)^\pm$ is a combination of four blocks
\begin{equation}
	(\mathbf{\Sigma}^*)^\pm=\begin{bmatrix}
		(\mathbf{A}^*)^\pm & (\mathbf{B}^*)^\pm\\ 
		(\mathbf{C}^*)^\pm & (\mathbf{D}^*)^\pm
	  \end{bmatrix},
\label{eq:block}
\end{equation}
where each block has entries as described below. What determines these entries is the elements $g_{ij}$ of the (binary or weighted) adjacency matrix $\mathbf{G}$ that different constraints have in common. The covariance between contraints that have no $g_{ij}$ in common is zero (as such constraints are independent), while the covariance between constraints that share a term $g_{ij}$ receives from that term a contribution equal to
\begin{equation}
  \textrm{Var}^\pm_{\vec{\alpha}^*, \vec{\beta}^*}[g_{ij}]=\frac{e^{-(\alpha^*_i+\beta^*_j)}}{[1\pm e^{-(\alpha^*_i+\beta^*_j)}]^2},
\label{Variance:both:two}
\end{equation}
obtained combining Eqs.~\eqref{Variance:binary:two} and~\eqref{Variance:weighted:two}.
\begin{itemize}
\item
Block $(\mathbf{A}^*)^\pm$ is the $n\times n$ covariance matrix between the row sums, with entries
\begin{eqnarray}
(A_{ij}^*)^{\pm}&=&{\textrm{Cov}^\pm_{\vec{\alpha}^*,\vec{\beta}^*}[r_i,r_j]}\nonumber\\
&=&\left.{\frac{\partial^2\ln Z^\pm(\vec{\alpha},\vec{\beta})}{{\partial{\alpha}_i}{\partial{\alpha}_j}}}\right|_{(\vec{\alpha}, \vec{\beta})=(\vec{\alpha}^*, \vec{\beta}^*)}\nonumber\\
&=&\delta_{ij}\sum_{k=1}^m\frac{e^{-(\alpha^*_i+\beta^*_k)}}{\left[1\pm e^{-(\alpha^*_i+\beta^*_k)}\right]^2}.
\label{a-ji}
\end{eqnarray}
Note that $(\mathbf{A}^*)^\pm$ is a diagonal matrix, since different row sums are all independent.
\item
Block $(\mathbf{B}^*)^\pm$ is the $n\times m$ matrix of covariances between row sums and column sums, with entries
\begin{eqnarray}
(B_{ij}^*)^{\pm}&=&{\textrm{Cov}^\pm_{\vec{\alpha}^*,\vec{\beta}^*}[r_i,c_j]}\nonumber\\
&=&\left.{\frac{\partial^2\ln Z^\pm(\vec{\alpha},\vec{\beta})}{{\partial{\alpha}_i}{\partial{\beta}_{j}}}}\right|_{(\vec{\alpha}, \vec{\beta})=(\vec{\alpha}^*, \vec{\beta}^*)}\nonumber\\
&=&\frac{e^{-(\alpha^*_i+\beta^*_{j})}}{\left[1\pm e^{-(\alpha^*_i+\beta^*_{j})}\right]^2},
\label{b-ji}
\end{eqnarray}
where we now see that the matrix is not diagonal, as reach row sum $r_i$ shares the entry $g_{ij}$ with the column sum $c_j$. 
\item
Similarly, block $(\mathbf{C}^*)^\pm$ is the $m\times n$ matrix of covariances between column sums and row sums, and is therefore the transpose of $(\mathbf{B}^*)^\pm$, as follows also from the fact that $(\mathbf{\Sigma}^*)^\pm$ must be symmetric. Indeed its entries are
\begin{eqnarray}
(C_{ij}^*)^{\pm}&=&{\textrm{Cov}^\pm_{\vec{\alpha}^*,\vec{\beta}^*}[c_i,r_j]}\nonumber\\
&=&\left.{\frac{\partial^2\ln Z^\pm(\vec{\alpha},\vec{\beta})}{{\partial{\alpha}_j}{\partial{\beta}_{i}}}}\right|_{(\vec{\alpha}, \vec{\beta})=(\vec{\alpha}^*, \vec{\beta}^*)}\nonumber\\
&=&\frac{e^{-(\alpha^*_j+\beta^*_{i})}}{\left[1\pm e^{-(\alpha^*_j+\beta^*_{i})}\right]^2}.
\label{c-ji}
\end{eqnarray}
\item
Finally, block $(\mathbf{D}^*)^\pm$ is the $m\times m$ matrix of covariances among the column sums, with entries
\begin{eqnarray}
(D_{ij}^*)^{\pm}&=&{\textrm{Cov}^\pm_{\vec{\alpha}^*,\vec{\beta}^*}[c_i,c_j]}\nonumber\\
&=&\left.{\frac{\partial^2\ln Z^\pm(\vec{\alpha},\vec{\beta})}{{\partial{\beta}_i}{\partial{\beta}_j}}}\right|_{(\vec{\alpha}, \vec{\beta})=(\vec{\alpha}^*, \vec{\beta}^*)}\nonumber\\
&=&\delta_{ij}\sum_{k=1}^n\frac{e^{-(\alpha^*_k+\beta^*_{j})}}{\left[1\pm e^{-(\alpha^*_k+\beta^*_{j})}\right]^2}.
\label{d-ji}
\end{eqnarray}
Like $(\mathbf{A}^*)^\pm$, $(\mathbf{D}^*)^\pm$ is a diagonal matrix, since different column sums are all independent.
\end{itemize}
Combining Eqs.~\eqref{eq:block},~\eqref{a-ji},~\eqref{b-ji},~\eqref{c-ji} and~\eqref{d-ji} proves Eq.~\eqref{eq:curlymain1} in the main text.

Now, in order to calculate the scaling of the determinant of $(\mathbf{\Sigma}^*)^\pm$, we follow the definition by Leibniz as
\begin{equation}
	\det[(\mathbf{\Sigma}^*)^\pm]=\sum_{{\sigma}\in \mathbf{Z}_{n+m}}\textrm{sgn}(\sigma)\prod_{l=1}^{n+m}(\Sigma^*_{l,\sigma_l})^{\pm},
\label{eq:Leibniz}
\end{equation}
where $\sigma$ is a permutation of the first $n+m$ integers that exchanges (without replacement) each of these integers $i$ with another such integer $j=\sigma_i$, $\mathbf{Z}_{n+m}$ is the set of all such $(n+m)!$ permutations, and the symbol $\textrm{sgn}(\sigma)$ represents the \emph{parity} of $\sigma$: $\textrm{sgn}(\sigma)=+1$ when $\sigma$ is an even permutation (i.e. obtained by combining an even number of pairwise exchanges of the type $j=\sigma_i$ and $i=\sigma_j$) and $\textrm{sgn}(\sigma)=-1$ when $\sigma$ is an odd permutation (i.e. obtained by combining an odd number of pairwise exchanges).
Let us call $\sigma^0$ the identity permutation, i.e. the one such that $\sigma^0_i=i$ for all $i$, and $\mathbf{Z}^0_{n+m}\equiv\mathbf{Z}_{n+m}\backslash\sigma^0$ the set of all other permutations. Clearly, $\textrm{sgn}(\sigma^0)=+1$ because $\sigma^0$ involves an even number (zero) of exchanges. We can therefore rewrite Eq.~\eqref{eq:Leibniz} as
\begin{equation}
\det[(\mathbf{\Sigma}^*)^\pm]=\Delta^0+\Delta'
\label{eq:delta01}
\end{equation}
where
\begin{equation}
\Delta^0=\prod_{l=1}^{n+m}(\Sigma^*_{l,\sigma^0_l})^{\pm}=
\prod_{i=1}^n(A^*_{ii})^{\pm}\prod_{j=1}^m(D^*_{jj})^{\pm}
\label{eq:delta0}
\end{equation} 
is the product of the diagonal entries of $(\mathbf{\Sigma}^*)^\pm$ and
\begin{equation}
\Delta'=\sum_{{\sigma}\in \mathbf{Z}^0_{n+m}}\textrm{sgn}(\sigma)\prod_{l=1}^{n+m}(\Sigma^*_{l,\sigma_l})^{\pm}.
\label{eq:delta1}
\end{equation}

We are going to show that $\Delta'$ is at most of the same order of $\Delta^0$. Setting $c_n=1/n$ in the sparse regime (for which we recall that $m=O(n)$ necessarily) and $c_n=1$ in the dense regime (for which $m$ can be either finite or $O(n)$), we note that each entry of the 
blocks $(\mathbf{B}^*)^\pm$ and $(\mathbf{C}^*)^\pm$ is of order $O(c_n)$, while each of the diagonal entries of block $(\mathbf{A}^*)^\pm$ is of order $O(c_n m)$ and each of the diagonal entries of block $(\mathbf{D}^*)^\pm$ is of order $O(c_n n)$. In general, the order of $\Delta^0$ is therefore 
\begin{equation}
\Delta^0=O\left((c_n n)^m (c_n m)^n\right)
\label{eq:delta0O}
\end{equation} 
as clear from Eq.~\eqref{eq:delta0}.

\begin{figure}
	\centering
	\includegraphics[width=8cm]{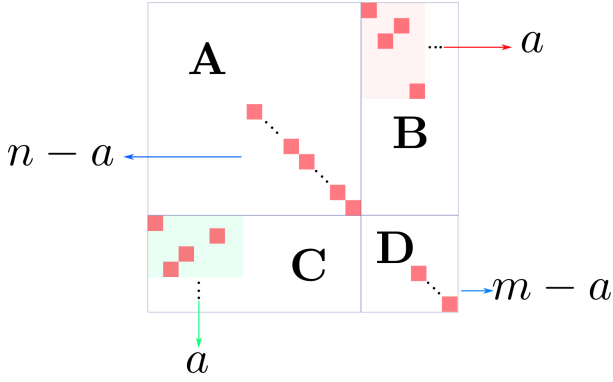}
	\caption{Illustration of the permutations producing the non-zero contributions to the determinant of the covariance matrix, as described in the text.}
	\label{Permutations_sigma}
\end{figure}

To this end, we note that each permutation $\sigma$ appearing in Eq.~\eqref{eq:delta1} can be expressed as a combination of a certain number (say $a>0$) of exchanges of pairs of the first $n+m$ integers. It is easy to see that all the ${n\choose 2}$ exchanges of pairs of the first $n$ integers give a zero contribution to $\Delta'$, because they lead to terms of the type $(\Sigma^*_{i,j})^\pm=0$ where $i,j\in[0,n]$ with $i\ne j$ (combination of exchanges that lead again to $i=j$ are such that $j=\sigma_i=i$ and therefore do not lead to new permutations: they are already accounted for in permutations with lower $a$). Similarly, all the ${m\choose 2}$ exchanges of pairs of the next $m$ integers give a zero contribution to $\Delta'$, because they lead to terms of the type $(\Sigma^*_{i,j})^\pm=0$ where $i,j\in[n+1,n+m]$ with $i\ne j$.
Therefore the only exchanges leading to nonzero contributions to $\Delta'$ are the $nm$ exchanges across the first $n$ integers and the next $m$ integers, i.e. those that lead to terms $(\Sigma^*_{i,j})^\pm>0$ where $i\in [0,n]$ and $j\in[n+1,n+m]$ or $j\in [0,n]$ and $i\in[n+1,n+m]$ in Eq.~\eqref{eq:delta1}. Each of these nontrivial contributing permutations involves $a$ (unrepeated) exchanges of integers, where $a\in[1,nm]$.
Compared with the identity $\sigma_0$, each of these permutations replaces $a$ of the first $n$ diagonal entries and $a$ of the next $m$ diagonal entries of $(\mathbf{\Sigma}^*)^\pm$ appearing in Eq.~\eqref{eq:delta0} with a number $2a$ of non-zero off-diagonal entries in blocks $(\mathbf{B}^*)^\pm$ and $(\mathbf{C}^*)^\pm$ (see Fig.~\ref{Permutations_sigma}).

Each such permutation therefore gives a contribution of order $(c_n n)^{m-a}(c_n m)^{n-a}$ to the summation in Eq.~\eqref{eq:delta1}. Individually, each such contribution is subleading with respect to the term $\Delta^0$. However, collectively all the contributions involving the same number $a$ of exchanges contribute a term of order $E_a (c_n n)^{m-a}(c_n m)^{n-a}$ where $E_a$ is the number of unrepeated exchanges of $a$ pairs. The order of $E_a$ is given by the number of distinct choices of $a$ exchanges out of the $nm$ possible ones, which is ${nm\choose a}$. This estimate does not control for the fact that, for each $a$, some of the exchanges reduce to simpler permutations already accounted for by smaller values of $a$, however the leading order is correct. 
Since $nm$ is large, we can apply Stirling's approximation to $(nm)!$ and estimate the order of $E_a$ as
\begin{equation}
E_a=O\left({nm\choose a}\right)=O\left(\frac{(nm)^a}{a!}\right)
\end{equation}
as in Eq.~\eqref{eq:ammazzateo}.
Therefore all the permutations realized by $a$ exchanges collectively give a contribution of order 
\begin{equation}
E_a(c_n n)^{m-a}(c_n m)^{n-a}=O\left(\frac{(c_n n)^m (c_n m)^n}{a!}\right)
\end{equation}
and sign $(-1)^a$ to the sum in Eq.~\eqref{eq:delta1}, so $\Delta'$ can be rewritten as a sum over $a$ (with $a=1,nm$) of terms of alternating sign. 
Qualitatively, and with an abuse of notation, the order of the entire sum defining $\det[(\mathbf{\Sigma}^*)^\pm]$ in Eq.~\eqref{eq:Leibniz} is (except for accidental cancellations due to particular combinations of values of the entries of $(\mathbf{\Sigma}^*)^\pm$) 
\begin{eqnarray}
O(\det[(\mathbf{\Sigma}^*)^\pm])&=&\sum_{a=1}^{nm}O\left(\frac{(-1)^a (c_n n)^m (c_n m)^n}{a!}\right)\nonumber\\
&=&O\left( (c_n n)^m (c_n m)^n e^{-1}\right)\nonumber\\
&=&O\left( (c_n n)^m (c_n m)^n \right).
\end{eqnarray}

We therefore see that the order of $\Delta'$ does not exceed that of $\Delta^0$, so the leading order of $\det[(\mathbf{\Sigma}^*)^\pm]$ is 
\begin{equation}
\det[(\mathbf{\Sigma}^*)^\pm]=O(\Delta^0)=O\left( (c_n n)^m (c_n m)^n \right).
\label{eq:sigmasigma}
\end{equation}
In other words, the off-diagonal terms of $(\mathbf{\Sigma}^*)^\pm$ do not alter the order obtained by multiplying the diagonal terms. 
For finite $m$, we therefore have
\begin{eqnarray}
{\alpha}_n^\pm&=&\ln\sqrt{\det\left[2\pi(\mathbf{\Sigma}^*)^\pm\right]}\nonumber\\
&=&O\left(m\ln(c_n n)+n\ln(c_n m)\right).
\label{eq:sobonituttino}
\end{eqnarray}
In the sparse case where $c_n=1/n$ and $m=O(n)$, we have
\begin{equation}
{\alpha}_n^\pm=O\left(n\right),
\end{equation}
while in the dense case with $c_n=1$ and $m=O(n)$ we have
\begin{equation}
{\alpha}_n^\pm=O\left(n\ln n\right),
\end{equation}
and finally in the dense case with $c_n=1$ and finite $m$ we have
\begin{equation}
{\alpha}_n^\pm=O\left(n\right),
\end{equation}
confirming the same scalings for the relative entropy obtained in Eqs.~\eqref{Rebothonesidesparse},~\eqref{eq:relativeonesidedbothapprox} and~\eqref{Rebothonesidedense} for the one-sided case.

\end{document}